\documentclass{SPM}
\usepackage{DHMmacros}
\usepackage{SPMcontents}

\begin{document}
\fontsize{11pt}{13.5pt}\selectfont

\title{\fontsize{12pt}{14.5pt}\selectfont
TASI 2011 lectures notes: two-component fermion notation\\ 
and supersymmetry}

\author{\fontsize{11pt}{13pt}\selectfont Stephen P. Martin}

\address{\large Physics Department\\ Northern Illinois University\\
DeKalb IL 60115, USA
%\\ $^*$E-mail: spmartin@niu.edu
}

\begin{abstract}\normalsize
\fontsize{11pt}{13pt}\selectfont
These notes, based on work with Herbi Dreiner and Howie Haber, 
discuss how to do practical calculations of cross sections 
and decay rates using two-component fermion notation, as appropriate for
supersymmetry and other beyond-the-Standard-Model theories.
Included are a list of two-component fermion
Feynman rules for the Minimal Supersymmetric Standard Model, 
and some example calculations.
\end{abstract}

\vspace{-0.3in}

\fontsize{12pt}{13.54pt}\selectfont

\tableofcontents\clearpage
\bodymatter

\fontsize{12pt}{15pt}\selectfont

\section{Introduction}\label{sec:introduction}
\renewcommand{\theequation}{\arabic{section}.\arabic{equation}}
\renewcommand{\thefigure}{\arabic{section}.\arabic{figure}}
\renewcommand{\thetable}{\arabic{section}.\arabic{table}}

For nearly the past four decades, the imaginations of particle theorists 
have been running unchecked, producing many clever ideas for what might 
be beyond the Standard Model of particle physics. Now the LHC is 
confronting these ideas. The era of ``clever" in model-building may soon 
come to an end, in favor of a new era where the emphasis is more on 
``true".

At TASI 2011 in Boulder, I talked about supersymmetry, which is probably the most popular 
class of new physics models. These are nominally 
the notes for those lectures. However, there are already many superb 
reviews on supersymmetry from diverse points of view 
[\refcite{HaberKane}]-[\refcite{DSB}]. Besides, everything I covered in 
Boulder is already presented in much more depth in my own attempt at an 
introduction to supersymmetry, {\it A supersymmetry primer} 
[\refcite{primer}], which was updated just after TASI, in September 2011. 
That latest version includes a new section on superspace and 
superfields. I'm not clever enough to think of better ways to say the 
same things, so perhaps you could just download that and read it instead. 
Then, when you are done, I'll continue with a complementary topic that I 
could have talked about at TASI 2011, but didn't for lack of time. So 
please go ahead and read the Primer now, as a prerequisite for the following. Take your 
time, I'll wait.

\begin{center} 
*~~~*~~~*
\end{center}
Good, you're back!

One of the most fundamental observations about physics at the weak scale 
is that it is chiral; the left-handed and right-handed components of 
fermions are logically distinct objects that have different gauge 
transformation properties. Despite this, textbooks on quantum field 
theory usually present calculations, such as cross-sections, decay rates, 
anomalies, and self-energy corrections, in the 4-component fermion 
language. One often hears that even though 2-component fermion language 
is better for devising many theories, including supersymmetry, it is 
somehow not practical for real calculations of physical observables. 
Herbi Dreiner, Howie Haber, and I decided to confront this erroneous 
notion by working out a complete formalism for doing such practical 
calculations, treating Dirac, Majorana and Weyl fermions in a unified 
way. The result went into a rather voluminous report [\refcite{DHM}]. In the 
rest of these notes, I will try to introduce our formalism in a more 
concise form, leaving out derivations and 
giving just enough examples to illustrate the main ideas.

\section{Notations and conventions}\label{sec:twocomp}
\renewcommand{\theequation}{\arabic{section}.\arabic{subsection}.\arabic{equation}}
\renewcommand{\thefigure}{\arabic{section}.\arabic{subsection}.\arabic{figure}}
\renewcommand{\thetable}{\arabic{section}.\arabic{subsection}.\arabic{table}}

\subsection{Two-component spinors}
\label{subsec:spinors}
\setcounter{equation}{0}
\setcounter{figure}{0}
\setcounter{table}{0}

First, the terrible issue of the sign of the metric. I use the correct 
(mostly plus sign) metric. Herbi and Howie both use the wrong metric (the 
one with mostly minus signs), but they won a 2-1 vote on which to use in 
the journal and arXiv versions of ref.~[\refcite{DHM}]. So far, the 
United States Supreme Court has declined to intervene on my behalf. 
However, we did devise a LaTeX macro to convert the sign of the metric, 
so there is an otherwise identical version of ref.~[\refcite{DHM}], with 
my metric, which you can get from the web page linked to in its arXiv 
abstract page. Conversely, if you don't like my metric sign, you can 
download a version of {\em these} notes from a web page linked to in the 
comments section of the arXiv abstract page. You can tell which version 
you are presently reading from this:
\beq
\label{signofmetric}
g_{\mu\nu}=
g^{\mu\nu}={\rm diag}(\BDplus 1 , \BDminus 1, \BDminus 1, \BDminus 1).
\eeq
Here $\mu, \nu= 0,1,2,3$ are spacetime vector indices.

Contravariant four-vectors (e.g. positions and momenta) are defined with
raised indices, and covariant four-vectors (e.g. derivatives) with
lowered indices:
\beqa
&&x^\mu = (t\,;\,\mathbold{\vec x}),\quad\qquad
p^\mu = (E\,;\,\mathbold{\vec p}),
\\
&&
\partial_\mu \equiv\frac{\partial}{\partial x^\mu}
= (\partial/\partial t\,;\,\mathbold{\vec \nabla})\,,
\eeqa
in units with $c=1$.  The totally antisymmetric pseudo-tensor
$\eps^{\mu\nu\rho\sigma}$ is defined such that
\beq
\eps^{0123}=-\eps\ls{0123}=+1\,.
\eeq

Two-component fermions transform in either the $(\half,0)$ (left-handed) 
or $(0, \half)$ (right-handed) spinor representations of the Lorentz 
group. By convention, the $(\half,0)$ rep carries an undotted spinor 
index $\alpha, \beta, \ldots$, and the $(0, \half)$ rep carries a dotted 
index $\dot\alpha, \dot\beta, \ldots$, each running from 1 to 2. If 
$\psi_\alpha$ is a left-handed Weyl spinor, then the Hermitian conjugate
\beq
\psi^\dagger_{\dot{\alpha}}\equiv (\psi_\alpha)^\dagger .
\label{eq:defbardagger}
\eeq
is a right-handed Weyl spinor. Therefore, any particular fermionic degrees of 
freedom
can be described equally well using a left-handed Weyl spinor or a right-handed 
one. 
By convention, the names of spinors are chosen so that right-handed spinors 
always carry daggers, and left-handed spinors do not. 
The spinor indices are raised and lowered using the 2-component antisymmetric
object 
\beq \label{epssign}
\eps^{12} = - \eps^{21} = \eps_{21} =
-\epsilon_{12} = 1\,,
\eeq
as follows:
\beq \label{epsalphabeta}
\psi_\alpha =
\epsilon_{\alpha\beta} \psi^\beta\,, \>\>\>\quad
\psi^\alpha =\epsilon^{\alpha\beta} \psi_\beta\,, \>\>\>\quad
\psi^\dagger_{\dot{\alpha}} = \epsilon_{\dot{\alpha}\dot{\beta}}
\psi^{\dagger\dot{\beta}}\,, \>\>\>\quad
\psi^{\dagger\dot{\alpha}} = \epsilon^{\dot{\alpha}\dot{\beta}}
\psi^\dagger_{\dot{\beta}}\,,
\eeq
with repeated indices summed over,
and 
\beq \label{eq:defbardagger2}
\psi^{\dagger\,\dot\alpha}\equiv(\psi^\alpha)^\dagger\,.
\eeq
When constructing Lorentz tensors from fermion fields, the heights of
spinor indices must be consistent in the sense that lowered
indices must only be contracted with raised indices.

To make contact with the (perhaps more familiar) 4-component language,
a Dirac spinor consists of two independent 2-component Weyl spinors, 
united into a 4-component object:
\beq
\Psi_{D} = \begin{pmatrix} \xi_\alpha \cr \chi^{\dagger{\dot\alpha}}
\end{pmatrix},
\eeq
while a 4-component Majorana spinor has the same form, but with the
two spinors identified through Hermitian conjugation:
\beq
\Psi_{\rm M} = \begin{pmatrix} \psi_\alpha \cr \psi^{\dagger{\dot\alpha}}
\end{pmatrix}.
\eeq
The free Dirac and Majorana Lagrangians are:
\beqa
{\cal L}_{\rm Dirac} &=& i \overline\Psi_D \gamma^\mu \partial_\mu \Psi_D
- m \overline \Psi_D \Psi_D;
\\
{\cal L}_{\rm Majorana} &=& \frac{i}{2} \overline\Psi_M \gamma^\mu 
\partial_\mu \Psi_M
- \frac{1}{2} m \overline \Psi_M \Psi_M,
\eeqa
where
\beq
\overline \Psi_D = \Psi_D^\dagger 
\begin{pmatrix} 0\phantom{x.} & 1 \cr 1\phantom{x.} & 0\end{pmatrix} 
=
( \chi^\alpha\>\>\, \xi^\dagger_{\dot \alpha}),
\eeq
and similarly for the Majorana 4-component spinor:
\beq
\overline \Psi_M = \Psi_M^\dagger 
\begin{pmatrix} 0\phantom{x.} & 1 \cr 1\phantom{x.} & 0\end{pmatrix} 
=
( \psi^\alpha\>\>\, \psi^\dagger_{\dot \alpha}),
\eeq
and the $4\times 4$ gamma matrices can be represented by
\beqa
\gamma^\mu = \begin{pmatrix} 0\phantom{x.} & \sigma^\mu \cr
                             \sigmabar^\mu\phantom{x.} & 0 \end{pmatrix},
\eeqa
where
\beqa
\sigma^0 &=& \sigmabar^0 = \begin{pmatrix} 1\phantom{xx} & 0 \cr 0 
\phantom{xx}& 1\end{pmatrix}
,
\qquad\>\>\>\>\>\>\quad
\sigma^1 = -\sigmabar^1 = \begin{pmatrix} 0\phantom{xx} & 1 \cr 1\phantom{xx} & 0\end{pmatrix}
,
\\ 
\sigma^2 &=& -\sigmabar^2 = \begin{pmatrix} 0\phantom{xx} & -i 
\cr i \phantom{xx}& 0\end{pmatrix}
,
\qquad\quad
\sigma^3 = -\sigmabar^3 = \begin{pmatrix} 1\phantom{xx} & 0 \cr 0\phantom{xx} & -1\end{pmatrix}.
\phantom{xxxx}
\eeqa
In order to include chiral interactions for these fermions in the 4-component language, one must define
$P_L$ and $P_R$ projection operators:
\beqa
P_L \Psi_D = \begin{pmatrix} \xi_\alpha \cr 0 \end{pmatrix}, \qquad\quad
P_R \Psi_D = \begin{pmatrix} 0\cr \chi^{\dagger \dot\alpha} 
\end{pmatrix}.
\eeqa

In the 2-component language, the Dirac Lagrangian is
\beqa
{\cal L}_{\rm Dirac} &=& i \xi^{\dagger}_{\dot \alpha}
(\sigmabar^\mu)^{\dot\alpha\beta} \partial_\mu \xi_\beta
+ i \chi^\alpha (\sigma^\mu)_{\alpha\dot\beta}
\partial_\mu \chi^{\dagger\dot\beta} - 
m (\xi^\dagger_{\dot\alpha} \chi^{\dagger \dot\alpha}
+ \chi^\alpha \xi_\alpha).
\phantom{xxxxxx}
\eeqa
This establishes that the sigma matrices
$(\sigmabar^\mu)^{\dot\alpha\beta}$ and
$(\sigma^\mu)_{\alpha\dot\beta}$ have the spinor indices with heights as indicated.
It is traditional and very convenient to suppress these indices wherever possible using the
convention that 
\textit{descending} contracted undotted indices and
\textit{ascending} contracted dotted indices,
\beq
{}^\alpha{}_\alpha\qquad\qquad {\rm and} \qquad\qquad
{}_{\dot{\alpha}}{}^{\dot{\alpha}}\> ,
\label{suppressionrule}
\eeq
can be omitted. Thus, the Dirac Lagrangian becomes simply
\beqa
{\cal L}_{\rm Dirac} &=& i \xi^{\dagger}
\sigmabar^\mu \partial_\mu \xi
+ i \chi \sigma^\mu
\partial_\mu \chi^\dagger - 
m (\xi^\dagger\chi^{\dagger}
+ \chi \xi).
\eeqa
More generally, in the index-free notation:
\beqa
\xi\chi &\equiv & \xi^\alpha \chi_\alpha ,\label{xieta}
\qquad\qquad\qquad\>\>\>\>\>\>
\xi^\dagger \chi^\dagger \equiv 
\xi^\dagger_{\dot\alpha} \chi^{\dagger \dot \alpha}
,\label{xidetad}
\\
\xi^\dagger\sigmabar^\mu\chi &\equiv &  \xi^\dagger_{\dot{\alpha}}
\sigmabar^{\mu\dot{\alpha}\beta}\chi_\beta ,
\qquad\quad\qquad
\xi\sigma^\mu \chi^\dagger \equiv  \xi^{{\alpha}}
\sigma^{\mu}_{\alpha \dot \beta} \chi^{\dagger\dot \beta} . \label{xisogeta}
\eeqa
Note that it is
useful to regard spinors like $\psi^\alpha$  and $\psi^\dagger_{\dot \alpha}$
as row vectors, and
$\psi^{\dagger\,\dot\alpha}$ and $\psi_\alpha$ as column vectors.
As an exercise, you can now show that, with $\xi$ and $\chi$ anticommuting spinors,
\beqa
&&\xi \chi = \chi \xi,\qquad\qquad\qquad \xi^\dagger \chi^\dagger = 
\chi^\dagger\xi^\dagger,
\\
&&\xi^\dagger \sigmabar^\mu \chi = -\chi \sigma^\mu \xi^\dagger.
\eeqa
For example, the Dirac Lagrangian can be rewritten yet again as:
\beqa
{\cal L}_{\rm Dirac} &=& 
i \xi^{\dagger} \sigmabar^\mu \partial_\mu \xi
+ i \chi^\dagger \sigmabar^\mu \partial_\mu \chi - 
m (\xi \chi + \xi^\dagger\chi^{\dagger}),
\label{eq:DiracL}
\eeqa
after discarding a total derivative. The Majorana Lagrangian is 
similarly:
\beqa
{\cal L}_{\rm Majorana} &=& 
i \psi^{\dagger} \sigmabar^\mu \partial_\mu \psi
- \frac{1}{2} m (\psi \psi + \psi^\dagger\psi^{\dagger} ).
\label{eq:MajoranaL}
\eeqa

Now, any theory of spin-$1/2$ fermions can be written in the 2-component
fermion notation, with kinetic terms:
\beqa
{\cal L} &=& i \psi^{\dagger i} \sigmabar^\mu \partial_\mu \psi_i
- \frac{1}{2} (M^{ij} \psi_i \psi_j + {\rm c.c.}),
\eeqa
where $i$ is a flavor and/or gauge label and $M^{ij}$ is a mass matrix,
and ``c.c."~denotes complex conjugation for classical fields.
In general, it can be shown that 
a unitary rotation on the indices $i$ will put the mass matrix
into a form where the only non-zero entries are diagonal entries $\mu_i$ and $2\times 2$ blocks $\begin{pmatrix} 0 & m_j\cr m_j & 0 \end{pmatrix}$, with 
$\mu_i$ and $m_j$ all real and non-negative:
\beqa
{\cal L} &=& i \psi^{\dagger i} \sigmabar^\mu \partial_\mu \psi_i 
- \half \mu_i (\psi_i \psi_i + \psi^{\dagger i} \psi^{\dagger i} )
\nonumber \\
&&
+ i \xi^{\dagger j} \sigmabar^\mu \partial_\mu \xi_j
+ i \chi^{\dagger}_j \sigmabar^\mu \partial_\mu \chi^j
- m_j (\xi_j \chi^j + \xi^{\dagger j} \chi^{\dagger}_j),
\eeqa
The resulting theory consists of Majorana fermions $\psi_i$ 
(for which diagonal mass terms are allowed by the symmetries), and Dirac 
fermions consisting of the pairs $(\xi_j, \chi^j)$.
 
The behavior of the spinor products under hermitian
conjugation (for quantum field operators) or complex conjugation (for
classical fields) is:
\beqa
&&(\xi\chi)^\dagger=\chi^\dagger \xi^\dagger\,,
\label{eq:conbil}
\\
&&(\xi\sigma^\mu\chi^\dagger)^\dagger=\chi\sigma^\mu\xi^\dagger ,
\label{eq:conbilsig}
\\
&&(\xi^\dagger \sigmabar^\mu \chi)^\dagger = \chi^\dagger \sigmabar^\mu \xi ,
\label{eq:conbilsigbar}
\\
&&(\xi\sigma^\mu\sigmabar^\nu\chi)^\dagger
=
\chi^\dagger \sigmabar^\nu\sigma^\mu \xi^\dagger\,.
\eeqa
More generally,
\beq \label{eq:conbilgen}
(\xi \Sigma \chi)^\dagger = \chi^\dagger \reversed{\Sigma} \xi^\dagger\,,
\qquad\qquad
(\xi \Sigma \chi^\dagger)^\dagger = \chi \reversed{\Sigma} \xi^\dagger\,,
\eeq
where in each case $\Sigma$ stands for any sequence of alternating
$\sigma$ and $\sigmabar$ matrices, and $\reversed{\Sigma}$
is obtained from $\Sigma$ by reversing the order of all of the
$\sigma$ and $\sigmabar$ matrices, since the sigma matrices are hermitian.
\Eqst{eq:conbil}{eq:conbilgen}
are applicable both to anticommuting and to
commuting spinors.

The following identities can be used to systematically simplify
expressions involving products of $\sigma$ and $\sigmabar$
matrices:
\beqa
&&
\sigma^\mu_{\alpha\dot{\alpha}} \sigmabar_\mu^{\dot{\beta}\beta}
= \BDpos 2
\delta_{\alpha}{}^{\beta} \delta^{\dot{\beta}}{}_{\dot{\alpha}}\,,
\label{mainfierz}
\\
&&
\sigma^\mu_{\alpha\dot{\alpha}} \sigma_{\mu\beta\dot{\beta}}
= \BDpos 2
\epsilon_{\alpha\beta} \epsilon_{\dot{\alpha}\dot{\beta}}\,,
\label{mainfierz2}
\\
&&
\sigmabar^{\mu\dot{\alpha}\alpha} \sigmabar_\mu^{\dot{\beta}\beta}
= \BDpos 2
\epsilon^{\alpha\beta} \epsilon^{\dot{\alpha}\dot{\beta}}\,,
\label{mainfierz3}
\\
&&
{(\sigma^\mu\sigmabar^\nu + \sigma^\nu \sigmabar^\mu )_\alpha}^\beta
= \BDpos 2\metric^{\mu\nu} \delta_{\alpha}{}^{\beta}\,,
\label{eq:ssbarsym}
\\
&&(\sigmabar^\mu\sigma^\nu + \sigmabar^\nu \sigma^\mu
)^{\dot{\alpha}}{}_{\dot{\beta}}
= \BDpos 2\metric^{\mu\nu} \delta^{\dot{\alpha}}{}_{\dot{\beta}}\,,
\label{eq:sbarssym}
\\
&& \sigma^\mu \sigmabar^\nu \sigma^\rho =
\BDpos \metric^{\mu\nu} \sigma^\rho
\BDminus \metric^{\mu\rho} \sigma^\nu
\BDplus \metric^{\nu\rho} \sigma^\mu
\BDplus i \epsilon^{\mu\nu\rho\kappa} \sigma_\kappa\,,
\label{eq:simplifyssbars}
\\
&& \sigmabar^\mu \sigma^\nu \sigmabar^\rho =
\BDpos \metric^{\mu\nu} \sigmabar^\rho
\BDminus \metric^{\mu\rho} \sigmabar^\nu
\BDplus \metric^{\nu\rho} \sigmabar^\mu
\BDminus i \epsilon^{\mu\nu\rho\kappa} \sigmabar_\kappa\,.
\label{eq:simplifysbarssbar}
\eeqa
Computations of cross-sections and decay rates generally require
traces of alternating products of $\sigma$ and $\sigmabar$ matrices:
\beqa
&&{\rm Tr}[1] = 2\,,
\label{traceone}
\\
&&{\rm Tr}[\sigma^\mu \sigmabar^\nu ] =
{\rm Tr}[\sigmabar^\mu \sigma^\nu ] = \BDpos 2 \metric^{\mu\nu} \,,
\label{trssbar}
\\
&&{\rm Tr}[\sigma^\mu \sigmabar^\nu \sigma^\rho \sigmabar^\kappa ] =
2 \left ( \metric^{\mu\nu} \metric^{\rho\kappa} - \metric^{\mu\rho}
\metric^{\nu\kappa} + \metric^{\mu\kappa} \metric^{\nu\rho} + i
\epsilon^{\mu\nu\rho\kappa} \right )\,, \qquad\phantom{xx}
\label{trssbarssbar}
\\
&&{\rm Tr}[\sigmabar^\mu \sigma^\nu \sigmabar^\rho \sigma^\kappa ] =
2 \left ( \metric^{\mu\nu} \metric^{\rho\kappa} - \metric^{\mu\rho}
\metric^{\nu\kappa} + \metric^{\mu\kappa} \metric^{\nu\rho} - i
\epsilon^{\mu\nu\rho\kappa} \right )\,.
\label{trsbarssbars}
\eeqa
Traces involving a larger even number of $\sigma$ and $\sigmabar$
matrices can be systematically obtained from
\eqst{traceone}{trsbarssbars} by repeated use of
\eqs{eq:ssbarsym}{eq:sbarssym} [and, if you are lucky, 
eqs.~(\ref{mainfierz})-(\ref{mainfierz3})], 
and the cyclic property of the trace.
Traces involving an odd number of $\sigma$ and $\sigmabar$ matrices
cannot arise, because there is no way to connect the spinor indices
consistently.

In addition to manipulating expressions containing anticommuting
fermion quantum fields, one often must deal with products of {\it commuting}
spinors that arise as external state wave functions in the Feynman rules.
In the following expressions, a generic commuting or anticommuting 
spinor is denoted by by $z_i$, with the notation:
\beq
(-1)^A\equiv\left\{\begin{array}{ll} +1\,, & \textrm{commuting~spinors,}\\[6pt]
-1\,, & \textrm{anticommuting~spinors}.\end{array}\right.
\eeq
The following identities hold for the $z_i$:
\beqa
&&z_1 z_2 = -(-1)^A z_2 z_1\,, \label{zonetwo}
\\
&&z^\dagger_1 z^\dagger_2 = -(-1)^A z^\dagger_2 z^\dagger_1\,,
\label{barzonetwo}
\\
&&z_1 \sigma^\mu z^\dagger_2 = (-1)^A z^\dagger_2 \sigmabar^\mu z_1\,,
\label{europeanvacation} \\
&&z_1 \sigma^\mu \sigmabar^\nu z_2 =
-(-1)^A z_2 \sigma^\nu \sigmabar^\mu z_1\,,
 \label{zsmunuz}\\
&&z^\dagger_1 \sigmabar^\mu \sigma^\nu z^\dagger_2 =
-(-1)^A z^\dagger_2 \sigmabar^\nu \sigma^\mu z^\dagger_1\,,
 \label{zsbarmunuz}\\
&&z^\dagger_1 \sigmabar^\mu \sigma^\rho \sigmabar^\nu z_2=(-1)^A
z_2 \sigma^\nu \sigmabar^\rho \sigma^\mu z^\dagger_1.
\label{zsssmunuz}
\eeqa
The hermiticity properties of the spinor products
in \eqst{eq:conbil}{eq:conbilgen} hold for both commuting
and anticommuting spinors.

Two-component spinor products can often be simplified by using
Fierz identities.
Using the antisymmetry of the suppressed two-index
epsilon symbol, you can show:
\beqa
(z_1 z_2)(z_3 z_4) &=& -(z_1 z_3) (z_4 z_2) - (z_1 z_4)(z_2 z_3)\,,
\label{eq:twocompfierzone}
\\
(z^\dagger_1 z^\dagger_2)(z^\dagger_3 z^\dagger_4) &=&
- (z^\dagger_1 z^\dagger_3) (z^\dagger_4 z^\dagger_2)
- (z^\dagger_1 z^\dagger_4) (z^\dagger_2 z^\dagger_3)\,,
\label{eq:twocompfierztwo}
\eeqa
where \eqs{zonetwo}{barzonetwo} have been used to eliminate
any residual factors of $(-1)^A$. 
Additional Fierz identities follow from \eqst{mainfierz}{mainfierz3}:
\beqa
(z_1 \sigma^\mu z^\dagger_2)(z^\dagger_3 \sigmabar_\mu z_4)
&=& \BDneg 2 (z_1 z_4) (z^\dagger_2 z^\dagger_3)\,,
\label{twocompfierza} \\
(z^\dagger_1 \sigmabar^\mu z_2)(z^\dagger_3 \sigmabar_\mu z_4)
&=&\BDpos 2 (z^\dagger_1 z^\dagger_3) (z_4 z_2)\,,
\label{twocompfierzb}
\\
(z_1 \sigma^\mu z^\dagger_2)(z_3 \sigma_\mu z^\dagger_4)
&=& \BDpos 2 (z_1 z_3) (z^\dagger_4 z^\dagger_2)\,.
\label{twocompfierzc}
\eeqa
\Eqst{eq:twocompfierzone}{twocompfierzc}
hold for both commuting and anticommuting spinors.

The preceding identities hold in the case that the number of spacetime 
dimensions is exactly 4. This is appropriate for tree-level computations,
but in calculations of radiative corrections one often makes use of
regularization by dimensional continuation to $d$
dimensions, where $d$ is infinitesimally different from 4.
For non-supersymmetric theories, the most common method is the 
classic dimensional regularization method
[\refcite{MSbar}], while in supersymmetry one 
uses some version of dimensional reduction 
[\refcite{DRbar}]
in order to avoid spurious
violations of supersymmetry due to a mismatch between the gaugino and
gauge boson degrees of freedom.

When using dimensional continuation regulators, some identities that 
would hold in unregularized four-dimensional theories are simply 
inconsistent and must not be used; other identities remain valid if $d$ 
replaces 4 in the appropriate spots; and still other identities hold 
without modification. Some important identities that do hold in $d\not=4$
dimensions are eqs.~(\ref{eq:ssbarsym}) and (\ref{eq:sbarssym}), and the
trace identities eqs.~(\ref{traceone}) and
(\ref{trssbar}). 

In contrast, the Fierz 
identities 
eqs.~(\ref{mainfierz}), (\ref{mainfierz2}), and
(\ref{mainfierz3}), and
eqs.~(\ref{twocompfierza}), (\ref{twocompfierzb}), 
and (\ref{twocompfierzc}), do not have a consistent, 
unambiguous meaning for $d\not=4$. 
However, the following 
identities
that are implied by these equations in $d=4$ do consistently generalize 
to $d\not=4$:
\beqa
&& [\sigma^\mu\sigmabar_\mu ]_\alpha{}^\beta =
\BDpos d \delta_\alpha^\beta\,,
\label{eq:genfierzone}
\\
&&
[\sigmabar^\mu\sigma_\mu ]^{\dot{\alpha}}{}_{\dot{\beta}}
= \BDpos d \delta^{\dot{\alpha}}_{\dot{\beta}}\,.
\label{eq:genfierztwo}
\eeqa
Taking these and repeatedly using
\eqs{eq:ssbarsym}{eq:sbarssym} then yields:
\beqa
&&
[\sigma^\mu\sigmabar^\nu \sigma_\mu ]_{\alpha\dot\beta} =
\BDneg (d-2) \sigma^\nu_{\alpha\dot\beta}\,,
\label{eq:genfierzthree}
\\
&&
[\sigmabar^\mu \sigma_\nu \sigmabar_\mu ]^{\dot\alpha\beta} =
\BDneg (d-2) \sigmabar_\nu^{\dot\alpha\beta}\,,
\label{eq:genfierzfour}
\\
&&
[\sigma^\mu \sigmabar^\nu \sigma^\rho\sigmabar_\mu
]_\alpha{}^\beta
= 4 \metric^{\nu\rho} \delta_\alpha^\beta
\BDminus (4-d) [\sigma^\nu \sigmabar^\rho]_\alpha{}^\beta\,,
\\
&&
[\sigmabar^\mu \sigma^\nu \sigmabar^\rho\sigma_\mu
]^{\dot{\alpha}}{}_{\dot{\beta}}
= 4 \metric^{\nu\rho} \delta^{\dot{\alpha}}_{\dot{\beta}}
\BDminus (4-d) [\sigmabar^\nu \sigma^\rho]^{\dot\alpha}{}_{\dot\beta} \,,
\\
&&
[\sigma^\mu \sigmabar^\nu
\sigma^\rho\sigmabar^\kappa \sigma_\mu]_{\alpha\dot\beta}
=
\BDneg 2 [\sigma^\kappa \sigmabar^\rho \sigma^\nu]_{\alpha\dot\beta}
\BDplus (4-d)[\sigma^\nu \sigmabar^\rho \sigma^\kappa]_{\alpha\dot\beta}
\,,\\
&&
[\sigmabar^\mu \sigma^\nu
\sigmabar^\rho\sigma^\kappa \sigmabar_\mu]^{\dot\alpha\beta}
=
\BDneg 2 [\sigmabar^\kappa \sigma^\rho \sigmabar^\nu]^{\dot\alpha\beta}
\BDplus (4-d)[\sigmabar^\nu \sigma^\rho
\sigmabar^\kappa]^{\dot\alpha\beta}\,.
\eeqa

Identities that involve the (explicitly and
inextricably four-dimensional)
$\epsilon^{\mu\nu\rho\kappa}$ symbol, such as 
eqs.~(\ref{eq:simplifyssbars}), (\ref{eq:simplifysbarssbar}),
(\ref{trssbarssbar}), and
(\ref{trsbarssbars}), 
are also only meaningful in exactly four dimensions.
This can lead to ambiguities or inconsistencies in loop computations where it is
necessary to perform the computation in $d\neq 4$ dimensions;
see, for example,
refs.~[\refcite{Siegel:1980qs}]-[\refcite{Jack:1997sr}].
Fortunately, in practice one typically finds that the above expressions
appear multiplied by the metric and/or other external tensors, such as
linearly dependent four-momenta appropriate to the problem at hand.  
In almost all such cases,
two of the indices appearing in the traces
are symmetrized, which eliminates the $\epsilon^{\mu\nu\rho\kappa}$
term, rendering the resulting expressions unambiguous.
For example, one can write 
\beq
{\rm Tr}[\sigma^\mu \sigmabar^\nu \sigma^\rho \sigmabar^\kappa
]+{\rm Tr}[\sigmabar^\mu \sigma^\nu \sigmabar^\rho \sigma^\kappa ] =
4 \left ( \metric^{\mu\nu} \metric^{\rho\kappa} - \metric^{\mu\rho}
\metric^{\nu\kappa} + \metric^{\mu\kappa} \metric^{\nu\rho} \right ).
\label{symmtracessss}
\eeq
unambiguously in $d\neq 4$ dimensions. In other cases, one can separate 
a calculation into a divergent part without ambiguities plus a convergent
part which would be ambiguous in $d\not=4$, but which can be safely 
evaluated in $d=4$. Sometimes this requires first combining the contributions of several 
Feynman diagrams. This is the case in the triangle anomaly calculation using 2-component
fermions, discussed in depth in ref.~[\refcite{DHM}].

\subsection{Fermion interaction vertices}
\setcounter{equation}{0}
\setcounter{figure}{0}
\setcounter{table}{0}

\begin{figure}[t!]
\begin{center}
\begin{picture}(200,64)(15,20)
\DashArrowLine(10,40)(60,40)5 
\ArrowLine(100,70)(60,40)
\ArrowLine(100,10)(60,40)
\Text(18,30)[]{$I$} 
\Text(75,10)[]{$k, \beta$}
\Text(75,70)[]{$j, \alpha$}
\Text(128,40)[l]{$-i Y^{Ijk}\delta_\alpha{}^\beta \quad
{\rm or} \quad -i Y^{Ijk}\delta_\beta{}^\alpha$}
\Text(-40,40)[]{(a)}
\end{picture}
\end{center}
%%%%%%%%%%%%%%%%%%%%%%%%%%%%
\begin{center}
\begin{picture}(200,75)(15,20)
\DashArrowLine(60,40)(10,40)5
\ArrowLine(60,40)(100,70)
\ArrowLine(60,40)(100,10)
\Text(18,30)[]{$I$}
\Text(75,10)[]{$k, \dot\beta$}
\Text(75,70)[]{$j, \dot\alpha$}
\Text(128,40)[l]{$-i Y_{Ijk}\delta^{\dot\alpha}{}_{\dot\beta}
\quad {\rm or} \quad
-i Y_{Ijk}\delta^{\dot\beta}{}_{\dot\alpha}$}
\Text(-40,40)[]{(b)}
\end{picture}
\end{center}
\caption{\label{fig:Yukawavertexrules}
  {Feynman rules for Yukawa couplings of scalars to 2-component
    fermions in a general field theory.  The choice of which rule to
    use depends on how the vertex connects to the rest of the
    amplitude. When spinor indices are suppressed, the Kronecker $\delta$'s are
    trivial in either case, so we will not show the
    explicit spinor indices in specific realizations of this rule below.}}
\end{figure}
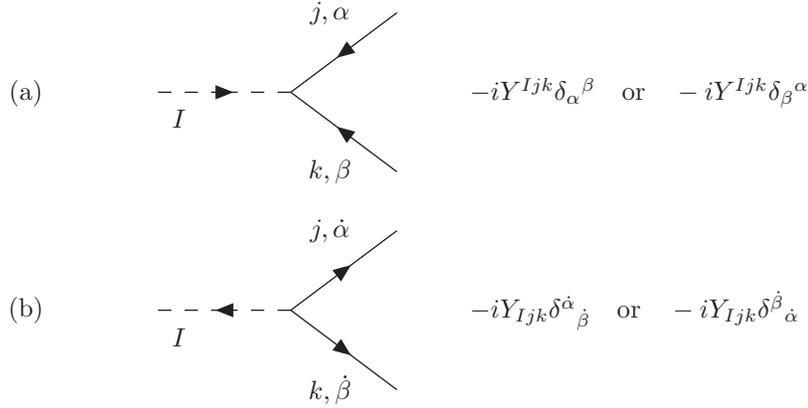

In renormalizable quantum field theories, fermions can interact either 
with scalars or with vector fields. In 2-component language, the 
scalar-fermion-fermion interactions can be written as:
\beq
{\cal L}_{\rm int} =
-\half Y^{Ijk} \phi_I \psi_j \psi_k
-\half Y_{Ijk} \phi^I \psi^{\dagger j} \psi^{\dagger k},
\eeq
where the fields are assumed to be mass eigenstates, and $Y^{Ijk}$ are 
dimensionless Yukawa
couplings with $Y_{Ijk} = \left (Y^{Ijk}\right )^*$. 
The 2-component fields $\psi_i$ make be 
either Majorana or parts of Dirac fermions, and
the spinor indices have been suppressed. 
The scalar fields
$\phi_I$ may be either real or complex, with $\phi^I \equiv (\phi_I)^*$.
The corresponding
Feynman rules are obtained as usual by multiplying the couplings in the 
Lagrangian by $i$, and are shown in Figure \ref{fig:Yukawavertexrules}.
%%%%%%%%%%%%%%%%%%%%%%%%%%%%
%%%%%%%%%%%%%%%%%%%%%%%%%%%%

In contrast to 4-component Feynman rules, the directions of arrows in 
2-component Feynman rules do {\it not} correspond to the flow of charge 
or fermion number. Instead, the arrows indicate the spinor index 
structure, with fields of undotted indices flowing into any vertex and 
fields of dotted indices flowing out of any vertex. This corresponds to 
the fact that the 2-component fields are distinguished by their Lorentz 
group transformation properties, rather than their status as particle or 
antiparticle as in 4-component notation. For the interactions of scalars 
in Figure \ref{fig:Yukawavertexrules}, the spinor indices are always just 
proportional to the identity matrix, and so can be trivially suppressed. 
For this reason, we will always just omit the spinor indices on 
scalar-fermion-fermion interaction Feynman rules in later sections.

\begin{figure}[tb!]
\begin{center}
\begin{picture}(200,55)(30,20)
\Photon(60,40)(10,40){3}{5}
\ArrowLine(60,40)(100,70)
\ArrowLine(100,10)(60,40)
\Text(24,25)[]{$a, \mu$}
\Text(70,18)[]{$j, \beta$}
\Text(70,66)[]{$i, \dot\alpha$}
\Text(123,40)[l]{
$\BDneg i (G^a)_i{}^{j}\, \sigmabar_\mu^{\dot{\alpha}\beta}$
\quad or \quad
$\BDpos i (G^a)_i{}^{j}\, \sigma_{\mu\beta\dot{\alpha}}$}
\end{picture}
\end{center}
%%%%%%%%%%%%%%%%%%%%%
\caption{\label{fig:Gaugevertexrules}
The Feynman rules for 2-component fermion interactions
with gauge bosons. Which one should be used
depends on how the vertex connects to the rest of the
diagram.  The $G^a$ are defined in \eq{eq:gaint}. In specific realizations of this
rule below, we will not explicitly show the spinor indices; it is understood that the
dotted index is associated with an outgoing line and the undotted with the incoming line.
Also, we will only show the $\sigmabar_\mu$ version of the rule; there is always 
another version of the rule with $\sigmabar_\mu \rightarrow -\sigma_\mu$.}
\end{figure}
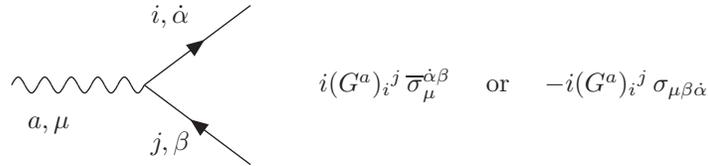

Next consider fermion interactions with vector fields.
The general form of the interactions of 2-component fermions with
vector bosons $A^\mu_a$ 
labeled by an index $a$ is:
\beq
{\cal L}_{\rm int} \,=\, \BDneg (G^a)_i{}^j A^\mu_a \psi^{\dagger i} 
\sigmabar_\mu \psi_j
.
\label{eq:gaint}
\eeq
Here $G^a$ is a dimensionless coupling matrix, which in the special case
that the fields are gauge eigenstates is given by 
$g_a \boldsymbol{T}^{\boldsymbol{a}}$, where $g_a$
and $\boldsymbol{T}^{\boldsymbol{a}}$ 
are the gauge coupling and fermion representation matrix of the 
theory. In general, the form of eq.~(\ref{eq:gaint}) is the result of 
diagonalizing
both the vector and fermion mass matrices. The corresponding
2-component fermion Feynman rules are shown in Figure 
\ref{fig:Gaugevertexrules}. 
%%%%%%%%%%%%%%%%%%%%%%%%%%%%%%%%%%%%%%%%%%%%%%%%%%%%%%%%%%%%%%%%%%%%%%
%%%%%%%%%%%%%%%%%%%%%%%%%%%%%%%%%%%%%%%%%%%%%%%%%%%%%%%%%%%%%%%%%%%%%%
Note that there are two different forms for the
Feynman rule, one proportional to $\sigmabar_\mu$ and the other to 
$-\sigma_\mu$.
Which one should be used depends 
on how the vertex is connect to the rest of the
amplitude; the spinor indices will connect in the only way possible. Below,
when presenting Feynman rules for specific vector-fermion-fermion interactions,
we will always have an outgoing arrow at the top (corresponding implicitly to
a dotted index $\dot\alpha$) and an incoming arrow at the bottom (for an undotted
index $\beta$), and simply write $\sigmabar_\mu$ rather than $\sigmabar_\mu^{\dot\alpha\beta}$,
and with the understanding that there is always a corresponding rule with 
$\sigmabar_\mu \rightarrow -\sigma_\mu$.
Thus the structure of each such Feynman rule will be 
exactly like Figure \ref{fig:Gaugevertexrules}, but with the indices suppressed for 
simplicity of presentation. 

\subsection{External wavefunctions for 2-component spinors}
\setcounter{equation}{0}
\setcounter{figure}{0}
\setcounter{table}{0}

\begin{figure}[t!]
\vspace{.5cm}
\begin{center}
\begin{picture}(300,150)(-150,-75)
\thicklines
\GCirc(0,0){21}{0.8}
\put(-60.5,42){$x$}
\ArrowLine(-57,57)(-14.3,14.3)
\ArrowLine(14.3,14.3)(57,57)
\put(51.5,40.5){$x^\dagger$}
\ArrowLine(-14.3,-14.3)(-57,-57)
\put(-60.5,-42){$y^\dagger$}
\ArrowLine(57,-57)(14.3,-14.3)
\put(51.5,-42){$y$}   
\put(-37,70){$L$~~($\half$, 0) fermion}
\put(-37,-77){$R$~~(0, $\half$) fermion}
\put (-138,0){Initial State}
\put (82,0){Final State}
\end{picture}
\vspace{.4cm}
\end{center}
\caption[0]{\label{fig:mnemonic} Mnemonic for the assignment of external wave function spinors, for initial and final state
left-handed \protect{$(\half,0)$} and right-handed
$(0,\half)$ fermions.}
\end{figure}
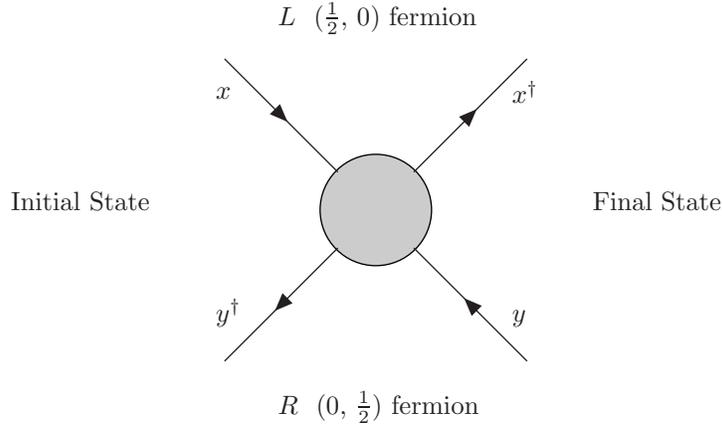

In the standard textbook calculations in 4-component spinor language,
one makes use of external wavefunction spinors $u$, $\bar v$, $\bar u$, $v$, 
for, respectively, initial state fermions, 
initial state anti-fermions,
final state fermions, and final state anti-fermions. Similarly, when doing
calculation in the 2-component formalism, 
one makes use of external wavefunction 2-component spinors:
\beq
x_\alpha(\vec{p}, s),\qquad  
y^\dagger_{\dot \alpha}(\vec{p}, s),\qquad
x^\dagger_{\dot\alpha}(\vec{p}, s),\qquad  
y_\alpha(\vec{p}, s), \qquad
\eeq
for, respectively, initial-state left-handed ($\half,0)$, 
initial-state right-handed ($0,\half$),
final-state left-handed ($\half,0$), and final-state right-handed
($0,\half$) states. See Figure \ref{fig:mnemonic} for a mnemonic.
%%%%%%%%%%%%%%%%%%%%%%%%%%%%
%%%%%%%%%%%%%%%%%%%%%%%%%%%%
These external wave function spinors 
are commuting (Grassmann-even) objects, despite carrying spinor indices, 
and are applicable to both Dirac and Majorana fermions.
They depend on the three-momentum
$\vec{p}$ and the spin $s$ of particle, and are related to the usual 
representation of the 4-component
$u$ and $v$ spinors by
\beq
u(\vec{p}, s) = \begin{pmatrix} x_\alpha (\vec{p}, s) \\[3pt] 
y^{\dagger\dot\alpha} (\vec{p}, s) \end{pmatrix},
\qquad
v(\vec{p}, s) = \begin{pmatrix} y_\alpha (\vec{p}, s) 
\\[3pt] x^{\dagger\dot\alpha} (\vec{p}, s)\end{pmatrix} .
\eeq
In the following, we will only
consider problems for which the spin states $s$ are summed over.
In that case, the explicit forms of the spinors $x,y,x^\dagger,y^\dagger$
are not needed.\footnote{See ref.~[\refcite{DHM}] for the explicit 
forms of $x,y,x^\dagger, y^\dagger$, and
examples with spins not summed.} 
Instead, one makes use of the spin-sum identities
\beqa
\sum_s x_\alpha({\boldsymbol{\vec p}},s)
x^\dagger_{\dot{\beta}}({\boldsymbol{\vec p}},s) =
\BDpos p\newcdot\sigma_{\alpha\dot{\beta}}
,
&\quad\>
&\sum_s x^{\dagger\dot{\alpha}}({\boldsymbol{\vec p}},s)
 x^{\beta}({\boldsymbol{\vec p}},s) =
\BDpos p\newcdot\sigmabar^{\dot{\alpha}\beta}
\!,\phantom{xxxxx}
\label{xxdagsummed}
\\
\sum_s y^{\dagger\dot{\alpha}}({\boldsymbol{\vec p}},s)
y^{\beta}({\boldsymbol{\vec p}},s)
= \BDpos p \newcdot \sigmabar^{\dot{\alpha}\beta}   
,
&\quad\>
&\sum_s y_\alpha({\boldsymbol{\vec p}},s)
y^\dagger_{\dot{\beta}}({\boldsymbol{\vec p}},s) =
\BDpos p \newcdot \sigma_{\alpha\dot{\beta}}
,
\label{yydagsummed}
\\
\sum_s x_\alpha({\boldsymbol{\vec p}},s)
y^\beta({\boldsymbol{\vec p}},s) = m \delta_\alpha{}^\beta
,
&\quad\>\>
&\sum_s y_\alpha({\boldsymbol{\vec p}},s)
x^\beta({\boldsymbol{\vec p}},s) = -m \delta_\alpha{}^\beta
,
\label{yxsummed}
\\
\sum_s y^{\dagger\dot{\alpha}}({\boldsymbol{\vec p}},s)
x^\dagger_{\dot{\beta}}({\boldsymbol{\vec p}},s) = m
\delta^{\dot{\alpha}}{}_{\dot{\beta}}
\,,
&\quad\>\>
&\sum_s x^{\dagger\dot{\alpha}}({\boldsymbol{\vec p}},s)
y^\dagger_{\dot{\beta}}({\boldsymbol{\vec p}},s) = - m
\delta^{\dot{\alpha}}{}_{\dot{\beta}}
,
\label{ydagxdagsummed}
\eeqa
where $m$ and $p^\mu$ are the mass and 4-momentum of the fermion.
They also obey useful reduction identities: 
\beqa
& (p\newcdot\sigmabar)^{\dot{\alpha}\beta} x_{\beta} =
\BDpos m y^{\dagger\dot{\alpha}} \, , \qquad\quad\qquad \,\,
&(p\newcdot\sigma)_{\alpha\dot{\beta}} y^{\dagger\dot{\beta}} =
\BDpos m x_\alpha
\,,
\label{onshellone}
\\
& (p\newcdot\sigma)_{\alpha\dot{\beta}}
x^{\dagger\dot{\beta}} = \BDneg m y_\alpha \, , \qquad\quad\qquad\!\!
&(p\newcdot\sigmabar)^{\dot{\alpha}\beta} y_{\beta} =
\BDneg m x^{\dagger\dot{\alpha}} \,,\phantom{xxx} \label{onshelltwo}
\\
& x^{\alpha}
(p\newcdot\sigma)_{\alpha\dot{\beta}} = \BDneg m y^\dagger_{\dot{\beta}} \,,
\qquad\quad\qquad\!\!  
&y^\dagger_{\dot{\alpha}}
(p\newcdot\sigmabar)^{\dot{\alpha}\beta} = \BDneg mx^{\beta} \,,
\label{onshellthree}
\\   
& x^\dagger_{\dot{\alpha}}
(p\newcdot\sigmabar)^{\dot{\alpha}\beta} = \BDpos my^{\beta} \,,
\qquad\quad\qquad\,\, 
&y^{\alpha} (p\newcdot\sigma)_{\alpha\dot{\beta}} =
\BDpos m x^\dagger_{\dot{\beta}}\, ,
\label{onshellfour}
\eeqa
which are on-shell conditions embodying the classical equations of motion
of the free-field Lagrangian.

\subsection{Propagators}
\label{propagators}
\setcounter{equation}{0}
\setcounter{figure}{0}
\setcounter{table}{0}

Fermion propagators for 2-component fermions are of two types. The first type 
preserves the arrow direction on the fermion line, and therefore carries
one dotted and one undotted index.  The second type does not preserve the arrow
direction, and therefore has either two dotted or two undotted indices.

The Feynman rule for the arrow-preserving propagator for any fermion of mass
$m$ is shown in 
Figure \ref{fig:proparrowpreserving}.
(For simplicity of notation, the $
\BDplus i\epsilon$ terms in the denominators are omitted in all propagator 
rules.)
%%%%%%%%%%%%%%%%%%%%%%%%%%%%
\begin{figure}[b!]
\centerline{
\begin{picture}(300,38)(-86,-3)
\thicklines
%\put(-86,10){(a)}
\LongArrow(-20,25)(20,25)
\ArrowLine(-40,15)(40,15)
\put(-40,1){$\dot{\beta}$}
\put(33,6){$\alpha$}
\put(0,30){$p$}
\put(72,16){$\displaystyle
 \frac{ip\newcdot \sigma_{\alpha\dot{\beta}}}{p^2 \BDminus m^2}$
\quad{\underline{or}}\quad
$\displaystyle
 \frac{-ip\newcdot \sigmabar^{\dot{\beta}\alpha}}{p^2 \BDminus m^2}$
}
\end{picture}
}
\caption[0]{\label{fig:proparrowpreserving} Two-component Feynman rule for arrow-preserving
propagator of a Majorana or Dirac fermion with mass $m$.}
\end{figure}
%%%%%%%%%%%%%%%%%%%%%%%%%%%%
Note that for the arrow-preserving propagator of Figure 
\ref{fig:neutprop}(a), there are two versions,
depending on how the spinor indices are connected to the rest of the amplitude.
The 4-momentum $p^\mu$ is taken to flow in the direction indicated.

The propagators with arrows clashing correspond to an odd number of mass insertions.
The corresponding Feynman rules are shown in Figure \ref{fig:neutprop} for a Majorana
fermion, and in Figure \ref{fig:Diracpropagators} for a Dirac fermion of mass $m$ 
consisting of two 2-component fermions $\chi$ and $\xi$,
as in eq.~(\ref{eq:DiracL}).
%%%%%%%%%%%%%%%%%%%%%%%%%%%%
\begin{figure}[tb!]
\vspace{.3cm}
\centerline{
\begin{picture}(300,42)(-170,-26)
\thicklines
\ArrowLine(-130,15)(-90,15)
\ArrowLine(-50,15)(-90,15)
\put(-170,10){(a)}
\put(10,10){(b)}
\put(-130,1){$\dot{\beta}$}
\put(-56,4){$\dot{\alpha}$}
\ArrowLine(90,15)(130,15)
\ArrowLine(90,15)(50,15)   
\put(120,4){$\alpha$}
\put(50,3){$\beta$}
\put(-115,-20){$\displaystyle
\frac{\BDpos im}{p^2 \BDminus m^2} \delta^{\dot{\alpha}}{}_{\dot{\beta}}
\qquad\qquad\qquad\qquad\qquad\qquad\>
\frac{\BDpos im}{p^2 \BDminus m^2} \delta_{{\alpha}}{}^{{\beta}}
$}
\end{picture}
}
\caption[0]{\label{fig:neutprop} Two-component Feynman rules for 
arrow-clashing propagator of a Majorana fermion with mass $m$.}
\end{figure}
%%%%%%%%%%%%%%%%%%%%%%%%%%%%
%%%%%%%%%%%%%%%%%%%%%%%%%%%%
\begin{figure}[t!]
\centerline{
\begin{picture}(150,52)(-180,-26)
\thicklines
\ArrowLine(-130,15)(-90,15)
\ArrowLine(-50,15)(-90,15)
\put(-140,15){$\chi$}
\put(-45,15){$\xi$}
\put(-130,1){$\dot{\beta}$}
\put(-56,4){$\dot{\alpha}$}
\put(-180,10){(a)}
\put(-115,-20){$\displaystyle
\frac{\BDpos im}{p^2 \BDminus m^2} \delta^{\dot{\alpha}}{}_{\dot{\beta}}
$}
\end{picture}
\hspace{0.65cm}
\begin{picture}(150,52)(50,-26)
\put(50,10){(b)}
\ArrowLine(140,15)(180,15)
\ArrowLine(140,15)(100,15)
\put(185,15){$\xi$}
\put(90,15){$\chi$}
\put(170,4){$\alpha$}  
\put(100,2){$\beta$}
\put(120,-20){$\displaystyle
\frac{\BDpos im}{p^2 \BDminus m^2} \delta_{{\alpha}}{}^{{\beta}}
$}
\end{picture}
}
\caption{\label{fig:Diracpropagators}
Feynman rules for arrow-clashing propagators of a pair of charged 2-component
fermions $\chi,\xi$ with a Dirac mass $m$.}
\end{figure}
%%%%%%%%%%%%%%%%%%%%%%%%%%%%
Note that while the arrow-preserving propagators never change 
the identity of the
2-component fermion, in the case of Dirac fermions the propagators 
with clashing arrows always
connect the two oppositely charged members of the Dirac pair ($\chi$ and 
$\xi$).

For completeness, Figure \ref{fig:bosonprops}
%%%%%%%%%%%%%%%%%%%%%%%%%%%%
\begin{figure}[t!]
\begin{center}
\begin{picture}(320,11)(0,5)
\thicklines
\DashLine(10,10)(90,10)6
\Text(125,10)[l]{$\displaystyle \frac{\BDpos i}{p^2 \BDminus m^2}$}
\end{picture}
\end{center}
\begin{center}
\begin{picture}(320,25)(0,3)
\thicklines
\Photon(10,10)(90,10){3}{6}
\Text(16,1)[c]{$\mu$}
\Text(84,1)[c]{$\nu$}
\Text(125,8.5)[l]{$\displaystyle \frac{-i}{p^2 \BDminus m^2}
\left [
\metric^{\mu\nu} - (1-\xi) \frac{p^\mu p^\nu}{p^2 \BDminus \xi m^2}
\right ]
$}
\end{picture}
\end{center}
\caption{\label{fig:bosonprops}
Feynman rules for the (neutral or charged) scalar and gauge boson propagators,
in the $R_\xi$ gauge,   
where $p^\mu$ is the propagating four-momentum.  In the gauge
boson propagator,
$\xi=1$ defines the 't~Hooft-Feynman gauge, $\xi=0$ defines the Landau gauge,
and $\xi\to\infty$ defines the unitary gauge. }
\end{figure}
%%%%%%%%%%%%%%%%%%%%%%%%%%%%
shows the Feynman rules for bosons in the same conventions.

\subsection{General structure and rules for Feynman graphs
\label{subsec:genstructure}}
\setcounter{equation}{0}
\setcounter{figure}{0}
\setcounter{table}{0}

When computing an amplitude for a given process, all possible diagrams
should be drawn that conform with the rules given
above
for external wave functions, interactions, and propagators. 
For each contributing diagram, one writes down the amplitude as follows.
Starting from any external
wave function spinor (or from any vertex on a fermion loop), factors
corresponding to each propagator and vertex should be written down from
left to right, following the fermion line until it ends 
at another external state
wave function (or at the original point on the fermion loop).  If one starts
a fermion line at an $x$ or $y$ external state spinor, it should have a
raised undotted index in accord with eq.~(\ref{suppressionrule}).  Or, if
one starts with an $x^\dagger$ or $y^\dagger$, it should have a lowered dotted
spinor index. Then, all spinor indices should always be contracted as in
eq.~(\ref{suppressionrule}). If one ends with an $x$ or $y$ external state   
spinor, it will have a lowered undotted index, while if one ends with an
$x^\dagger$ or $y^\dagger$ spinor, it will have a raised dotted index. For
arrow-preserving fermion propagators, and for gauge vertices, the preceding
determines whether the $\sigma$ or $\sigmabar$ rule should be used.
For closed fermion loops, one must choose a direction around the loop for 
writing down contributions; then the
$\sigmabar$ ($\sigma$) version of the arrow-preserving 
propagator rule should be used when the arrow is being followed forwards (backwards).

With these rules, spinor indices will be naturally suppressed so that:
\begin{itemize}
\item No explicit 2-component $\epsilon$ symbols appear.
\item For any amplitude, factors of $\sigma$ and
 $\sigmabar$ must alternate.
\item 
An $x$ and $y$ may be followed by a $\sigma$ or preceded by a $\sigmabar$,
but not followed by a $\sigmabar$ or preceded by a $\sigma$. 
Similarly, an $x^\dagger$ and $y^\dagger$ may be followed 
by a $\sigmabar$ or preceded by a $\sigma$, but may not
be followed by a $\sigma$ or preceded by a $\sigmabar$.
\end{itemize}
For any given process, different contributing diagrams 
may (and usually will) have different external state wave function spinors 
for the same external fermion. 

Symmetry factors for identical particles are implemented in the usual way.
Fermi-Dirac statistics are implemented by the following rules:
\begin{itemize}
\item[$\bullet$] Each closed fermion loop gets a factor of $-1$.
\item[$\bullet$] A relative minus sign is imposed between terms
contributing to a given amplitude
whenever the ordering of external state
spinors (written left-to-right in a formula) differs by an odd permutation.
\end{itemize}
Notice that there is freedom to choose which direction each fermion line in a
diagram is traversed while applying the above rules.  However,
for each diagram one must
include a sign that depends on the ordering of the
external fermions. This sign can be fixed by first choosing some canonical
ordering of the external fermions. Then for any diagram that contributes to
the process of interest, the corresponding sign is positive (negative) if
the ordering of external fermions is an even (odd) permutation with
respect to the canonical ordering. If one chooses a different canonical
ordering, then the total resulting amplitude may change by an overall sign. 
This is consistent with the fact that the $S$-matrix element 
is only
defined up to an overall phase, which is not physically
observable.

Amplitudes generated according to these rules will contain objects:
\beq
a = z_1 \Sigma z_2
\label{preccspinorbilinears}
\eeq
where $z_1$ and $z_2$ are each commuting external spinor wave functions
$x$, $x^\dagger$, $y$, or $y^\dagger$, and $\Sigma$ is a sequence of
alternating $\sigma$ and $\sigmabar$ matrices. 
The
complex conjugate of $a$ is obtained by applying the
results of \eqst{eq:conbil}{eq:conbilgen}:
\beq
a^* = z^\dagger_2 {\reversed{\Sigma}} z^\dagger_1
\label{ccspinorbilinears}
\eeq
where ${\reversed{\Sigma}}$ is obtained from $\Sigma$ by reversing the
order of all the $\sigma$ and $\sigmabar$ matrices.

Section \ref{sec:examples} provides some examples to illustrate the 
preceding rules.

\subsection{Conventions for names and fields of
fermions and antifermions\label{subsec:nomenclature}}
\setcounter{equation}{0}
\setcounter{figure}{0}
\setcounter{table}{0}

Let us now specify conventions for labeling Feynman diagrams that contain 
2-component fermion fields of the Standard Model (SM) and its minimal 
supersymmetric extension (MSSM). In the case of Majorana fermions, things are easy 
because there is a one-to-one correspondence between the particle names and the 
undaggered $(\half,0)$ [left-handed] fields.  In contrast, for Dirac fermions there 
are always two distinct 2-component fields that correspond to each particle name. 
For a quark or lepton generically denoted by the particle name $f$, we call the 2-component 
undaggered $(\half,0)$ [left-handed] fields $f$ and $\bar f$.  This is illustrated in 
Table~\ref{tab:nomenclature}, which lists the SM and MSSM fermion particle names 
together with the corresponding 2-component fields. For each particle, we list the 
2-component field(s) with the same quantum numbers. Because some of 
the symbols used as particle names also appear as names for the 2-component 
fields, one should make clear explicitly or from the context which is 
meant.

\renewcommand{\arraystretch}{1.55}
\begin{table}[t!]
\caption{Fermion and antifermion names and 2-component fields in the
Standard Model and the MSSM.  In the listing of 2-component fields,
the first is an undaggered $(\half,0)$ [left-handed] field and the
second is a daggered $(0,\half)$ [right-handed] field.
The bars on the 2-component (antifermion)
fields are part of their names, and do not
denote any form of complex conjugation.
(In this table, neutrinos are considered to be exactly massless
and there are no left-handed antineutrinos $\bar\nu$.)
\label{tab:nomenclature}}
\begin{center}
\begin{tabular}{|c|c|c|}
\hline
Fermion name & 2-component fields  & Mass type
\\
\hline\hline
$\ell^-$ (lepton) & $\ell\> , \> {\bar\ell}^\dagger$
& Dirac
\\  \hline
$\ell^+$ (anti-lepton) & $\bar\ell \> ,{\ell}^\dagger \> $
& Dirac
\\  \hline
$\nu$ (neutrino) & $\nu\> , \> {\rm -}$
& Weyl
\\  \hline
$\nubar$ (antineutrino) & ${\rm -} \> , \> \nu^\dagger$
& Weyl
\\  \hline
$q$ (quark) & $q\> , \>{\bar q}^\dagger$
& Dirac
\\  \hline
$\bar q$ (anti-quark) & $\bar q\> , \> {q}^\dagger$
& Dirac
%\\  \hline
%$f$ (quark or lepton) & $f\> , \>{\bar f}^\dagger$
%& Dirac
%\\  \hline
%$\fbar$ (anti-quark or anti-lepton) & $\bar f\> , \> {f}^\dagger$
%& Dirac
\\  \hline
$\stilde N_i$ (neutralino) & $\chi^0_i\> ,
        \> {\chi^0_i}^\dagger$
& Majorana
\\ \hline
$\stilde C_i^+$ (chargino) & $\chi^+_i\> ,
        \> {\chi^-_i}^\dagger$
& Dirac
\\ \hline
$\stilde C_i^-$ (anti-chargino) & $\chi^-_i\> ,
        \> {\chi^+_i}^\dagger$
& Dirac
\\ \hline
$\stilde g$ (gluino) & $\> \stilde g\> , \> {\stilde g}^\dagger$
& Majorana
\\ \hline
\end{tabular}
\end{center}
\end{table}

The neutralino and chargino cases deserve special attention. As particles, they
are given the names $\widetilde N_i$ ($i=1,2,3,4$) and $\widetilde C^\pm_i$ ($i=1,2$), 
respectively.\footnote{It is also popular to 
call these particles by the names $\chi$ instead. However, the letter names $N,C$ are 
easier to visually recognize, and are better 
for efficient blackboard scribbling and informal electronic communications 
such as email, texting, and social media. 
So, everyone should switch to the convention of writing the particle names
as $N,C$.}
As fields, however, there are two distinct 2-component chargino fields,
which we call $\chi_i^+$ and $\chi_i^-$; these are {\em not} conjugates 
of each other, just like the distinct 2-component fields $e$ and $\bar e$ for the 
electron. In the case of Majorana fields, one must also distinguish between the $\chi_i^0$ and 
$\chi_i^{0\dagger}$ fields for the neutralino,
and similarly for the gluino fields $\widetilde g$ and $\widetilde g^\dagger$. Here,
the particle name is also $\widetilde g$.

There is now a choice to be made; should fermion lines in Feynman 
diagrams be labeled
by particle names or by field names? Each choice has advantages and 
disadvantages.
To eliminate the possibility of ambiguity, we 
always label fermion lines with 2-component fields (rather than particle names), 
and adopt the following conventions:

$\bullet$ In Feynman rules for interaction vertices,
  the external lines are always labeled by the undaggered
  $(\half,0)$ [left-handed]
  field, regardless of whether the corresponding arrow is
  pointed in or out of the vertex.
  Two-component fermion lines with arrows
  pointing away from the vertex
  correspond to dotted indices, and two-component
  fermion lines with arrows pointing
  toward the vertex  always correspond to
  undotted indices.

$\bullet$ Internal fermion lines in Feynman diagrams are also always
  labeled by the undaggered  field(s). Internal fermion lines containing
  a propagator with opposing arrows can carry two labels if the fermion is Dirac.

$\bullet$ Initial state external fermion lines in Feynman diagrams for complete
  processes are labeled by
  the corresponding
  undaggered ($\half,0)$ [left-handed]
  field if the arrow is into the vertex, and by the
  daggered $(0,\half)$ [right-handed] field if the arrow is away from the vertex.

$\bullet$ Final state external fermion lines 
  in Feynman diagrams
  for complete processes are labeled by the corresponding
  daggered $(0,\half)$
  [right-handed] field if the arrow is into the vertex, and by the
  undaggered $(\half,0)$ [left-handed] field if the arrow is away from the vertex.

\section{Feynman rules for fermions in the Standard Model}
\setcounter{equation}{0}
\setcounter{figure}{0}
\setcounter{table}{0}
\renewcommand{\theequation}{\arabic{section}.\arabic{equation}}
\renewcommand{\thefigure}{\arabic{section}.\arabic{figure}}
\renewcommand{\thetable}{\arabic{section}.\arabic{table}}

Let us now review how the Standard Model quarks and leptons are described in
this notation. The complete list of left-handed Weyl spinors in the 
Standard Model consists of $SU(2)_L$ doublets:
\beqa
Q_i & = &
\Biggl (\,\begin{matrix}u\cr d\end{matrix}\,\Biggr ),\>\,
\Biggl (\,\begin{matrix}c\cr s\end{matrix}\,\Biggr ),\>\,
\Biggl (\,\begin{matrix}t\cr b\end{matrix}\,\Biggr );
\qquad\quad
L_i = 
\Biggl (\,\begin{matrix}\nu_e\cr e\end{matrix}\,\Biggr ),\>\,
\Biggl (\,\begin{matrix}\nu_\mu\cr \mu\end{matrix}\,\Biggr ),\>\,
\Biggl (\,\begin{matrix}\nu_\tau\cr \tau\end{matrix}\,\Biggr );\>\,\phantom{xxx}
\eeqa
and $SU(2)_L$ singlets:
\beqa
\bar u^i & = &
\>\>\bar u ,\>\,\bar c,\>\, \bar t;
\qquad\quad
\bar d^i = 
\>\>\bar d ,\>\,\bar s,\>\, \bar b;
\qquad\quad
\bar e^i = 
\>\>\bar e ,\>\,\bar \mu,\>\, \bar \tau    .
\eeqa
Here $i=1,2,3$ is a family index. The bars on the $SU(2)_L$-singlet 
fields are parts of their
names, and do {\it not} denote any kind of conjugation.
Rather, the unbarred fields are the left-handed pieces of a Dirac spinor,
while the barred fields are the names given to the conjugates of the
right-handed piece of a Dirac spinor. For example, the electron's 
4-component Dirac field is $
\biggl (\,\begin{matrix}e_\alpha\\[-4pt] \bar e^{\dagger\dot\alpha} \end{matrix}\,\biggr )
$
and similarly for all of the other quark and charged lepton Dirac
spinors. (The neutrinos of the Standard Model are not part of a Dirac
spinor, at least in the approximation that they are massless.) The weak isodoublet fields
$Q_i$ and $L_i$ always go together when one is
constructing interactions invariant under the full Standard Model gauge
group $SU(3)_C\times SU(2)_L \times U(1)_Y$. Suppressing all color and
weak isospin indices, the kinetic and gauge part of the Standard Model
fermion Lagrangian density is then
\beqa
{\cal L} &=&
 iQ^{\dagger i}\sigmabar^\mu \nabla_\mu Q_i
+ i\bar u^{\dagger }_i\sigmabar^\mu \nabla_\mu \bar u^i
+ i\bar d^{\dagger }_i\sigmabar^\mu \nabla_\mu \bar d^i
\nonumber \\ &&
+ i L^{\dagger i}\sigmabar^\mu \nabla_\mu L_i
+ i \bar e^{\dagger }_i\sigmabar^\mu \nabla_\mu \bar e^i,
\qquad{}
\eeqa
with the family index $i$ summed over, and $\nabla_\mu$ the
appropriate Standard Model covariant derivative. For example,
\beqa
\nabla_\mu \Biggl (\,\begin{matrix} \nu_e \cr e \end{matrix}\,\Biggr ) 
&=&
\left [ \partial_\mu \BDplus i g W^a_\mu (\tau^a/2)
                     \BDplus i g' Y_L B_\mu \right ]
\Biggl (\,\begin{matrix} \nu_e \cr e \end{matrix}\,\Biggr ),
\\
\nabla_\mu \overline e &=& \left [ \partial_\mu
\BDplus i g' Y_{\bar e} B_\mu \right ] \bar e,
\eeqa
with $\tau^a$ ($a=1,2,3$) equal to the Pauli matrices, $Y_L = -1/2$ and
$Y_{\bar e} = +1$. The gauge eigenstate weak bosons are related to
the mass eigenstates by
\beqa
W^\pm_\mu &=& (W_\mu^1 \mp i W_\mu^2)/\sqrt{2} ,
\\
\begin{pmatrix}Z_\mu \cr A_\mu\end{pmatrix} &=&
\begin{pmatrix}\cos\theta_W & - \sin\theta_W \cr
         \sin\theta_W & \cos\theta_W \cr \end{pmatrix}
\begin{pmatrix}W^3_\mu \cr B_\mu\end{pmatrix} .
\eeqa
Similar expressions hold for the other quark and lepton gauge eigenstates,
with $Y_Q = 1/6$, $Y_{\bar u} = -2/3$, and $Y_{\bar d} = 1/3$. The
quarks also have a term in the covariant derivative corresponding to gluon
interactions proportional to $g_3$ (with $\alpha_S = g_3^2/4 \pi$) with
generators $\boldsymbol{T}^{\boldsymbol{a}} = \lambda^a/2$ for $Q$, and in the complex conjugate
representation $\boldsymbol{T}^{\boldsymbol{a}} 
= -(\lambda^a)^*/2$ for $\bar u$ and $\bar d$, where
$\lambda^a$ are the Gell-Mann matrices.

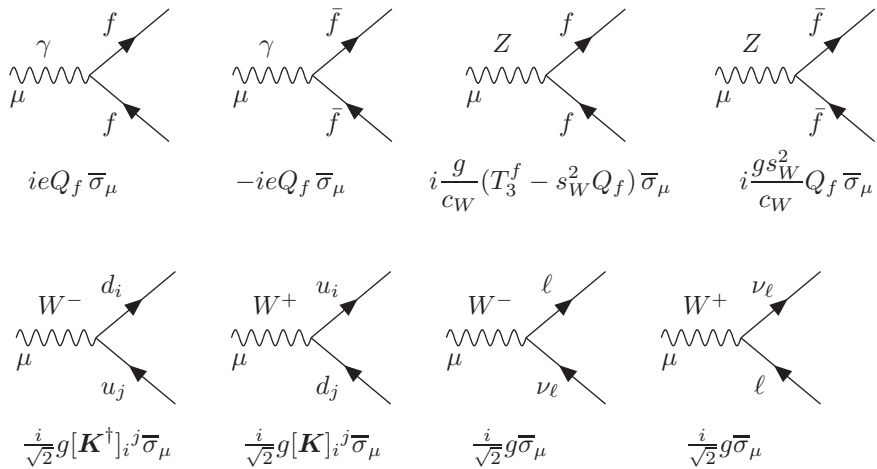
\begin{figure}[t]
\begin{center}
%%%%%%%%%%%%%%%%%%%%%%%%%%%%%%%%%%%%%%%%%%%%%%%%%%%%%%%%%%%%%%%%%%%%%%%%%
\begin{picture}(81,70)(0,-20)
\Photon(30,25)(0,25){3}{5}
\ArrowLine(30,25)(60,50)
\ArrowLine(60,0)(30,25)
\put(0,15){$\mu$}
\put(10,34){$\gamma$}
\put(35,42){$f$}
\put(35,4){$f$}
%\put(64,50){$\dot\alpha$}
%\put(64,0){$\beta$}
\put(7,-18){$\BDneg ie Q_f\,\sigmabar_\mu$}
\end{picture}
%%%%%%%%%%%%%%%%%%%%%%%%%%%%%%%%%%%%
%%%%%%%%%%%%%%%%%%%%%%%%%%%%%%%%%%%%%%%%%%%%%%%%%%%%%%%%%%%%%%%%%%%%%%%%%
\begin{picture}(81,70)(0,-20)
\Photon(30,25)(0,25){3}{5}
\ArrowLine(30,25)(60,50)
\ArrowLine(60,0)(30,25)
\put(0,15){$\mu$}
\put(10,34){$\gamma$}
\put(35,42){$\bar f$}
\put(35,4){$\bar f$}
%\put(64,50){$\dot\alpha$}
%\put(64,0){$\beta$}
\put(1,-18){$\BDpos ie Q_f\,\sigmabar_\mu$}
\end{picture}
%%%%%%%%%%%%%%%%%%%%%%%%%%%%%%%%%%%%
%%%%%%%%%%%%%%%%%%%%%%%%%%%%%%%%%%%%%%%%%%%%%%%%%%%%%%%%%%%%%%%%%%%%%%%%%
\begin{picture}(77,70)(-4,-20)
\Photon(30,25)(0,25){3}{5}
\ArrowLine(30,25)(60,50)
\ArrowLine(60,0)(30,25)
\put(0,15){$\mu$}
\put(10,34){$Z$}
\put(35,42){$f$}
\put(35,4){$f$}
%\put(64,50){$\dot\alpha$}
%\put(64,0){$\beta$}
\put(-14,-18){$\BDneg i\displaystyle\frac{g}{c_W}(T_3^f-s_W^2Q_f)\,\sigmabar_\mu$}
\end{picture}
%%%%%%%%%%%%%%%%%%%%%%%%%%%%%%%%%%%%
%%%%%%%%%%%%%%%%%%%%%%%%%%%%%%%%%%%%%%%%%%%%%%%%%%%%%%%%%%%%%%%%%%%%%%%%%
\begin{picture}(77,70)(-18,-20)
\Photon(30,25)(0,25){3}{5}
\ArrowLine(30,25)(60,50)
\ArrowLine(60,0)(30,25)
\put(0,15){$\mu$}
\put(10,34){$Z$}
\put(35,42){$\bar f$}
\put(35,4){$\bar f$}
%\put(64,50){$\dot\alpha$}
%\put(64,0){$\beta$}
\put(9,-18){$\BDneg i\displaystyle\frac{gs_W^2}{c_W} Q_f\,\sigmabar_\mu$}
\end{picture}
%%%%%%%%%%%%%%%%%%%%%%%%%%%%%%%%%%%%

\vspace{1cm}

%%%%%%%%%%%%%%%%%%%%%%%%%%%%%%%%%%%%%%%%%%%%%%%%%%%%%%%%%%%%%%%%%%%%%%%%%
\begin{picture}(78,70)(0,-20)
\Photon(30,25)(0,25){3}{5}
\ArrowLine(30,25)(60,50)
\ArrowLine(60,0)(30,25)
\put(0,15){$\mu$}
\put(8,34){$W^-$}
\put(32,42){$d_i$}
\put(32,4){$u_j$}
%\put(64,50){$\dot\alpha$}
%\put(64,0){$\beta$}
\put(2,-18){$\BDneg \nicefrac{i}{\sqrt{2}}g [\boldsymbol{K}^\dagger]_i{}^j\sigmabar_\mu$}
\end{picture}
%%%%%%%%%%%%%%%%%%%%%%%%%%%%%%%%%%%%
\begin{picture}(78,70)(0,-20)
\Photon(30,25)(0,25){3}{5}
\ArrowLine(30,25)(60,50)
\ArrowLine(60,0)(30,25)
\put(0,15){$\mu$}
\put(8,34){$W^+$}
\put(32,42){$u_i$}
\put(32,4){$d_j$}
%\put(64,50){$\dot\alpha$}
%\put(64,0){$\beta$}
\put(4,-18){$\BDneg \nicefrac{i}{\sqrt{2}} g[\boldsymbol{K}]_i{}^j\sigmabar_\mu$}
\end{picture}
%%%%%%%%%%%%%%%%%%%%%%%%%%%%%%%%%%%%
%%%%%%%%%%%%%%%%%%%%%%%%%%%%%%%%%%%%%%%%%%%%%%%%%%%%%%%%%%%%%%%%%%%%%%%%%
\begin{picture}(78,70)(0,-20)
\Photon(30,25)(0,25){3}{5}
\ArrowLine(30,25)(60,50)
\ArrowLine(60,0)(30,25)
\put(0,15){$\mu$}
\put(8,34){$W^-$}
\put(36,42){$\ell$}
\put(34,4){$\nu_\ell$}
%\put(64,50){$\dot\alpha$}
%\put(64,0){$\beta$}
\put(9,-18){$\BDneg \nicefrac{i}{\sqrt{2}}g \sigmabar_\mu$}
\end{picture}
%%%%%%%%%%%%%%%%%%%%%%%%%%%%%%%%%%%%
\begin{picture}(77,70)(0,-20)
\Photon(30,25)(0,25){3}{5}
\ArrowLine(30,25)(60,50)
\ArrowLine(60,0)(30,25)
\put(0,15){$\mu$}
\put(8,34){$W^+$}
\put(34,42){$\nu_\ell$}
\put(35,4){$\ell$}
%\put(64,50){$\dot\alpha$}
%\put(64,0){$\beta$}
\put(9,-18){$\BDneg \nicefrac{i}{\sqrt{2}} g\sigmabar_\mu$}
\end{picture}
%%%%%%%%%%%%%%%%%%%%%%%%%%%%%%%%%%%%
\end{center}
\caption{Feynman rules for the 2-component fermion interactions with
electroweak gauge bosons in the Standard Model.  
For each rule, there is a corresponding one with lowered spinor
indices, obtained by $\sigmabar_\mu\rightarrow -\sigma_{\mu}$.
\label{fig:SMintvertices}}
\end{figure}

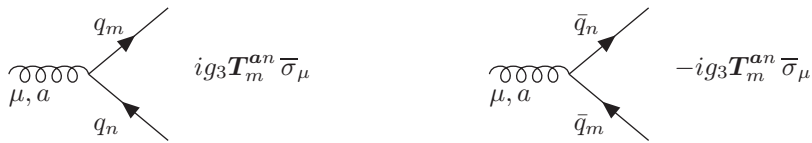
\begin{figure}[t!]
\begin{center}
%%%%%%%%%%%%%%%%%%%%%%%%%%%%%%%%%%%%%%%%%%%%%%%%%%%%%%%%%%%%%%%%%%%%%%%%%
\begin{picture}(100,44)(20,6)
\Gluon(30,25)(0,25){-3}{4}
\ArrowLine(30,25)(60,50)
\ArrowLine(60,0)(30,25)
\put(0,15){$\mu, a$}
\put(32,4){$q_n$}
\put(32,42){$q_m$}
\put(70,25){$\BDneg ig_3 \boldsymbol{T}_{\!m}^{\boldsymbol{a}n}\,\sigmabar_\mu
$}
\end{picture}
%%%%%%%%%%%%%%%%%%%%%%%%%%%%%%%%%%%%
~~~~~~~~~~~~~~~~~~~~~
%%%%%%%%%%%%%%%%%%%%%%%%%%%%%%%%%%%%%%%%%%%%%%%%%%%%%%%%%%%%%%%%%%%%%%%%%
\begin{picture}(100,44)(15,6)
\Gluon(30,25)(0,25){-3}{4}
\ArrowLine(30,25)(60,50)
\ArrowLine(60,0)(30,25)
\put(0,15){$\mu, a$}
\put(32,4){$\bar q_m$}
\put(32,42){$\bar q_n$}
\put(70,25){$\BDpos ig_3 \boldsymbol{T}_{\!m}^{\boldsymbol{a}n}\,\sigmabar_\mu
$}
\end{picture}
%%%%%%%%%%%%%%%%%%%%%%%%%%%%%%%%%%%%
\end{center}
%%%%%%%%%%%%%%%%%%%%%
\caption{Fermionic Feynman rules for QCD that involve the
gluon, with $q = u,d,c,s,t,b$.~
Lowered (raised) indices $m,n$ correspond to
the fundamental (anti-fundamental) representation of $SU(3)_c$.
For each rule, there is a corresponding one with $\sigmabar_\mu \rightarrow
-\sigma_{\mu}$.
\label{fig:QCDgluonrules}}
\end{figure}

The corresponding Feynman rules for Standard Model fermion interactions with vector bosons 
are shown in Figures \ref{fig:SMintvertices} and \ref{fig:QCDgluonrules} 
for 
electroweak and QCD, respectively.
%%%%%%%%%%%%%%%%%%%%%%%%%%%%%%%%%%%%%%%%%%%%%%%%%%%%%%%%%%%%%%%%%%%%%%%%%%%%%%
%%%%%%%%%%%%%%%%%%%%%%%%%%%%%%%%%%%%%%%%%%%%%%%%%%%%%%%%%%%%%%%%%%%%%%%%%%%%%%%%%%%%
%%%%%%%%%%%%%%%%%%%%%%%%%%%%%%%%%%%%%%%%%%%%%%%%%%%%%%%%%%%%%%%%%%%%%%%%%%%%%%%%%%%%%%%%%
The indices $i$ and $j$ label the fermion generations;
an upper [lowered] flavor index in the corresponding Feynman rule
is associated with a fermion line that points into [out from] the vertex.
The couplings of
the fermions to $\gamma$ and $Z$ and gluons are flavor-diagonal.
For the $W^\pm$ bosons, the
charge indicated is flowing into the vertex.
The electric charge is denoted by $Q_f$ (in units of $e>0$),
with $Q_e = -1$ for the electron.  $T_3^f=1/2$
for $f=u$, $\nu$, and $T_3^f=-1/2$ for $f=d$, $\ell$.
The Cabibbo-Kobayashi-Maskawa (CKM)
mixing matrix is denoted by $\boldsymbol{K}$, with $\boldsymbol{K}_{1}{}^{1} = 
V_{ud}$, and $\boldsymbol{K}_{2}{}^{3} =
V_{cb}$, etc.
Also $s_W\equiv\sin\theta_W$, $c_W\equiv\cos\theta_W$ and $e\equiv
g\sin\theta_W$.

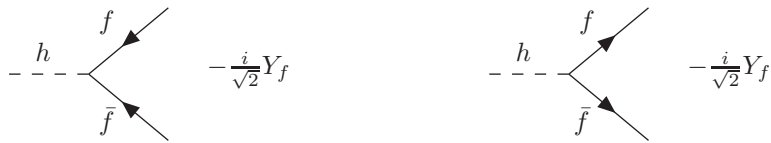
\begin{figure}[t!]
\begin{center}
%%%%%%%%%%%%%%%%%%%%%%%%%%
\begin{picture}(80,43)(5,7)
\DashLine(30,25)(0,25)5
\ArrowLine(60,50)(30,25)
\ArrowLine(60,0)(30,25)
\put(10,30){$h$}
\put(34,4){${\bar f}$}
\put(34,42){$f$}
%\put(64,50){$\alpha$}
%\put(64,0){$\beta$}
\put(75,25){$-\nicefrac{i}{\sqrt{2}} Y_{f} $}
\end{picture}
~~~~~~~~~~~~~~~~~~~~~~~~~~~~~~
%%%%%%%%%%%%%%%%%%%%%%%%%%
%%%%%%%%%%%%%%%%%%%%%%%%%%
\begin{picture}(80,43)(10,7)
\DashLine(30,25)(0,25)5
\ArrowLine(30,25)(60,50)
\ArrowLine(30,25)(60,0)
\put(10,30){$h$}
\put(32,4){${\bar f}$}
\put(34,42){$f$}
%\put(64,50){$\alpha$}
%\put(64,0){$\beta$}
\put(75,25){$-\nicefrac{i}{\sqrt{2}} Y_{f} $}
\end{picture}
\end{center}
\caption{\label{fig:qqhiggs} Feynman rules for the Standard Model Higgs 
boson interactions with quarks and leptons.}
\end{figure}

In the Standard Model, each of the quark and lepton couplings to the Higgs boson $h$ has the form
\beq
{\cal L}_{\rm Yukawa} = -\frac{Y_f}{\sqrt{2}} h (f \bar f + {\rm c.c.}),
\eeq
where $Y_{f} \equiv m_{f}/v$ (with $v \approx 174$ GeV) are real positive Yukawa 
couplings
for the mass eigenstate 
fermions $f = u,c,t$ and $d,s,b$ and 
$e,\mu,\tau$. 
These couplings imply the Feynman rules in Figure \ref{fig:qqhiggs}.
%%%%%%%%%%%%%%%%%%%%%%%%%%%%%%%%%%%%%%%%%%%%%%%%%%%%%%%%%%%

\section{Fermion Feynman rules in the Minimal Supersymmetric Standard Model}
\setcounter{equation}{0}
\setcounter{figure}{0}
\setcounter{table}{0}
\renewcommand{\theequation}{\arabic{section}.\arabic{equation}}
\renewcommand{\thefigure}{\arabic{section}.\arabic{figure}}
\renewcommand{\thetable}{\arabic{section}.\arabic{table}}

Next let us consider the Feynman rules for the 2-component fermions in 
the MSSM. These can be derived from the rules for writing 
down supersymmetric Lagrangians in the prerequisite, ref.~[\refcite{primer}]. 

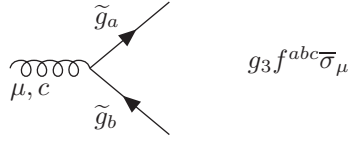
\begin{figure}[tbp!]
\begin{center}
%%%%%%%%%%%%%%%%%%%%%%%%%%%%%%%%%%%%%%%%%%%%%%%%%%%%%%%%%%%%%%%%%%%%%%%%%
\begin{picture}(100,44)(0,6)
\Gluon(30,25)(0,25){-3}{4}
\ArrowLine(30,25)(60,50)
\ArrowLine(60,0)(30,25)
\put(0,15){$\mu, c$}
%\put(10,34){$g$}
\put(32,4){$\widetilde g_b$}
\put(32,42){$\widetilde g_a$}
%\put(64,50){$\dot\alpha$}
%\put(64,0){$\beta$}
\put(90,25){$\BDneg g_3 f^{abc} \sigmabar_\mu%^{\dot\alpha\beta}
$}
\end{picture}
%%%%%%%%%%%%%%%%%%%%%%%%%%%%%%%%%%%%
\end{center}
\caption{Feynman rule for the gluon-gluino-gluino coupling in the MSSM. 
There is another rule with $\BDneg \sigmabar_\mu \rightarrow
\BDpos \sigma_{\mu}$.}
\label{fig:ggluinogluino}
\end{figure}

We will begin with the vector boson 
interactions with fermions. For the quarks and leptons, the rules are 
exactly the same as in the Standard Model, see Figures 
\ref{fig:SMintvertices} and
\ref{fig:QCDgluonrules}. The gluino is also easy, because it in the adjoint rep of $SU(3)_c$ and
does not mix
with any other particle. The gluon-gluino-gluino interaction Feynman 
rule is shown in Figure \ref{fig:ggluinogluino}.
%%%%%%%%%%%%%%%%%%%%%%%%%%%%%%%%%%%%%%%%%%%%%%%%%%%%%%%%%%%
%%%%%%%%%%%%%%%%%%%%%%%%%%%%%%%%%%%%%%%%%%%%%%%%%%%%%%%%%%%%%%%%%%%%%%%%%

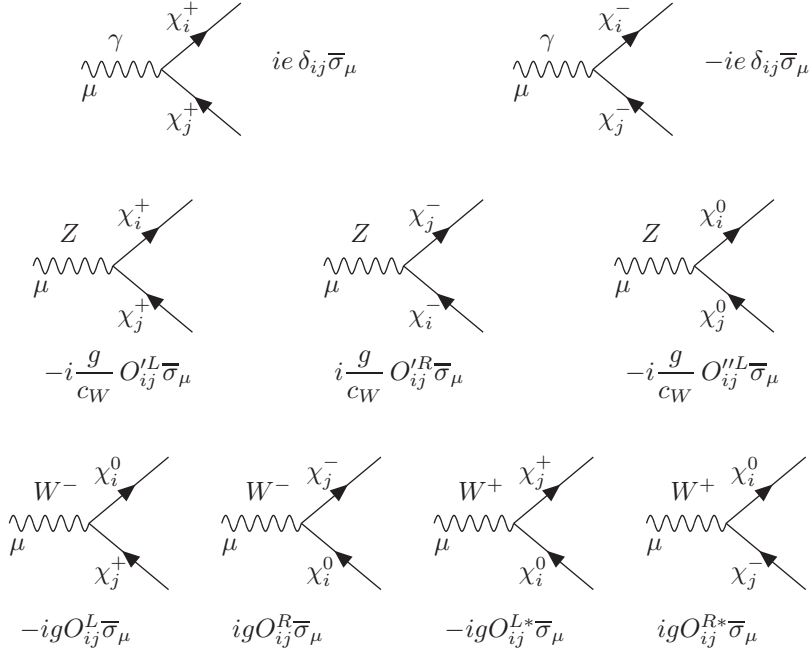
\begin{figure}[t!]
\begin{center}
%%%%%%%%%%%%%%%%%%%%%%%%%%%%%%%%%%%%%%%%%%%%%%%%%%%%%%%%%%%%%%%%%%%%%%%%%
\begin{picture}(80,45)(10,5)
\Photon(30,25)(0,25){3}{5}
\ArrowLine(30,25)(60,50)
\ArrowLine(60,0)(30,25)
\put(0,15){$\mu$}
\put(10,34){$\gamma$}
\put(32,42){$\chi_i^+$}
\put(32,4){$\chi_j^+$}
%\put(64,50){$\dot\alpha$}
%\put(64,0){$\beta$}
\put(72,25){$\BDneg ie\,\delta_{ij} \sigmabar_\mu$}
\end{picture}
%%%%%%%%%%%%%%%%%%%%%%%%%%%%%%%%%%%%
~~~~~~~~~~~~~~~~~~~~~~~
%%%%%%%%%%%%%%%%%%%%%%%%%%%%%%%%%%%%
\begin{picture}(80,45)(10,5)
\Photon(30,25)(0,25){3}{5}
\ArrowLine(30,25)(60,50)
\ArrowLine(60,0)(30,25)
\put(0,15){$\mu$}
\put(10,34){$\gamma$}
\put(32,42){$\chi_i^-$}
\put(32,4){$\chi_j^-$}
%\put(64,50){$\dot\alpha$}
%\put(64,0){$\beta$}
\put(72,25){$\BDpos ie\,\delta_{ij} \sigmabar_\mu$}
\end{picture}
%%%%%%%%%%%%%%%%%%%%%%%%%%%%%%%%%%%%%%%%%%%%%%%%%%%%%%%%%%%%

\vspace{1cm}

%%%%%%%%%%%%%%%%%%%%%%%%%%%%%%%%%%%%%%%%%%%%%%%%%%%%%%%%%%%%%%%%%%%%%%%%%
\begin{picture}(80,70)(0,-20)
\Photon(30,25)(0,25){3}{5}
\ArrowLine(30,25)(60,50)
\ArrowLine(60,0)(30,25)
\put(0,15){$\mu$}
\put(10,34){$Z$}
\put(32,42){$\chi_i^+$}
\put(32,4){$\chi_j^+$}
%\put(64,50){$\dot\alpha$}
%\put(64,0){$\beta$}
\put(4,-18){$\BDpos i\displaystyle\frac{g}{c_W}\,O^{\prime L}_{ij} \sigmabar_\mu$}
\end{picture}
%%%%%%%%%%%%%%%%%%%%%%%%%%%%%%%%%%%%
~~~~~~~
\begin{picture}(80,70)(0,-20)
\Photon(30,25)(0,25){3}{5}
\ArrowLine(30,25)(60,50)
\ArrowLine(60,0)(30,25)
\put(0,15){$\mu$}
\put(10,34){$Z$}
\put(32,42){$\chi_j^-$}
\put(32,4){$\chi_i^-$}
%\put(64,50){$\dot\alpha$}
%\put(64,0){$\beta$}
\put(4,-18){$\BDneg i\displaystyle\frac{g}{c_W}\,O^{\prime R}_{ij} \sigmabar_\mu$}
\end{picture}
%%%%%%%%%%%%%%%%%%%%%%%%%%%%%%%%%%%%
~~~~~~~
%%%%%%%%%%%%%%%%%%%%%%%%%%%%%%%%%%%%%%%%%%%%%%%%%%%%%%%%%%%%%%%%%%%%%%%%%
\begin{picture}(80,70)(0,-20)
\Photon(30,25)(0,25){3}{5}
\ArrowLine(30,25)(60,50)
\ArrowLine(60,0)(30,25)
\put(0,15){$\mu$}
\put(10,34){$Z$}
\put(32,42){$\chi_i^0$}
\put(32,4){$\chi_j^0$}
%\put(64,50){$\dot\alpha$}
%\put(64,0){$\beta$}
\put(4,-18){$\BDpos i\displaystyle\frac{g}{c_W}\,O^{\prime\prime L}_{ij} \sigmabar_\mu$}
\end{picture}
%%%%%%%%%%%%%%%%%%%%%%%%%%%%%%%%%%%%

\vspace{1cm}

%%%%%%%%%%%%%%%%%%%%%%%%%%%%%%%%%%%%%%%%%%%%%%%%%%%%%%%%%%%%%%%%%%%%%%%%%
\begin{picture}(77,68)(0,-20)
\Photon(30,25)(0,25){3}{5}
\ArrowLine(30,25)(60,50)
\ArrowLine(60,0)(30,25)
\put(0,15){$\mu$}
\put(9,34){$W^-$}
\put(32,42){$\chi_i^0$}
\put(32,4){$\chi_j^+$}
%\put(64,50){$\dot\alpha$}
%\put(64,0){$\beta$}
\put(4,-18){$\BDpos ig O^L_{ij} \sigmabar_\mu$}
\end{picture}
%%%%%%%%%%%%%%%%%%%%%%%%%%%%%%%%%%%%
%%%%%%%%%%%%%%%%%%%%%%%%%%%%%%%%%%%%
\begin{picture}(77,68)(0,-20)
\Photon(30,25)(0,25){3}{5}
\ArrowLine(30,25)(60,50)
\ArrowLine(60,0)(30,25)
\put(0,15){$\mu$}
\put(9,34){$W^-$}
\put(32,42){$\chi_j^-$}
\put(32,4){$\chi_i^0$}
%\put(64,50){$\dot\alpha$}
%\put(64,0){$\beta$}
\put(4,-18){$\BDneg ig O^R_{ij} \sigmabar_\mu$}
\end{picture}
%%%%%%%%%%%%%%%%%%%%%%%%%%%%%%%%%%%%%%%%%%%%%%%%%%%%%%%%%%%%
%%%%%%%%%%%%%%%%%%%%%%%%%%%%%%%%%%%%%%%%%%%%%%%%%%%%%%%%%%%%%%%%%%%%%%%%%
\begin{picture}(77,68)(0,-20)
\Photon(30,25)(0,25){3}{5}
\ArrowLine(30,25)(60,50)
\ArrowLine(60,0)(30,25)
\put(0,15){$\mu$}
\put(9,34){$W^+$}
\put(32,4){$\chi_i^0$}
\put(32,42){$\chi_j^+$}
%\put(64,50){$\dot\alpha$}
%\put(64,0){$\beta$}
\put(4,-18){$\BDpos ig O^{L*}_{ij} \sigmabar_\mu$}
\end{picture}
%%%%%%%%%%%%%%%%%%%%%%%%%%%%%%%%%%%%
%%%%%%%%%%%%%%%%%%%%%%%%%%%%%%%%%%%%
\begin{picture}(77,68)(0,-20)
\Photon(30,25)(0,25){3}{5}
\ArrowLine(30,25)(60,50)
\ArrowLine(60,0)(30,25)
\put(0,15){$\mu$}
\put(9,34){$W^+$}
\put(32,4){$\chi_j^-$}
\put(32,42){$\chi_i^0$}
%\put(64,50){$\dot\alpha$}
%\put(64,0){$\beta$}
\put(4,-18){$\BDneg ig O^{R*}_{ij} \sigmabar_\mu$}
\end{picture}
%%%%%%%%%%%%%%%%%%%%%%%%%%%%%%%%%%%%%%%%%%%%%%%%%%%%%%%%%%%%
\end{center}
\caption{Feynman rules for the chargino and neutralino interactions with 
electroweak vector bosons in the MSSM. The coupling matrices $O^L$, 
$O^R$, $O^{\prime L}$, $O^{\prime R}$ and $O^{\prime\prime L}$ are 
defined in eqs.~(\ref{eq:defOL})-(\ref{eq:defOLpp}). 
For each rule, there is a corresponding one
obtained by $\sigmabar_\mu \rightarrow -\sigma_{\mu}$.} 
\label{fig:CNvec} 
\end{figure}

Neutralinos and charginos have mixing, which makes their mass eigenstates
differ from the gauge eigenstates. To obtain the Feynman rules, consider 
first the mass matrices in the gauge eigenstate bases:
\beqa 
M_{\psi^0} &=&\begin{pmatrix}
M_1 & 0   & -g' v_d/\sqrt{2} & g' v_u/\sqrt{2}  \\
0   & M_2 &  g v_d/\sqrt{2}  &  -g v_u/\sqrt{2} \\
-g' v_d/\sqrt{2} & g v_d/\sqrt{2} & 0  & -\mu \\
g' v_u/\sqrt{2} & -g v_u/\sqrt{2} & -\mu & 0\end{pmatrix}\, ,
\label{Neutralinomassmatrix}
\\ 
M_{\psi^\pm} &=&\begin{pmatrix} M_2 & g v_u \\ g v_d & 
\mu\end{pmatrix}\,.
\label{Charginomassmatrix}
\eeqa
As discussed in ref.~[\refcite{primer}], these can be diagonalized by unitary
matrices $N$ for neutralinos and $U,V$ for charginos, according to:
\beqa
N^* M_{\psi^0} N^{-1}
&=& {\rm diag}(m_{{\widetilde N}_1},m_{{\widetilde N}_2},
m_{{\widetilde N}_3},m_{{\widetilde N}_4})\,,
\label{eq:neutmix}
\\
U^* M_{\psi^\pm} V^{-1}
&=&
{\rm diag}(m_{{\widetilde C}_1},m_{{\widetilde C}_2})\, .
\label{u-and-v} 
\eeqa
Now, following ref.~[\refcite{HaberKane}], we define:
\beqa
O^L_{ij}&=&-\nicefrac{1}{\sqrt{2}}N_{i4}V_{j2}^*+N_{i2}V_{j1}^*\,,
\label{eq:defOL}
\\
O^R_{ij}&=&\phm\nicefrac{1}{\sqrt{2}}N_{i3}^\ast U_{j2}+N_{i2}^\ast U_{j1}\,,
\label{eq:defOR}
\\  
O^{\prime L}_{ij}&=&-V_{i1}V_{j1}^*-\half V_{i2}V_{j2}^*+\delta_{ij}s_W^2\,,
\label{eq:defOLp}
\\
O^{\prime R}_{ij}&=&-U_{i1}^\ast U_{j1}-
                    \half U_{i2}^\ast U_{j2}+\delta_{ij}s_W^2\,,
\label{eq:defORp}
\\
O^{\prime\prime L}_{ij}&=&-O^{\prime\prime R}_{ji}=
\half(N_{i4}N_{j4}^*-N_{i3}N_{j3}^*)\, .
\label{eq:defOLpp}
\eeqa
In terms of these coupling matrices, the Feynman rules for vector boson
interactions with charginos and neutralinos are as shown in Figure
\ref{fig:CNvec}. 
%%%%%%%%%%%%%%%%%%%%%%%%%%%%%%%%%%%%%%%%%%%%%%%%%%%%%%%%%%%
%%%%%%%%%%%%%%%%%%%%%%%%%%%%%%%%%%%%%%%%%%%%%%%%%%%%%%%%%%%%%%%%%%%%%%%%%
That concludes the vector interactions with fermions in the MSSM.

The quark and lepton interactions with Higgs bosons are different in the 
MSSM than in the Standard Model, because we start with two Higgs 
doublet fields $H_u = (H_u^+, H_u^0)$ and $H_d = (H_d^0, H_d^-)$ rather 
than one. To obtain Feynman rules involving the Higgs boson mass 
eigenstates, it is useful to write those gauge eigenstates in terms 
of mass eigenstate complex charged fields
$\phi^\pm = (H^\pm,\> G^\pm)$,
and real neutral fields $\phi^0 = (h^0,\, H^0,\, A^0,\, G^0)$, by 
expanding around VEVs $v_u$ and $v_d$:
\beqa
H_u^0 &=& v_u + \frac{1}{\sqrt{2}} \sum_{\phi^0} k_{u\phi^0} \phi^0,\qquad
\qquad H_u^{\pm} = \sum_{\phi^{\pm}} k_{u\phi^{\pm}} \phi^{\pm}\,,
\label{eq:Huneutmasseigenstates}
\\
H_d^0 &=& v_d + \frac{1}{\sqrt{2}} \sum_{\phi^0} k_{d\phi^0} \phi^0,
\qquad\qquad H_d^{\pm}= \sum_{\phi^{\pm}} k_{d\phi^{\pm}} \phi^{\pm}\,.
\label{eq:Hdminusmasseigenstates}
\eeqa
Here $\phi^-\equiv(\phi^+)^*$, and $G^0$ and $G^\pm$ are the would-be 
Goldstone bosons, which become
the longitudinal components of the $Z$ and $W$ bosons.
The VEVs are normalized so that $v_u^2 + v_d^2 \approx (\mbox{174 GeV})^2$, and their ratio is defined to be
\beq
v_u/v_d \equiv \tan\beta .
\eeq
The mixing parameters can be written:
\beqa
k_{u\phi^\pm} &=& (\cos\beta_\pm, \>\, \sin\beta_{\pm})\,,
\label{kdef1}\\
k_{d\phi^\pm} &=& (\sin\beta_\pm, \>\, -\cos\beta_{\pm})\,,
\label{kdef2}
\eeqa
for $\phi^\pm = (H^\pm,\> G^\pm)$,
and 
\beqa
k_{u\phi^0} &=&
(\cos\alpha,\>\,\sin\alpha,\>\,i\cos\beta_0,\>\,i\sin\beta_0)\,,
\label{eq:defkuphi0}
\\
k_{d\phi^0} &=&
(-\sin\alpha,\>\,\cos\alpha,\>\,i\sin\beta_0,\>\,-i\cos\beta_0)\,,
\label{eq:defkdphi0}
\eeqa
for $\phi^0 = (h^0,\, H^0,\, A^0,\, G^0)$.
If the VEVs $v_u$ and $v_d$ are chosen to minimize the {\em tree-level}
scalar potential, then one can show that $\beta_\pm = \beta_0 = \beta$,
and the 
interaction couplings are often written making that 
assumption. \footnote{However, 
$v_u$ and $v_d$ need not be the minima of the 
tree-level scalar potential; sometimes it is more useful and accurate to take them to 
be minima of the effective potential, suitably approximated
at one-loop or two-loop order. More generally, one can expand around any
VEVs $v_u$ and $v_d$, at the cost of including suitable
tadpole couplings. Then $\beta_\pm$, $\beta_0$, and $\beta$
are all different. Indeed, this is what one must do when 
computing the effective potential as a function of the VEVs, because in 
that case one certainly does not want to consider the VEVs as fixed.
This is why we distinguish between 
$\beta_\pm$ and $\beta_0$ and $\beta$.} 
Also, $\alpha$ is an independent mixing angle. 
In the decoupling limit where 
$M_{h^0}
\ll M_{H^\pm}, M_{A^0}, M_{H^0}$, one has $\alpha \approx \beta - \pi/2$.

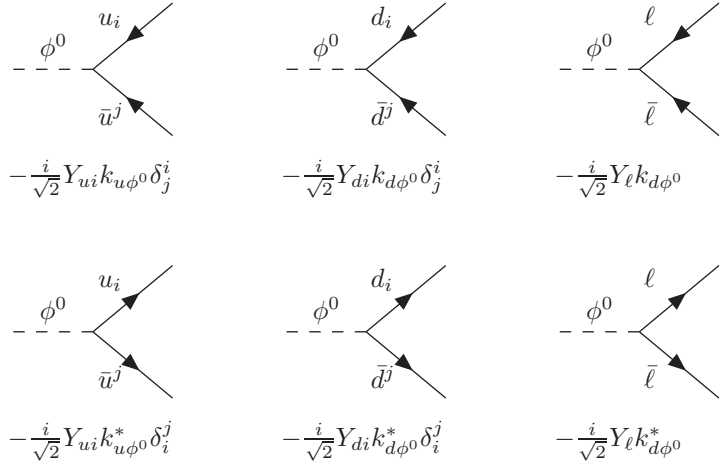
\begin{figure}[tp!]
\begin{center}
%%%%%%%%%%%%%%%%%%%%%%%%%%
\begin{picture}(80,70)(0,-20)
\DashLine(30,25)(0,25)5
\ArrowLine(60,50)(30,25)
\ArrowLine(60,0)(30,25)
\put(10,30){$\phi^0$}
\put(32,4){${\bar u}^{j}$}
\put(32,42){$u_i$}
%\put(64,50){$\alpha$}
%\put(64,0){$\beta$}
\put(-2,-18){$-\nicefrac{i}{\sqrt{2}} Y_{ui} k_{u\phi^0}
              \delta_j^i \, %\delta_{\alpha}{}^{\beta} 
              $}
\end{picture}
~~~~~
%%%%%%%%%%%%%%%%%%%%%%%%%%
\begin{picture}(80,70)(0,-20)
\DashLine(30,25)(0,25)5
\ArrowLine(60,50)(30,25)
\ArrowLine(60,0)(30,25)
\put(10,30){$\phi^0$}
\put(32,4){${\bar d}^{j}$}
\put(32,42){$d_i$}
%\put(64,50){$\alpha$}
%\put(64,0){$\beta$}
\put(-2,-18){$-\nicefrac{i}{\sqrt{2}} Y_{di} k_{d\phi^0}
              \delta_j^i \, %\delta_{\alpha}{}^{\beta} 
              $}
\end{picture}
%%%%%%%%%%%%%%%%%%%%%%%%%%
~~~~~
%%%%%%%%%%%%%%%%%%%%%%%%%%
\begin{picture}(80,70)(0,-20)
\DashLine(30,25)(0,25)5
\ArrowLine(60,50)(30,25)
\ArrowLine(60,0)(30,25)
\put(10,30){$\phi^0$}
\put(32,4){${\bar \ell}$}
\put(32,42){$\ell$}
%\put(64,50){$\alpha$}
%\put(64,0){$\beta$}
\put(-2,-18){$-\nicefrac{i}{\sqrt{2}} Y_{\ell} k_{d\phi^0}
              \, %\delta_{\alpha}{}^{\beta} 
              $}
\end{picture}
%%%%%%%%%%%%%%%%%%%%%%%%%%

\vspace{1cm}

%%%%%%%%%%%%%%%%%%%%%%%%%%
\begin{picture}(80,70)(0,-20)
\DashLine(30,25)(0,25)5
\ArrowLine(30,25)(60,50)
\ArrowLine(30,25)(60,0)
\put(10,30){$\phi^0$}
\put(32,4){${\bar u}^{j}$}
\put(32,42){$u_i$}
%\put(64,50){$\dot\alpha$}
%\put(64,0){$\dot\beta$}
\put(-2,-18){$-\nicefrac{i}{\sqrt{2}} Y_{ui} k_{u\phi^0}^* \delta_i^j
              \, 
              %\delta^{\dot{\alpha}}{}_{\dot{\beta}}
              $}
\end{picture}
%%%%%%%%%%%%%%%%%%%%%%%%%%
~~~~~
%%%%%%%%%%%%%%%%%%%%%%%%%%
\begin{picture}(80,70)(0,-20)
\DashLine(30,25)(0,25)5
\ArrowLine(30,25)(60,50)
\ArrowLine(30,25)(60,0)
\put(10,30){$\phi^0$}
\put(32,4){${\bar d}^{j}$}
\put(32,42){$d_i$}
%\put(64,50){$\dot\alpha$}
%\put(64,0){$\dot\beta$}
\put(-2,-18){$-\nicefrac{i}{\sqrt{2}} Y_{di} k_{d\phi^0}^* \delta_i^j
              \, %\delta^{\dot{\alpha}}{}_{\dot{\beta}}
              $}
\end{picture}
%%%%%%%%%%%%%%%%%%%%%%%%%%
~~~~~
%%%%%%%%%%%%%%%%%%%%%%%%%%
\begin{picture}(80,70)(0,-20)
\DashLine(30,25)(0,25)5
\ArrowLine(30,25)(60,50)
\ArrowLine(30,25)(60,0)
\put(10,30){$\phi^0$}
\put(32,4){${\bar \ell}$}
\put(32,42){$\ell$}
%\put(64,50){$\dot\alpha$}
%\put(64,0){$\dot\beta$}
\put(-2,-18){$-\nicefrac{i}{\sqrt{2}} Y_{\ell} k_{d\phi^0}^* \, 
              %\delta^{\dot{\alpha}}{}_{\dot{\beta}}
              $}
\end{picture}
\end{center}
%%%%%%%%%%%%%%%%%%%%%%%%%%
\caption{\label{nehiggsqq}Feynman rules for the interactions of neutral
Higgs bosons $\phi^0 = (h^0, H^0, A^0, G^0)$ with fermion-antifermion
pairs in the MSSM. The repeated index $i$ is not summed.}
\end{figure}

\begin{figure}[tbp!]
\begin{center}
%%%%%%%%%%%%%%%%%%%%%%%%%%
\begin{picture}(80,75)(0,-20)
\DashArrowLine(0,25)(30,25)5
\ArrowLine(60,50)(30,25)
\ArrowLine(60,0)(30,25)
\put(12,32){$\phi^+$}
\put(32,4){$d_i$}
\put(32,42){$\bar u^j$}
%\put(64,50){$\alpha$}
%\put(64,0){$\beta$}
\put(-2,-18){$i Y_{uj} [\mathbold{K}]_j{}^i
k_{u\phi^\pm}
\, %\delta_{\alpha}{}^{\beta} 
$}
\end{picture}
~~~~~
\begin{picture}(80,75)(0,-20)
\DashArrowLine(0,25)(30,25)5
\ArrowLine(60,50)(30,25)
\ArrowLine(60,0)(30,25)
\put(12,32){$\phi^-$}
\put(32,4){$\bar d^j$}
\put(32,42){$u_i$}
%\put(64,50){$\alpha$}
%\put(64,0){$\beta$}
\put(-2,-18){$iY_{dj}[\mathbold{K}^\dagger]_j{}^i k_{d\phi^\pm}
%\,\delta_{\alpha}{}^{\beta}
$}
\end{picture}
~~~~~
\begin{picture}(80,75)(0,-20)
\DashArrowLine(0,25)(30,25)5
\ArrowLine(60,50)(30,25)
\ArrowLine(60,0)(30,25)
\put(12,32){$\phi^-$}
\put(32,4){$\bar \ell$}
\put(32,42){$\nu_\ell$}
%\put(64,50){$\alpha$}
%\put(64,0){$\beta$}
\put(5,-18){$iY_{\ell} k_{d\phi^\pm}
%\,\delta_{\alpha}{}^{\beta}
$}
\end{picture}
%%%%%%%%%%%%%%%%%%%%%%%%%%

\vspace{1cm}

%%%%%%%%%%%%%%%%%%%%%%%%%%
\begin{picture}(80,70)(0,-20)
\DashArrowLine(0,25)(30,25)5
\ArrowLine(30,25)(60,50)
\ArrowLine(30,25)(60,0)
\put(11,32){$\phi^-$}
\put(32,4){$d_i$}
\put(32,42){$\bar u^j$}
%\put(64,50){$\dot\alpha$}
%\put(64,0){$\dot\beta$}
\put(-2,-18){$i Y_{uj} [\mathbold{K}^\dagger]_i{}^j
k_{u\phi^\pm}
%\, \delta^{\dot\alpha}{}_{\dot\beta} 
$}
\end{picture}
~~~~~
\begin{picture}(80,70)(0,-20)
\DashArrowLine(0,25)(30,25)5
\ArrowLine(30,25)(60,50)
\ArrowLine(30,25)(60,0)
\put(10,32){$\phi^+$}
\put(32,4){$\bar d^j$}
\put(32,42){$u_i$}
%\put(64,50){$\dot\alpha$}
%\put(64,0){$\dot\beta$}
\put(-2,-18){$iY_{dj}[\mathbold{K}]_i{}^j k_{d\phi^\pm}
%\, \delta^{\dot\alpha}{}_{\dot\beta}
$}
\end{picture}
~~~~~
\begin{picture}(80,70)(0,-20)
\DashArrowLine(0,25)(30,25)5
\ArrowLine(30,25)(60,50)
\ArrowLine(30,25)(60,0)
\put(10,32){$\phi^+$}
\put(32,4){$\bar \ell$}
\put(32,42){$\nu_\ell$}
%\put(64,50){$\dot\alpha$}
%\put(64,0){$\dot\beta$}
\put(5,-18){$iY_{\ell} k_{d\phi^\pm}
%\, \delta^{\dot\alpha}{}_{\dot\beta}
$}
\end{picture}
%%%%%%%%%%%%%%%%%%%%%
\end{center}
\caption{Feynman rules for the interactions of charged Higgs bosons
$\phi^\pm = (H^\pm,G^\pm)$ with fermion-antifermion pairs in the MSSM.
The meaning of the arrows on 
the scalar lines is that the $\phi^\pm$ line carry charges  $\pm 1$ into 
the vertex.
The repeated index $j$ is not summed.}
\label{fig:chiggsqq}
\end{figure}
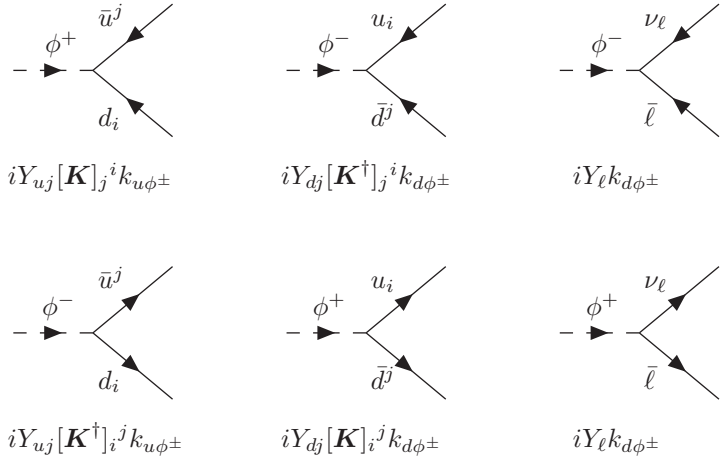

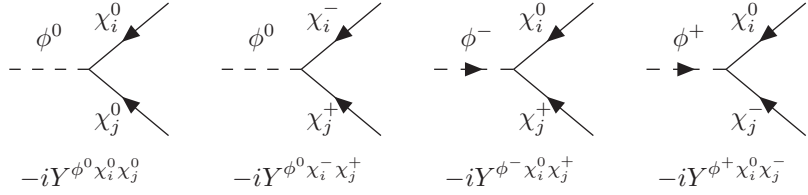
\begin{figure}[tbp!]
\begin{center}
%%%%%%%%%%%%%%%%%%%%%%%%%%%%%%%%%%%%%%%%%%%%%%%%%%%%%%%%%%%%%%%%%%%%%%%%%
\begin{picture}(77,70)(0,-20)
\DashLine(30,25)(0,25){5}
\ArrowLine(60,50)(30,25)
\ArrowLine(60,0)(30,25)
\put(10,34){$\phi^0$}
\put(32,42){$\chi_i^0$}
\put(32,4){$\chi_j^0$}
%\put(64,50){$\dot\alpha$}
%\put(64,0){$\beta$}
\put(4,-19){$-iY^{\phi^0 \chi_i^0 \chi_j^0}$}
\end{picture}
%%%%%%%%%%%%%%%%%%%%%%%%%%%%%%%%%%%%
\begin{picture}(77,70)(0,-20)
\DashLine(30,25)(0,25){5}
\ArrowLine(60,50)(30,25)
\ArrowLine(60,0)(30,25)
\put(10,34){$\phi^0$}
\put(32,42){$\chi_i^-$}
\put(32,4){$\chi_j^+$}
%\put(64,50){$\dot\alpha$}
%\put(64,0){$\beta$}
\put(4,-19){$-iY^{\phi^0 \chi_i^- \chi_j^+}$}
\end{picture}
%%%%%%%%%%%%%%%%%%%%%%%%%%%%%%%%%%%%
%%%%%%%%%%%%%%%%%%%%%%%%%%%%%%%%%%%%%%%%%%%%%%%%%%%%%%%%%%%%%%%%%%%%%%%%%
\begin{picture}(77,70)(0,-20)
\DashArrowLine(0,25)(30,25){5}
\ArrowLine(60,50)(30,25)
\ArrowLine(60,0)(30,25)
\put(10,34){$\phi^-$}
\put(32,42){$\chi_i^0$}
\put(32,4){$\chi_j^+$}
%\put(64,50){$\dot\alpha$}
%\put(64,0){$\beta$}
\put(4,-19){$-iY^{\phi^- \chi_i^0 \chi_j^+}$}
\end{picture}
%%%%%%%%%%%%%%%%%%%%%%%%%%%%%%%%%%%%
\begin{picture}(77,70)(0,-20)
\DashArrowLine(0,25)(30,25){5}
\ArrowLine(60,50)(30,25)
\ArrowLine(60,0)(30,25)
\put(10,34){$\phi^+$}
\put(32,42){$\chi_i^0$}
\put(32,4){$\chi_j^-$}
%\put(64,50){$\dot\alpha$}
%\put(64,0){$\beta$}
\put(4,-19){$-iY^{\phi^+ \chi_i^0 \chi_j^-}$}
\end{picture}
%%%%%%%%%%%%%%%%%%%%%%%%%%%%%%%%%%%%
\end{center}
\caption{Feynman rules for the chargino and neutralino interactions with 
Higgs bosons in the MSSM. The couplings are defined in 
eqs.~(\ref{eq:defOL})-(\ref{eq:defOLpp}).  For each rule, there is a corresponding one with all arrows 
reversed, and the $Y$ coupling (without the explicit $i$) replaced by its 
complex conjugate.} \label{Higgschichi}
\end{figure}

Using the above notation, the interactions of quarks and leptons with
the neutral Higgs bosons are as shown in Figures \ref{nehiggsqq} and
\ref{fig:chiggsqq}.
%%%%%%%%%%%%%%%%%%%%%%%%%%%%%%%%%%%%%%%%%%%%%%%%%%%%%%%%%%%
%%%%%%%%%%%%%%%%%%%%%%%%%%%%%%%%%%%%%%%%%%%%%%%%%%%%%%%%%%%
%%%%%%%%%%%%%%%%%%%%%%%%%%%%%%%%%%%%%%%%%%%%%%%%%%%%%%%%%%%%%%%%%%%%%%%%%
The rules for Higgs boson couplings to neutralinos and charginos are 
shown 
in Figure \ref{Higgschichi}, in terms of
\beqa
Y^{\phi^0\chi^0_i \chi^0_j} &=&
\frac{1}{2} (k_{d\phi^0}^* N_{i3}^*  -  k_{u\phi^0}^* N_{i4}^*)
(g N_{j2}^* - g' N_{j1}^*) + (i \leftrightarrow j)\,,
\label{higgs-gauginos1}
\\
Y^{\phi^0\chi^-_i \chi^+_j} &=&
\frac{g}{\sqrt{2}} (
   k_{u\phi^0}^* U_{i1}^* V_{j2}^*
  +k_{d\phi^0}^* U_{i2}^* V_{j1}^*  )\,,\label{higgs-gauginos2}
\eeqa\beqa
Y^{\phi^+\chi_i^0\chi^-_j} &=&
k_{d\phi^\pm} \bigl [
g (N_{i3}^* U_{j1}^*  - \frac{1}{\sqrt{2}} N_{i2}^* U_{j2}^* )
-\frac{g'}{\sqrt{2}} N_{i1}^* U_{j2}^* \bigr ]\,,\label{higgs-gauginos3}
\\
Y^{\phi^-\chi_i^0\chi^+_j} &=&
k_{u\phi^\pm} \bigl [
g (N_{i4}^* V_{j1}^*  + \frac{1}{\sqrt{2}} N_{i2}^* V_{j2}^* )
+\frac{g'}{\sqrt{2}} N_{i1}^* V_{j2}^*  \bigr ]\,,
\label{higgs-gauginos4}   
\eeqa
for $\phi^0=h^0, H^0, A^0, G^0$ and $\phi^\pm=H^\pm, G^\pm$.
%%%%%%%%%%%%%%%%%%%%%%%%%%%%%%%%%%%%%%%%%%%%%%%%%%%%%%%%%%%
%%%%%%%%%%%%%%%%%%%%%%%%%%%%%%%%%%%%%%%%%%%%%%%%%%%%%%%%%%%%%%%%%%%%%%%%%

\begin{figure}[tbp!]
\begin{center}
%%%%%%%%%%%%%%%%%%%%%%%%%%
\begin{picture}(78,70)(0,-20)
\DashArrowLine(30,25)(0,25)5
\ArrowLine(60,50)(30,25)
\ArrowLine(60,0)(30,25)
\put(10,30){$\widetilde d_{Lj}$}
\put(32,42){$\chi_i^-$}
\put(32,4){$u_k$}
\put(-2,-18){$-i g U^*_{i1} [\mathbold{K}^\dagger]_j{}^k$}
\end{picture}
%%%%%%%%%%%%%%%%%%%%%%%%%%
\begin{picture}(78,70)(0,-20)
\DashArrowLine(30,25)(0,25)5
\ArrowLine(60,50)(30,25)
\ArrowLine(60,0)(30,25)
\put(10,30){$\widetilde u_{Lj}$}
\put(32,42){$\chi_i^+$}
\put(32,4){$d_k$}
\put(-2,-18){$-i g V^*_{i1}[\mathbold{K}]_j{}^k$}
\end{picture}
%%%%%%%%%%%%%%%%%%%%%%%%%%
\begin{picture}(78,70)(0,-20)
\DashArrowLine(0,25)(30,25)5
\ArrowLine(60,50)(30,25)
\ArrowLine(60,0)(30,25)
\put(10,30){$\widetilde d_{Lj}$}
\put(32,42){$\chi_i^+$}
\put(32,4){$\bar u_k$}
\put(-2,-18){$iV_{i2}^*[\mathbold{K}]_k{}^j Y_{uk}$}
\end{picture}
%%%%%%%%%%%%%%%%%%%%%%%%%%
\begin{picture}(76,70)(0,-20)
\DashArrowLine(0,25)(30,25)5
\ArrowLine(60,50)(30,25)
\ArrowLine(60,0)(30,25)
\put(10,30){$\widetilde u_{Lj}$}
\put(32,42){$\chi_i^-$}
\put(32,4){$\bar d_k$}
\put(-2,-18){$i U_{i2}^* [\mathbold{K}^\dagger]_k{}^j Y_{dj}$}
\end{picture}
%%%%%%%%%%%%%%%%%%%%%%%%%%%

\vspace{1cm}

%%%%%%%%%%%%%%%%%%%%%%%%%%
\begin{picture}(76,70)(0,-20)
\DashArrowLine(30,25)(0,25)5
\ArrowLine(60,50)(30,25)
\ArrowLine(60,0)(30,25)
\put(10,30){$\widetilde d_{Rj}$}
\put(32,42){$\chi_i^-$}
\put(32,4){$u_k$}
\put(-2,-18){$i
U_{i2}^* [\mathbold{K}^\dagger]_j{}^k Y_{dj}$}
\end{picture}
~~~~~~~~~~~~
%%%%%%%%%%%%%%%%%%%%%%%%%%
\begin{picture}(76,70)(0,-20)
\DashArrowLine(30,25)(0,25)5
\ArrowLine(60,50)(30,25)
\ArrowLine(60,0)(30,25)
\put(10,30){$\widetilde u_{Rj}$}
\put(32,42){$\chi_i^+$}
\put(32,4){$d_k$}
\put(-2,-18){$i V_{i2}^* [\mathbold{K}]_j{}^k Y_{uj}$}
\end{picture}
%%%%%%%%%%%%%%%%%%%%%%%%%%

\vspace{1cm}

%%%%%%%%%%%%%%%%%%%%%%%%%%
\begin{picture}(76,70)(0,-20)
\DashArrowLine(30,25)(0,25)5
\ArrowLine(60,50)(30,25)
\ArrowLine(60,0)(30,25)
\put(10,30){$\widetilde \ell_{L}$}
\put(32,42){$\chi_i^-$}
\put(32,4){$\nu_\ell$}
\put(8,-18){$-i g U^*_{i1}$}
\end{picture}
%%%%%%%%%%%%%%%%%%%%%%%%%%
\begin{picture}(76,70)(0,-20)
\DashArrowLine(30,25)(0,25)5
\ArrowLine(60,50)(30,25)
\ArrowLine(60,0)(30,25)
\put(10,30){$\widetilde \nu_{\ell}$}
\put(32,42){$\chi_i^+$}
\put(32,4){$\ell$}
\put(8,-18){$-i g V^*_{i1}$}
\end{picture}
%%%%%%%%%%%%%%%%%%%%%%%%%%
\begin{picture}(76,70)(0,-20)
\DashArrowLine(30,25)(0,25)5
\ArrowLine(60,50)(30,25)
\ArrowLine(60,0)(30,25)
\put(10,30){$\widetilde \ell_{R}$}
\put(32,42){$\chi_i^-$}
\put(32,4){$\nu_\ell$}
\put(11,-18){$i U_{i2}^* Y_\ell$}
\end{picture}
%%%%%%%%%%%%%%%%%%%%%%%%%%
\begin{picture}(76,70)(0,-20)
\DashArrowLine(0,25)(30,25)5
\ArrowLine(60,50)(30,25)
\ArrowLine(60,0)(30,25)
\put(10,30){$\widetilde \nu_{\ell}$}
\put(32,42){$\chi_i^-$}
\put(32,4){$\bar \ell$}
\put(12,-18){$i U_{i2}^* Y_\ell$}
\end{picture}
%%%%%%%%%%%%%%%%%%%%%%%%%%
\end{center}
\caption{Feynman rules for charginos interactions with
fermion/sfermion pairs in the MSSM.  
The fermions are taken to be
in a mass eigenstate basis, and the sfermions are in a basis
whose elements are the supersymmetric partners of the fermions.
This is usually considered to be a good approximation for the squarks and sleptons
of the first two families.
For each rule, there is a corresponding one
with all arrows reversed and the coupling (without the explicit $i$)
replaced by its complex conjugate.
}
\label{fig:cqsq}
\end{figure}
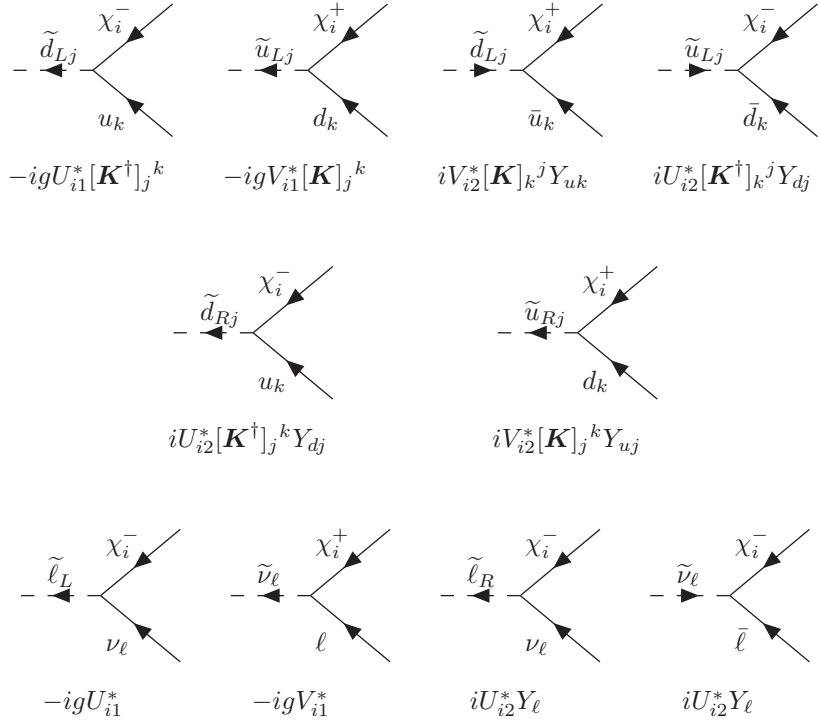

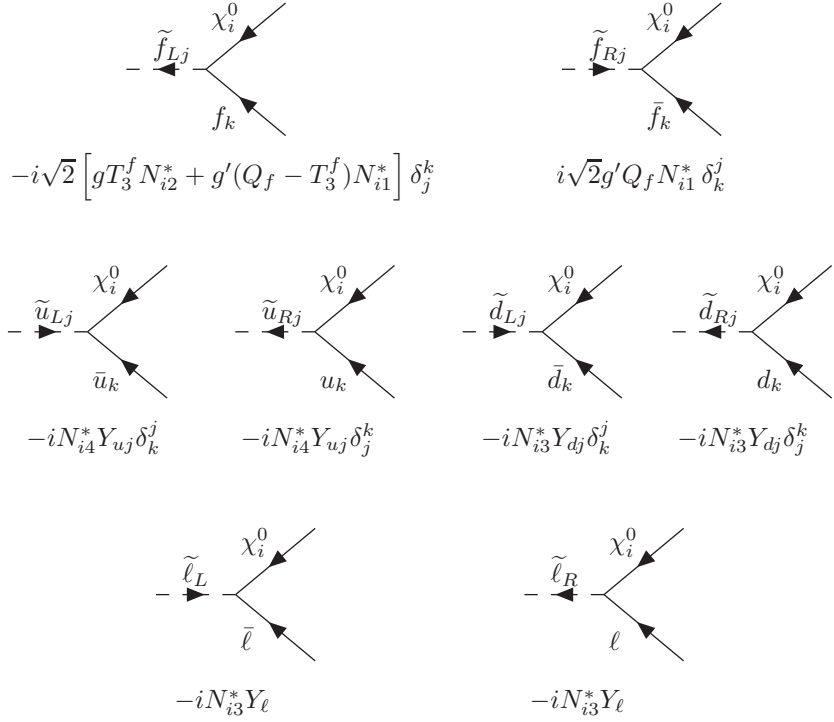
\begin{figure}[tbp!]
\begin{center}
%%%%%%%%%%%%%%%%%%%%%%%%%%
\begin{picture}(78,70)(0,-20)
\DashArrowLine(30,25)(0,25)5
\ArrowLine(60,50)(30,25)
\ArrowLine(60,0)(30,25)
\put(10,30){$\widetilde f_{Lj}$}
\put(32,42){$\chi_i^0$}
\put(32,4){$f_k$}
\put(-44,-18){$-i \sqrt{2} \left [ g T_3^f N_{i2}^* + g' (Q_f - T_3^f) N_{i1}^* \right ]
\delta_j^k$}
\end{picture}
%%%%%%%%%%%%%%%%%%%%%%%%%%
~~~~~~~~~~~~~~~~~~~~~~~~
%%%%%%%%%%%%%%%%%%%%%%%%%%
\begin{picture}(73,70)(0,-20)
\DashArrowLine(0,25)(30,25)5
\ArrowLine(60,50)(30,25)
\ArrowLine(60,0)(30,25)
\put(10,30){$\widetilde f_{Rj}$}
\put(32,42){$\chi_i^0$}
\put(32,4){$\bar f_k$}
\put(-2,-18){$i \sqrt{2} g' Q_f N_{i1}^*\,\delta_k^j$}
\end{picture}
%%%%%%%%%%%%%%%%%%%%%%%%%%

\vspace{1cm}

%%%%%%%%%%%%%%%%%%%%%%%%%%
\begin{picture}(76,70)(0,-20)
\DashArrowLine(0,25)(30,25)5
\ArrowLine(60,50)(30,25)
\ArrowLine(60,0)(30,25)
\put(10,30){$\widetilde u_{Lj}$}
\put(32,42){$\chi_i^0$}
\put(32,4){$\bar u_k$}
\put(7,-18){$-i N_{i4}^* Y_{uj} \delta_k^j$}
\end{picture}
~
%%%%%%%%%%%%%%%%%%%%%%%%%%
\begin{picture}(76,70)(0,-20)
\DashArrowLine(30,25)(0,25)5
\ArrowLine(60,50)(30,25)
\ArrowLine(60,0)(30,25)
\put(10,30){$\widetilde u_{Rj}$}
\put(32,42){$\chi_i^0$}
\put(32,4){$u_k$}
\put(2,-18){$-i N_{i4}^* Y_{uj} \delta_j^k$}
\end{picture}
~
%%%%%%%%%%%%%%%%%%%%%%%%%%
\begin{picture}(76,70)(0,-20)
\DashArrowLine(0,25)(30,25)5
\ArrowLine(60,50)(30,25)
\ArrowLine(60,0)(30,25)
\put(10,30){$\widetilde d_{Lj}$}
\put(32,42){$\chi_i^0$}
\put(32,4){$\bar d_k$}
\put(7,-18){$-i N_{i3}^* Y_{dj} \delta_k^j$}
\end{picture}
%%%%%%%%%%%%%%%%%%%%%%%%%%
%%%%%%%%%%%%%%%%%%%%%%%%%%
\begin{picture}(76,70)(0,-20)
\DashArrowLine(30,25)(0,25)5
\ArrowLine(60,50)(30,25)
\ArrowLine(60,0)(30,25)
\put(10,30){$\widetilde d_{Rj}$}
\put(32,42){$\chi_i^0$}
\put(32,4){$d_k$}
\put(2,-18){$-i N_{i3}^* Y_{dj} \delta_j^k$}
\end{picture}
%%%%%%%%%%%%%%%%%%%%%%%%%%

\vspace{1cm}

%%%%%%%%%%%%%%%%%%%%%%%%%%
\begin{picture}(76,70)(0,-20)
\DashArrowLine(0,25)(30,25)5
\ArrowLine(60,50)(30,25)
\ArrowLine(60,0)(30,25)
\put(10,30){$\widetilde \ell_{L}$}
\put(32,42){$\chi_i^0$}
\put(32,4){$\bar \ell$}
\put(7,-18){$-i N_{i3}^* Y_{\ell}$}
\end{picture}
%%%%%%%%%%%%%%%%%%%%%%%%%%
~~~~~~~~~~~~~~~~~
%%%%%%%%%%%%%%%%%%%%%%%%%%
\begin{picture}(76,70)(0,-20)
\DashArrowLine(30,25)(0,25)5
\ArrowLine(60,50)(30,25)
\ArrowLine(60,0)(30,25)
\put(10,30){$\widetilde \ell_{R}$}
\put(32,42){$\chi_i^0$}
\put(32,4){$\ell$}
\put(2,-18){$-i N_{i3}^* Y_\ell$}
\end{picture}
%%%%%%%%%%%%%%%%%%%%%%%%%%

%%%%%%%%%%%%%%%%%%%%%
\end{center}
\caption{Feynman rules for the interactions of neutralinos with first and second family
fermion/sfermion pairs in the MSSM.
The comments on Figure \ref{fig:cqsq} also apply here.
%The fermions are taken to be
%in a mass eigenstate basis, and the sfermions are in a basis
%whose elements are the supersymmetric partners of them.
%For each rule, there is a corresponding one
%with all arrows reversed, undotted indices changed to dotted indices with
%the opposite height, and the coupling (without the explicit $i$)
%replaced by its complex conjugate.
%The case of more general sfermion mixing can be obtained by taking linear combinations of these rules.
}
\label{fig:nqsq}
\end{figure}

\begin{figure}[tbp!]
\begin{center}
%%%%%%%%%%%%%%%%%%%%%%%%%%
\begin{picture}(79,70)(0,-20)
\DashArrowLine(30,25)(0,25)5
\ArrowLine(60,50)(30,25)
\ArrowLine(60,0)(30,25)
\put(8,32){$\widetilde q_{Lm}$}
\put(32,42){$q_n$}
\put(32,4){$\widetilde g_a$}
\put(0,-18){$-i \sqrt{2} g_3 \boldsymbol{T}_{\!m}^{\boldsymbol{a}n}$}
\end{picture}
%%%%%%%%%%%%%%%%%%%%%%%%%%
\begin{picture}(79,70)(0,-20)
\DashArrowLine(0,25)(30,25)5
\ArrowLine(30,25)(60,50)
\ArrowLine(30,25)(60,0)
\put(8,32){$\widetilde q_{Ln}$}
\put(32,42){$q_m$}
\put(32,4){$\widetilde g_a$}
\put(0,-18){$-i \sqrt{2} g_3  \boldsymbol{T}_{\!m}^{\boldsymbol{a}n}$}
\end{picture}
%%%%%%%%%%%%%%%%%%%%%%%%%%
%%%%%%%%%%%%%%%%%%%%%%%%%%
\begin{picture}(79,70)(0,-20)
\DashArrowLine(30,25)(0,25)5
\ArrowLine(60,50)(30,25)
\ArrowLine(60,0)(30,25)
\put(8,32){$\widetilde q^{*n}_{R}$}
\put(32,42){$\bar q_m$}
\put(32,4){$\widetilde g_a$}
\put(1,-18){$i \sqrt{2} g_3 \boldsymbol{T}_{\!m}^{\boldsymbol{a}n}$}
\end{picture}
%%%%%%%%%%%%%%%%%%%%%%%%%%
\begin{picture}(78,70)(0,-20)
\DashArrowLine(0,25)(30,25)5
\ArrowLine(30,25)(60,50)
\ArrowLine(30,25)(60,0)
\put(8,32){$\widetilde q^{*m}_{R}$}
\put(32,42){$\bar q_n$}
\put(32,4){$\widetilde g_a$}
\put(1,-18){$i \sqrt{2} g_3 \boldsymbol{T}_{\!m}^{\boldsymbol{a}n} $}
\end{picture}
%%%%%%%%%%%%%%%%%%%%%%%%%%%%
\end{center}
\caption{Feynman rules for gluino interactions with first and second family
quark/squark pairs in the MSSM. The indices $m,n$ are for the fundamental
representation of $SU(3)_c$, and $a$ is an adjoint representation index. 
The comments on Figure \ref{fig:cqsq} also apply here.
%The quarks are taken to be
%in a mass eigenstate basis, and the squarks are in a basis
%whose elements are the supersymmetric partners of the quarks.
%For each rule, there is a corresponding one
%with all arrows reversed and the coupling (without the explicit $i$)
%replaced by its complex conjugate.
%The case of more general squark mixing can be obtained by 
%taking linear combinations of these rules.
}
\label{fig:gluinoqsq}
\end{figure}
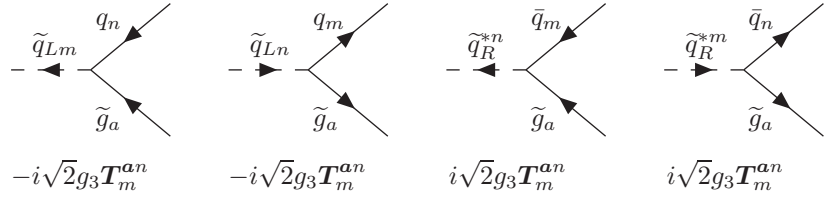

Feynman rules for sfermion-fermion in interactions with charginos, neutralinos, and the gluino 
in the MSSM appear in 
Figures \ref{fig:cqsq}, \ref{fig:nqsq}, and \ref{fig:gluinoqsq}, 
respectively. 
%%%%%%%%%%%%%%%%%%%%%%%%%%%%%%%%%%%%%%%%%%%%%%%%%%%%%%%%%%%%%%%%%%%%%%%%%%%%
%%%%%%%%%%%%%%%%%%%%%%%%%%%%%%%%%%%%%%%%%%%%%%%%%%%%%%%%%%%%%%%%%%%%%%%%%
%%%%%%%%%%%%%%%%%%%%%%%%%%%%%%%%%%%%%%%%%%%%%%%%%%%%%%%%%%%
%%%%%%%%%%%%%%%%%%%%%%%%%%%%%%%%%%%%%%%%%%%%%%%%%%%%%%%%%%%
%%%%%%%%%%%%%%%%%%%%%%%%%%%%%%%%%%%%%%%%%%%%%%%%%%%%%%%%%%%%%%%%%%%%%%%%%
In these rules, the Standard Model quarks and leptons are 
assumed to be in the mass eigenstate bases, and the squarks and sleptons 
are assumed to be in the basis defined by superpartners of the fermion 
mass eigenstates.

However, in principle all sfermions with a given 
electric charge can mix with each other. There is a popular,
and perhaps phenomenologically and theoretically favored, approximation
in which
only the sfermions of the third family have significant mixing. For $f
= t,b,\tau$, one can then write the relationship between the gauge
eigenstates $\widetilde f_L$, $\widetilde f_R$ and the mass eigenstates
$\widetilde f_1$, $\widetilde f_2$ as
\beq
\begin{pmatrix}
\widetilde f_R
\\
\widetilde f_L
\end{pmatrix}
= X_{\widetilde f}
\begin{pmatrix}
\widetilde f_1
\\
\widetilde f_2
\end{pmatrix}
\,,\qquad\quad
X_{\widetilde f} \equiv
\begin{pmatrix}
R_{\widetilde f_1} & R_{\widetilde f_2}
\\
L_{\widetilde f_1} & L_{\widetilde f_2}
\end{pmatrix}\,,
\label{eq:sfermionmix}
\eeq
where $X$ is a $2\times 2$ unitary matrix. 

[One can choose
$R_{\widetilde f_1} = L^*_{\widetilde f_2} = c_{\widetilde f}$,
and
$L_{\widetilde f_1} = -R^*_{\widetilde f_2} = s_{\widetilde f}$,
with
%\beqa
$
|c_{\widetilde f}|^2 + |s_{\widetilde f}|^2 = 1
%\label{eq:cfsfunitary}
%\eeqa
$.
If there is no CP violation, then
$c_{\widetilde f}$ and $s_{\widetilde f}$ can be taken real,
and they are the cosine
and sine of a sfermion mixing angle. This convention for
$c_{\widetilde f}, s_{\widetilde f}$ has the nice property that for zero mixing angle,
$\stilde f_1 = \stilde f_R$ and $\stilde f_2 = \stilde f_L$.
Various other conventions are found in the literature. 
We use $R_{\widetilde f_i}$ and $L_{\widetilde f_i}$ in the Feynman rules,
rather than
$c_{\widetilde f}$ and $s_{\widetilde f}$,
to make it easier to compare to your favorite mixing angle
convention using eq.~(\ref{eq:sfermionmix}).]

The resulting Feynman rules for chargino, neutralino, and gluino interactions with
third-family squarks and sleptons that mix
within each generation are shown in Figures \ref{fig:cqsqmixed}, \ref{fig:nqsqmixed},
and \ref{fig:gluinoqsqmixed}. 
%%%%%%%%%%%%%%%%%%%%%%%%%%%%%%%%%%%%%%%%%%%%%%%%%%%%%%%%%%%%%%%%%%%%%%%%%%%%
\begin{figure}[tbp!]
\begin{center}
%%%%%%%%%%%%%%%%%%%%%%%%%%
\begin{picture}(76.5,70)(0,-20)
\DashArrowLine(0,25)(30,25)5
\ArrowLine(60,50)(30,25)
\ArrowLine(60,0)(30,25)
\put(10,32){$\widetilde t_j$}
\put(32,42){$\chi_i^-$}
\put(32,4){$\bar b$}
\put(8,-18){$
i Y_b U_{i2}^* L_{\widetilde t_j}
$}
\end{picture}
%%%%%%%%%%%%%%%%%%%%%%%%%%
\begin{picture}(79.5,70)(0,-20)
\DashArrowLine(30,25)(0,25)5
\ArrowLine(60,50)(30,25)
\ArrowLine(60,0)(30,25)
\put(10,32){$\widetilde t_{j}$}
\put(32,42){$\chi_i^+$}
\put(32,4){$b$}
\put(-9.5,-18){$
i (Y_t V_{i2}^* R_{\widetilde t_j}^*\!\! -g V^*_{i1} L_{\widetilde t_j}^*\!)
$}
\end{picture}
%%%%%%%%%%%%%%%%%%%%%%%%%%
\begin{picture}(76.5,70)(0,-20)
\DashArrowLine(0,25)(30,25)5
\ArrowLine(60,50)(30,25)
\ArrowLine(60,0)(30,25)
\put(10,32){$\widetilde b_j$}
\put(32,42){$\chi_i^+$}
\put(32,4){$\bar t$}
\put(10,-18){$i Y_t V_{i2}^* L_{\widetilde b_j}$}
\end{picture}
%%%%%%%%%%%%%%%%%%%%%%%%%%%
\begin{picture}(78,70)(0,-20)
\DashArrowLine(30,25)(0,25)5
\ArrowLine(60,50)(30,25)
\ArrowLine(60,0)(30,25)
\put(10,32){$\widetilde b_j$}
\put(32,42){$\chi_i^-$}
\put(32,4){$t$}
\put(-12.5,-18){$i (Y_b U_{i2}^* R_{\widetilde b_j}^*\!\!-g U^*_{i1} L_{\widetilde 
b_j}^*\!)$}
\end{picture}
%%%%%%%%%%%%%%%%%%%%%%%%%%

\vspace{1cm}

%%%%%%%%%%%%%%%%%%%%%%%%%%
\begin{picture}(76,70)(0,-20)
\DashArrowLine(30,25)(0,25)5
\ArrowLine(60,50)(30,25)
\ArrowLine(60,0)(30,25)
\put(10,32){$\widetilde \nu_\tau$}
\put(32,42){$\chi_i^+$}
\put(32,4){$\tau$}
\put(9,-18){$-i g V^*_{i1}$}
\end{picture}
%%%%%%%%%%%%%%%%%%%%%%%%%%
~~~~
\begin{picture}(76,70)(0,-20)
\DashArrowLine(0,25)(30,25)5
\ArrowLine(60,50)(30,25)
\ArrowLine(60,0)(30,25)
\put(10,32){$\widetilde \nu_{\tau}$}
\put(32,42){$\chi_i^-$}
\put(32,4){$\bar \tau$}
\put(10,-18){$i Y_\tau U_{i2}^* $}
\end{picture}
%%%%%%%%%%%%%%%%%%%%%%%%%%
~~~~
\begin{picture}(76,70)(0,-20)
\DashArrowLine(30,25)(0,25)5
\ArrowLine(60,50)(30,25)
\ArrowLine(60,0)(30,25)
\put(10,32){$\widetilde \tau_{j}$}
\put(32,42){$\chi_i^-$}
\put(32,4){$\nu_\tau$}
\put(-4,-18){$i(Y_\tau U_{i2}^*R_{\widetilde \tau_j}^*
- g U^*_{i1}L_{\widetilde \tau_j}^* )$}
\end{picture}
%%%%%%%%%%%%%%%%%%%%%%%%%%
\end{center}
\caption{Feynman rules for chargino interactions with
third-family fermion/sfermion pairs.  
The fermions are taken to be
in a mass eigenstate basis, and the sfermions are in the mass 
eigenstate basis of eq.~(\ref{eq:sfermionmix}).
%  The
%  corresponding rules for the first and second families with the
%  approximation of no mixing and vanishing fermion masses can be
% obtained from these by setting $Y_f = 0$ and $L_{\widetilde f_2} =
%  R_{\widetilde f_1} = 1$ and $L_{\widetilde f_1} = R_{\widetilde f_2} = 0$ (so
%  that $\stilde f_1 = \stilde f_R$ and $\stilde f_2 = \stilde f_L$).
For each rule, there is a corresponding one
with all arrows reversed and the coupling (without the explicit $i$)
replaced by its complex conjugate.
}
\label{fig:cqsqmixed}
\end{figure}
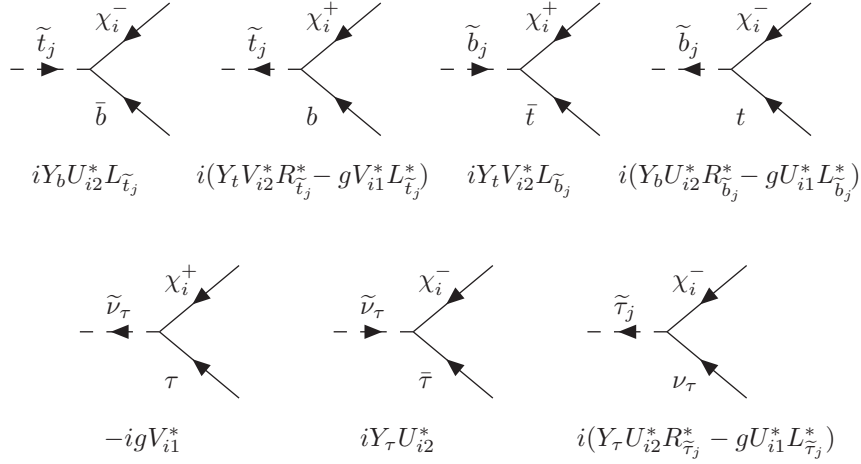
%%%%%%%%%%%%%%%%%%%%%%%%%%%%%%%%%%%%%%%%%%%%%%%%%%%%%%%%%%%%%%%%%%%%%%%%%
%%%%%%%%%%%%%%%%%%%%%%%%%%%%%%%%%%%%%%%%%%%%%%%%%%%%%%%%%%%
\begin{figure}[tbp!]
\begin{center}
%%%%%%%%%%%%%%%%%%%%%%%%%%
\begin{picture}(78,70)(0,-20)
\DashArrowLine(30,25)(0,25)5
\ArrowLine(60,50)(30,25)
\ArrowLine(60,0)(30,25)
\put(8,32){$\widetilde f_j$}
\put(32,42){$\chi^0_i$}
\put(32,4){$f$}
\put(2,-19){$-iY^{\widetilde f_j^*\! f \chi_i^0}$}
\end{picture}
%%%%%%%%%%%%%%%%%%%%%%%%%%
~~~~~~
%%%%%%%%%%%%%%%%%%%%%%%%%%
\begin{picture}(78,70)(0,-20)
\DashArrowLine(0,25)(30,25)5
\ArrowLine(60,50)(30,25)
\ArrowLine(60,0)(30,25)
\put(8,32){$\widetilde f_j$}
\put(32,42){$\chi^0_i$}
\put(32,4){$\bar f$}
\put(3,-19){$-iY^{\widetilde f_j \bar f \chi_i^0}$}
\end{picture}
%%%%%%%%%%%%%%%%%%%%%%%%%%
~~~~~~~
%%%%%%%%%%%%%%%%%%%%%%%%%%
\begin{picture}(78,70)(0,-20)
\DashArrowLine(30,25)(0,25)5
\ArrowLine(60,50)(30,25)
\ArrowLine(60,0)(30,25)
\put(8,32){$\widetilde \nu_\tau$}
\put(32,42){$\chi^0_i$}
\put(32,4){$\nu_\tau$}
\put(-5,-18){$-\frac{i}{\sqrt{2}} (g N_{i2}^* - g' N_{i1}^*)$}
\end{picture}
%%%%%%%%%%%%%%%%%%%%%%%%%%
\caption{Feynman rules for neutralino interactions with
third-family fermion/sfermion pairs in the MSSM. 
Here $f = t, b, \tau$, with couplings $Y^{\widetilde f_j^*\! f \chi_i^0}$
and $Y^{\widetilde f_j \bar f \chi_i^0}$
given in 
eqs.~(\ref{eq:stoptbarN})-(\ref{eq:stautaubarN}).
The comments on Figure \ref{fig:cqsqmixed} also apply here.}
\label{fig:nqsqmixed}
\end{center}
\end{figure}
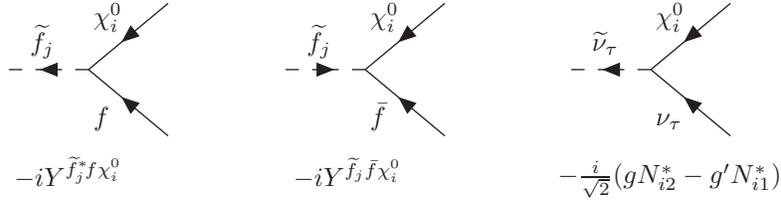
%%%%%%%%%%%%%%%%%%%%%%%%%%%%%%%%%%%%%%%%%%%%%%%%%%%%%%%%%%%%%%%%%%%%%%%%%
%%%%%%%%%%%%%%%%%%%%%%%%%%%%%%%%%%%%%%%%%%%%%%%%%%%%%%%%%%%
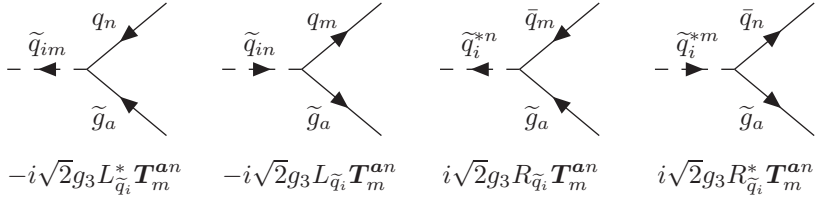
\begin{figure}[tbp!]
\begin{center}
%%%%%%%%%%%%%%%%%%%%%%%%%%
\begin{picture}(78,70)(0,-20)
\DashArrowLine(30,25)(0,25)5
\ArrowLine(60,50)(30,25)
\ArrowLine(60,0)(30,25)
\put(8,32){$\widetilde q_{im}$}
\put(32,42){$q_n$}
\put(32,4){$\widetilde g_a$}
\put(0,-18){$-i \sqrt{2} g_3 L^*_{\widetilde q_i} \boldsymbol{T}_{\!m}^{\boldsymbol{a}n}$}
\end{picture}
%%%%%%%%%%%%%%%%%%%%%%%%%%
\begin{picture}(79,70)(0,-20)
\DashArrowLine(0,25)(30,25)5
\ArrowLine(30,25)(60,50)
\ArrowLine(30,25)(60,0)
\put(8,32){$\widetilde q_{in}$}
\put(32,42){$q_m$}
\put(32,4){$\widetilde g_a$}
\put(0,-18){$-i \sqrt{2} g_3 L_{\widetilde q_i} \boldsymbol{T}_{\!m}^{\boldsymbol{a}n}$}
\end{picture}
%%%%%%%%%%%%%%%%%%%%%%%%%%
%%%%%%%%%%%%%%%%%%%%%%%%%%
\begin{picture}(78,70)(0,-20)
\DashArrowLine(30,25)(0,25)5
\ArrowLine(60,50)(30,25)
\ArrowLine(60,0)(30,25)
\put(8,32){$\widetilde q^{*n}_{i}$}
\put(32,42){$\bar q_m$}
\put(32,4){$\widetilde g_a$}
\put(1,-18){$i \sqrt{2} g_3 R_{\widetilde q_i} \boldsymbol{T}_{\!m}^{\boldsymbol{a}n} $}
\end{picture}
%%%%%%%%%%%%%%%%%%%%%%%%%%
\begin{picture}(78,70)(0,-20)
\DashArrowLine(0,25)(30,25)5
\ArrowLine(30,25)(60,50)
\ArrowLine(30,25)(60,0)
\put(8,32){$\widetilde q^{*m}_{i}$}
\put(32,42){$\bar q_n$}
\put(32,4){$\widetilde g_a$}
\put(1,-18){$i \sqrt{2} g_3 R^*_{\widetilde q_i}\boldsymbol{T}_{\!m}^{\boldsymbol{a}n} $}
\end{picture}
%%%%%%%%%%%%%%%%%%%%%%%%%%%%
\end{center}
\caption{Feynman rules for gluino interactions with
third-family quark/squark pairs in the MSSM. 
The indices $m,n$ are for the fundamental
representation of $SU(3)_c$, and $a$ is an adjoint representation index.
The index $i=1,2$ runs over the mass eigenstates.
The comments on Figure \ref{fig:nqsqmixed} also apply here.}
\label{fig:gluinoqsqmixed}
\end{figure}
%%%%%%%%%%%%%%%%%%%%%%%%%%%%%%%%%%%%%%%%%%%%%%%%%%%%%%%%%%%%%%%%%%%%%%%%%
The neutralino interaction rules in 
Figure \ref{fig:nqsqmixed} make use of the following couplings:
\beqa
Y^{\widetilde t_j^* t \chi_i^0} &=&  
Y_t N_{i4}^* R_{\widetilde t_j}^*
+\frac{1}{\sqrt{2}} (g N_{i2}^* + \third g'
N_{i1}^*) L_{\widetilde t_j}^* ,
\label{eq:stoptbarN}
\\
Y^{\widetilde t_j \bar t \chi_i^0} &=&
Y_t N_{i4}^*L_{\widetilde t_j}
- \frac{2\sqrt{2}}{3} g' N_{i1}^* R_{\widetilde t_j} ,
\\
Y^{\widetilde b_j^* b \chi_i^0} &=&
Y_b N_{i3}^* R_{\widetilde b_j}^*
+\frac{1}{\sqrt{2}} (-g N_{i2}^* + \third g'
N_{i1}^*) L_{\widetilde b_j}^* ,
\\
Y^{\widetilde b_j \bar b \chi_i^0} &=&
Y_b N_{i3}^* L_{\widetilde b_j}
 + \frac{\sqrt{2}}{3} g' N_{i1}^* R_{\widetilde b_j} ,
\\
Y^{\widetilde \tau_j^* \tau \chi_i^0} &=&
Y_\tau N_{i3}^* R_{\widetilde \tau_j}^*
-\frac{1}{\sqrt{2}} (g N_{i2}^* + g'
N_{i1}^*) L_{\widetilde \tau_j}^* .
\\
Y^{\widetilde \tau_j \bar \tau \chi_i^0} &=&
Y_\tau N_{i3}^* L_{\widetilde \tau_j}
+ \sqrt{2} g' N_{i1}^* R_{\widetilde \tau_j} .
\label{eq:stautaubarN}
\eeqa
The rules in Figures \ref{fig:cqsqmixed}, \ref{fig:nqsqmixed},
and \ref{fig:gluinoqsqmixed} can be obtained from the preceding three diagrams
by simply taking the appropriate linear combinations of Feynman rules for unmixed
squarks and sleptons. Conversely,
for the charged sfermions of the first two families,
($\widetilde{f}=
\widetilde{u},\widetilde{d},\widetilde{c},\widetilde{s},
\widetilde{e},\widetilde\mu$), one can use the same notation as in Figures \ref{fig:cqsqmixed}, \ref{fig:nqsqmixed},
and \ref{fig:gluinoqsqmixed}, and take $Y_f=0$ and
$L_{\widetilde f_2} = R_{\widetilde f_1} = 1$ and
$L_{\widetilde f_1} = R_{\widetilde f_2} = 0$.

\fontsize{12pt}{14.7pt}\selectfont

\clearpage

\section{Examples}\label{sec:examples}
\renewcommand{\theequation}{\arabic{section}.\arabic{subsection}.\arabic{equation}}
\renewcommand{\thefigure}{\arabic{section}.\arabic{subsection}.\arabic{figure}}
\renewcommand{\thetable}{\arabic{section}.\arabic{subsection}.\arabic{table}}

\subsection[Top quark decay: $t\ra b W^+$]{Top quark decay:
$\boldsymbol{t\ra b W^+}$}
\label{tdecay}
\setcounter{equation}{0}
\setcounter{figure}{0}
\setcounter{table}{0}

We begin by calculating the decay width of a top quark into a bottom
quark and $W^+$ vector boson. Let the four-momenta and
helicities of these particle be $(p_t,\lam_t)$, $(k_b,\lam_b)$ and
$(k\ls{W},\lam_ W)$, respectively. Then $p_t^2 = \BDpos m_t^2$ and $k_b^2
= \BDpos m_b^2$ and $k\ls{W}^2 = \BDpos m\ls{W}^2$ and
\beqa
2 p_t \newcdot k\ls{W} &=& \BDpos m_t^2 \BDminus m_b^2 \BDplus  
m_W^2\,,\\ 2 p_t \newcdot k_b &=& \BDpos m_t^2 \BDplus m_b^2 \BDminus
m_W^2\,,\\ 2 k\ls{W} \newcdot k_b &=& \BDpos m_t^2 \BDminus m_b^2
\BDminus m_W^2 \,.
\eeqa
Because only left-handed top quarks couple to the $W$ boson, the only
Feynman diagram for $t\ra b W^+$ is the one shown in
\fig{fig:topdecay}.
\begin{figure}[t!]
\begin{picture}(400,80)(50,0)
\ArrowLine(155,40)(210,40)
\Photon(210,40)(250,70){3}{5}
\ArrowLine(210,40)(250,10)
\Text(132,40)[]{$t(p_t,\lambda_t)$}
\Text(290,70)[]{$W^+(k_W,\lambda_W)$}
\Text(280,10)[]{$b(k_b,\lambda_b)$}
\end{picture}
\caption{The Feynman diagram for $t\ra b W^+$ at tree level.}
\label{fig:topdecay}
\end{figure}
The corresponding amplitude can be read off of the fifth Feynman rule of
\fig{fig:SMintvertices}. 
Here the initial state top quark is
a 2-component field $t$ going into the vertex and the final state  
bottom quark is created by a 2-component field $b^\dagger$. Therefore
the amplitude is given by, using $\boldsymbol{K}_{3}{}^{3} = V_{tb}$:
\beq 
i {\cal M} = \BDneg i \frac{g}{\sqrt{2}} V_{tb}^*\, \varepsilon_\mu^*
x^\dagger_b \sigmabar^\mu x_t\,,
\eeq
where $\varepsilon_\mu^* \equiv \varepsilon_\mu (k_W,\lambda_W)^*$ is
the polarization vector of the $W^+$, and $x^\dagger_b \equiv  x^\dagger(
\boldsymbol {\vec k}_b,\lambda_b)$ and
$x_t \equiv x(\boldsymbol{\vec p}_t,\lambda_t)$
are the external state wave functions for the
bottom and top quark. Squaring this amplitude
using \eq{eq:conbilsigbar} yields:   
\beq
|{\cal M}|^2 = \frac{g^2}{2} |V_{tb}|^2
\varepsilon_\mu^* \varepsilon_\nu
(x^\dagger_b \sigmabar^\mu x_t)\,
(x^\dagger_t \sigmabar^\nu x_b) \, .
\eeq
Next, we can
average over the top quark spin polarizations using
\eq{xxdagsummed}:
\beq
\frac{1}{2} \sum_{\lambda_t} |{\cal M}|^2
=
\frac{g^2}{4} |V_{tb}|^2\,\varepsilon_\mu^* \varepsilon_\nu  x^\dagger_b
\sigmabar^\mu \,p_t \newcdot \sigma\, \sigmabar^\nu x_b\, .
\eeq
Summing over the bottom quark spin polarizations in the same way
yields a trace over spinor indices:   
\beqa
\frac{1}{2} \sum_{\lambda_t,\lambda_b} |{\cal M}|^2 
&=&
\frac{g^2}{4}|V_{tb}|^2\, \varepsilon_\mu^* \varepsilon_\nu \,
{\rm Tr}[\sigmabar^\mu p_t \newcdot \sigma \,\sigmabar^\nu k_b \newcdot
\sigma] 
\\ =
\frac{g^2}{2}|V_{tb}|^2&&\!\!\!\!\!\! \varepsilon_\mu^*  
\varepsilon_\nu \left ( p_t^\mu k_b^\nu + k_b^\mu p_t^\nu -
\metric^{\mu\nu} p_t \newcdot k_b
- i \epsilon^{\mu\rho\nu\kappa} p_{t\rho} k_{b\kappa}
\right ) \, ,\phantom{xxx}
\eeqa
where we have used eq.~(\ref{trsbarssbars}).  From here, the calculation is unaffected 
by the treatment of the fermionic Feynman rules. One sums over
the $W^+$ polarizations according to:
\beq
\sum_{\lambda_W}
\varepsilon_\mu^* \varepsilon_\nu = \BDneg \metric_{\mu\nu}
+ (k\ls{W})_\mu (k\ls{W})_\nu/m\ls{W}^2\,.
\eeq
The end result is:
\beqa
\frac{1}{2} \sum_{%\lambda_W\lambda_t\lambda_b
\mbox{\small spins}} |{\cal M}|^2 &=&
\frac{g^2}{2}|V_{tb}|^2\, \left [ \BDpos p_t \newcdot k_b +   
2 (p_t \newcdot k\ls{W})(k_b \newcdot k\ls{W})/m\ls{W}^2 \right ]
\,.\phantom{xxx}
\eeqa 
After performing the phase space integration, one obtains:
\beqa
&& \hspace{-0.15in}
\Gamma (t \ra b W^+) = \frac{|V_{tb}|^2}{16 \pi m_t^3}
\lambda^{1/2} (m_t^2 ,m\ls{W}^2, m_b^2 )
\Bigl ( \frac{1}{2} \sum_{\mbox{\small spins}} |{\cal M}|^2 \Bigr )
\\[4pt]
&&
= \frac{g^2|V_{tb}|^2}{64\pi m_W^2 m_t^3}
\lambda^{1/2}(m_t^2,m\ls{W}^2, m_b^2)
\Bigl [
(m_t^2 + 2m_W^2)(m_t^2 - m_W^2)
\nonumber \\[4pt]
&&
\quad + m_b^2 (m_W^2 - 2 m_t^2) + m_b^4
\Bigr ]
,\phantom{xxxx}   
\label{eq:gammat}
\eeqa
where the kinematic triangle 
function $\lambda^{1/2}$ is defined as usual by: 
\beq
\lambda(x,y,z) \equiv x^2 + y^2 + z^2 - 2xy-2xz-2yz .
\label{eq:deftrianglefunction}
\eeq
In the approximation $m_b \ll m\ls{W}, m_t$, one obtains
the well-known result
\beq
\Gamma
(t \ra b W^+) = \frac{m_t g^2 |V_{tb}|^2}{64 \pi}
\left ( 2 + \frac{m_t^2}{m\ls{W}^2} \right ) \left
( 1 - \frac{m\ls{W}^2}{m_t^2} \right )^2\,,
\eeq
exhibiting the Nambu-Goldstone enhancement factor 
$(m_t^2/m\ls{W}^2)$ for the longitudinal $W$ contribution compared to the two  
transverse $W$ contributions.

%%%%%%%%%%%%%%%%%%%%%%%%%%%%%%%%%%%%%%%%%%%%%%%%%%%%%%%
\subsection[$Z^0$ vector boson decay:
$Z^0\ra f \fbar$]{$\boldsymbol{Z^0}$ vector boson decay:
$\boldsymbol{Z^0\ra f \fbar}$}
\label{zff}
\setcounter{equation}{0}
\setcounter{figure}{0}
\setcounter{table}{0}

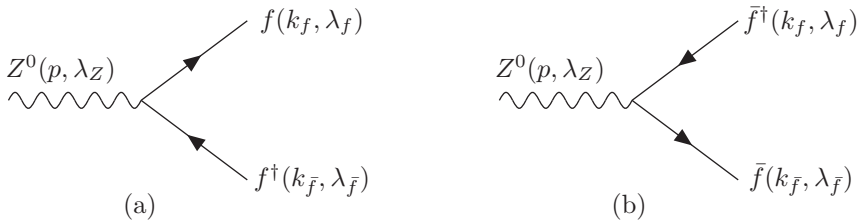
\begin{figure}[b!]
\begin{picture}(400,80)(55,0)
\Photon(110,40)(60,40){3}{5}
\ArrowLine(110,40)(150,70)
\ArrowLine(150,10)(110,40)
\Text(80,52)[]{$Z^0(p,\lambda_Z)$}
\Text(175,70)[]{$f(k_f,\lambda_f)$}
\Text(175,10)[]{$f^\dagger(k_{\fbar},\lambda_{\fbar})$}
\Text(110,0)[]{(a)}   
\Photon(295,40)(245,40){3}{5}
\ArrowLine(335,70)(295,40)
\ArrowLine(295,40)(335,10)
\Text(265,52)[]{$Z^0(p,\lambda_Z)$}
\Text(360,70)[]{${\bar f}^\dagger(k_f,\lambda_f)$}
\Text(360,10)[]{$\bar f(k_{\fbar},\lambda_{\fbar})$}
\Text(295,0)[]{(b)}  
\end{picture}
\caption{The Feynman diagrams for $Z^0$ decay into a
fermion-antifermion pair.  Fermion lines are labeled according to the
2-component fermion field labeling convention established in
\sec{subsec:nomenclature}.}
\label{fig:Zdecay}
\end{figure}

Consider the partial decay width of the $Z^0$ boson into a Standard
Model fermion-antifermion pair.  There are two contributing Feynman
diagrams, shown in \fig{fig:Zdecay}.
In diagram (a), the fermion particle $f$
in the final state is created by
a 2-component field $f$ in the Feynman rule, and the
antifermion particle $\fbar$ by a 2-component field $f^\dagger$.  In
diagram (b), the fermion particle $f$ in the final state is
created by a
2-component field $\bar f$, and the antifermion particle $\fbar$
by a 2-component field ${\bar f}^\dagger$.
Denote the initial $Z^0$ four-momentum and helicity ($p$, $\lam_Z$)
and the final state fermion ($f$) and antifermion ($\fbar$)
momentum and helicities
($k_f,\lambda_f$) and ($k_{\fbar}, \lambda_{\fbar}$),
respectively.  Then, $k_f^2 = k_{\fbar}^2
= \BDpos m_f^2$ and $p^2 =
\BDpos m\ls{Z}^2$, and
\beqa
&& k_f \newcdot  k_{\fbar}
= \BDpos \frac{1}{2} m\ls{Z}^2 \BDminus m_f^2
\,,
\label{Zffkintwo}\\
&& p \newcdot k_f = p\newcdot k_{\fbar}  = \BDpos\half m\ls{Z}^2
\> . \label{Zffkinthree}
\eeqa
According to the third and fourth rules of \fig{fig:SMintvertices}, the matrix elements 
for
the two Feynman graphs are:
\beqa
i {\cal M}_{a} &\,=\,& \BDneg i \frac{g}{c\ls{W}}(T_3^f - s\ls{W}^2 Q_f) \,
\varepsilon_\mu  x^\dagger_f \sigmabar^\mu y_{\fbar}\,, \label{zffamp1}
\\
i {\cal M}_{b} &\,=\,&
\BDpos i g \frac{s\ls{W}^2}{c\ls{W}} Q_f\, \varepsilon_\mu 
y_f \sigma^\mu  x^\dagger_{\fbar}\,, \label{zffamp2}
\eeqa
where $x_i\equiv x(\boldsymbol{\vec k}_i,\lambda_i)$ and
$y_i\equiv y(\mathbold{\vec k_i},\lambda_i)$,
for $i=f,\fbar$, and  
$\varepsilon_\mu\equiv\varepsilon_\mu(p,\lambda_Z)$.

It is convenient to define:
\beq 
a_f \equiv T_3^f - Q_f s\ls{W}^2 \,, \qquad\qquad
b_f \equiv  -Q_f s\ls{W}^2\,.
\eeq
Then the squared matrix element is, using eqs.~(\ref{eq:conbilsig}) and (\ref{eq:conbilsigbar}),
\beq
|{\cal M}|^2 =\frac{g^2}{c^2\ls{W}}
\varepsilon_\mu\varepsilon^*_\nu
\left ( a_f  x^\dagger_f \sigmabar^\mu y_{\fbar} +
b_f y_f \sigma^\mu  x^\dagger_{\fbar} \right )
\left ( a_f  y^\dagger_{\fbar} \sigmabar^\nu x_f +  
b_f x_{\fbar} \sigma^\nu  y^\dagger_f \right ) .\>\phantom{x}
\eeq
Summing over the antifermion helicity using
\eqst{xxdagsummed}{ydagxdagsummed} gives:
\beqa
\sum_{\lambda_{\fbar}}|{\cal M}|^2 &=& \frac{g^2}{c^2\ls{W}} 
\varepsilon_\mu \varepsilon^*_\nu
\Bigl (
a_f^2  x^\dagger_f \sigmabar^\mu k_{\fbar} \newcdot \sigma
\sigmabar^\nu
x_f+
b_f^2 y_f \sigma^\mu k_{\fbar} \newcdot \sigmabar
\sigma^\nu  y^\dagger_f
\cr && \quad
- m_f a_f b_f  x^\dagger_f \sigmabar^\mu \sigma^\nu  y^\dagger_f
- m_f a_f b_f y_f \sigma^\mu\sigmabar^\nu x_f
\Bigr ) \, .
\eeqa
Next, we sum over the fermion helicity:
\beqa
\sum_{\lambda_f, \lambda_{\fbar}} |{\cal M}|^2
&=& \frac{g^2}{c^2\ls{W}}
\varepsilon_\mu \varepsilon^*_\nu
\Bigl (
a_f^2 {\rm Tr}[\sigmabar^\mu k_{\fbar} \newcdot \sigma \sigmabar^\nu
k_f\newcdot\sigma ]
+
b_f^2  {\rm Tr}[\sigma^\mu k_{\fbar} \newcdot \sigmabar \sigma^\nu
k_f\newcdot\sigmabar ]
\cr
&& \qquad
- m^2_f a_f b_f {\rm Tr}[\sigmabar^\mu \sigma^\nu ]
- m^2_f a_f b_f {\rm Tr}[\sigma^\mu\sigmabar^\nu ]
\Bigr )\, .
\eeqa
Averaging over the $Z^0$ polarization using
\beq
\frac{1}{3} \sum_{\lambda_Z}
\varepsilon_\mu \varepsilon_{\nu}^* = \frac{1}{3} \left (
\BDneg \metric_{\mu\nu} + \frac{p_\mu p_\nu}{m\ls{Z}^2}
\right )\,,
\eeq
and applying \eqst{trssbar}{trsbarssbars}, one gets:
\beqa
\frac{1}{3} \sum_{\rm spins} |{\cal M}|^2 &=& \frac{g^2}{3 c^2\ls{W}}
\left [
(a_f^2 + b_f^2) \left (
\BDpos 2 k_f \newcdot k_{\fbar}
+ 4 \, k_f \newcdot p \, k_{\fbar} \newcdot p/m\ls{Z}^2
\right ) + 12 a_f b_f m_f^2 \right
] \nonumber
\\
 &=&
\frac{2 g^2}{3 c^2\ls{W}}
\left [
(a_f^2 + b_f^2) (m\ls{Z}^2 - m_f^2 ) + 6 a_f b_f m_f^2 \right
]\, ,
\eeqa
where we have used \eqs{Zffkintwo}{Zffkinthree}. After the
standard phase space integration, we obtain the well-known result:
\beqa
&&\!\!\!\!\!\!\!\!\!\!\!
\Gamma(Z^0 \ra f \fbar) = \frac{N_c^f}{16 \pi m\ls{Z}} \left (
1 - \frac{4m_f^2}{m\ls{Z}^2}\right )^{1/2}\, \left (
\frac{1}{3} \sum_{\rm spins} |{\cal M}|^2 \right ) 
\nonumber \\[5pt] 
&&\!\!\!\!\!\!\!\!\!\!\!
=
\frac{N_c^f g^2 m\ls{Z}}{24 \pi c^2\ls{W}}
\left (
1 - \frac{4m_f^2}{m\ls{Z}^2}\right )^{1/2}
\left [
(a_f^2 + b_f^2) \left(1 - \frac{m_f^2}{m\ls{Z}^2}\right)
+ 6 a_f b_f \frac{m_f^2}{m\ls{Z}^2}
\right ]\! . 
\phantom{xxx}
\eeqa
Here we have also included a factor of $N_c^f$ (equal to $1$ for
leptons and $3$ for quarks) for the sum over colors.
Since the $Z^0$ is a color singlet,
the color factor is simply equal to the dimension of the
color representation of the final-state fermions.

%%%%%%%%%%%%%%%%%%%%%%%%%%%%%%%%%%%%%%%%%%%%%%%%%%%%%%%%%%%%

\subsection[Bhabha scattering: $e^- e^+ \ra e^- e^+$]
{Bhabha scattering: $\boldsymbol{e^- e^+ \ra e^- e^+}$
\label{subsec:Bhabha}}
\setcounter{equation}{0}
\setcounter{figure}{0}
\setcounter{table}{0} 

\begin{figure}[bp!]
\vspace{0.2in}
\centerline{
\begin{picture}(300,155)(-135,-25)
\thicklines
\Photon(-110,105)(-50,105){3}{7}
\ArrowLine(-140,135)(-110,105)
\ArrowLine(-110,105)(-140,75)
\ArrowLine(-20,135)(-50,105)
\ArrowLine(-50,105)(-20,75)
\put(-132,134){$e$}
\put(-132,70){$e^\dagger$}
\put(-37,133){${\bar e}^\dagger$}
\put(-37,70){$\bar e$}
\Photon(60,105)(120,105){3}{7}
\ArrowLine(60,105)(30,135)
\ArrowLine(30,75)(60,105) 
\ArrowLine(150,135)(120,105)
\ArrowLine(120,105)(150,75)
\put(38,134){${\bar e}^\dagger$}
\put(38,70){$\bar e$}
\put(133,133){${\bar e}^\dagger$}
\put(133,70){$\bar e$}
\Photon(-110,15)(-50,15){3}{7}
\ArrowLine(-140,45)(-110,15)
\ArrowLine(-110,15)(-140,-15)
\ArrowLine(-50,15)(-20,45)
\ArrowLine(-20,-15)(-50,15)
\put(-132,44){$e$}
\put(-132,-20){$e^\dagger$}
\put(-37,43){$e$}
\put(-37,-20){$e^\dagger$}
\Photon(60,15)(120,15){3}{7}
\ArrowLine(60,15)(30,45)
\ArrowLine(30,-15)(60,15) 
\ArrowLine(120,15)(150,45)
\ArrowLine(150,-15)(120,15) 
\put(38,44){${\bar e}^\dagger$}
\put(38,-20){$\bar e$}
\put(133,43){$e$}
\put(133,-20){$e^\dagger$}
\end{picture}
}
\caption{Tree-level $s$-channel Feynman diagrams for $e^+ e^-\to e^+ e^-$.}
\label{fig:Bhabhalabels}
\end{figure}
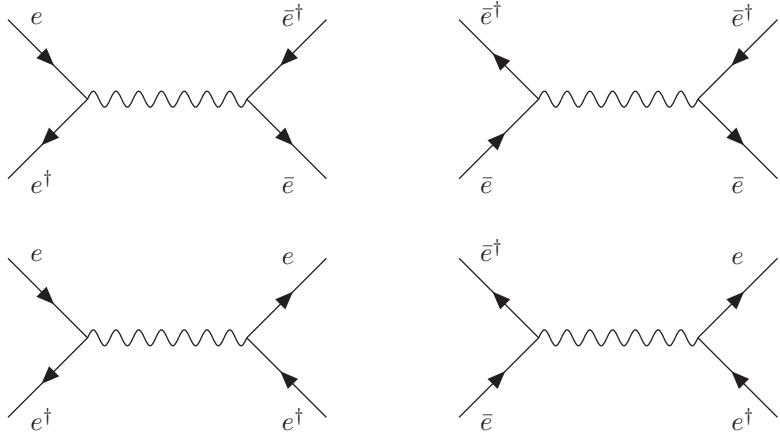

\begin{figure}[t!]
\centerline{
\begin{picture}(300,190)(-135,-55)
\thicklines
\Photon(-80,135)(-80,75){3}{5}
\ArrowLine(-140,75)(-80,75)
\ArrowLine(-140,135)(-80,135)
\ArrowLine(-80,75)(-20,75)
\ArrowLine(-80,135)(-20,135)
\put(-132,60){$\bar e$}
\put(-132,140){$e$}
\put(-37,60){$\bar e$}
\put(-37,140){$e$}
\Photon(90,135)(90,75){3}{5}
\ArrowLine(150,75)(90,75)
\ArrowLine(90,135)(150,135)
\ArrowLine(90,75)(30,75)
\ArrowLine(30,135)(90,135)
\put(38,60){${e}^\dagger$}
\put(38,140){$e$}
\put(133,60){${e}^\dagger$}
\put(133,140){$e$}
\Photon(-80,15)(-80,-45){3}{5}
\ArrowLine(-140,-45)(-80,-45)  
\ArrowLine(-80,15)(-140,15)
\ArrowLine(-80,-45)(-20,-45) 
\ArrowLine(-20,15)(-80,15)
\put(-132,-60){$\bar e$}
\put(-132,20){${\bar e}^\dagger$}
\put(-37,-60){$\bar e$}
\put(-37,20){${\bar e}^\dagger$}
\Photon(90,15)(90,-45){3}{5}
\ArrowLine(150,-45)(90,-45)  
\ArrowLine(90,15)(30,15) 
\ArrowLine(90,-45)(30,-45) 
\ArrowLine(150,15)(90,15)
\put(38,-60){${e}^\dagger$}
\put(38,20){${\bar e}^\dagger$}
\put(133,-60){${e}^\dagger$}
\put(133,20){${\bar e}^\dagger$}
\end{picture}
}
\caption{Tree-level $t$-channel Feynman diagrams for $e^- e^+\to e^- e^+$,
  with the external lines labeled according to the 2-component field
  names.  The momentum flow of the external particles is from left to
  right.}
\label{fig:Bhabhatchannel} 
\end{figure}
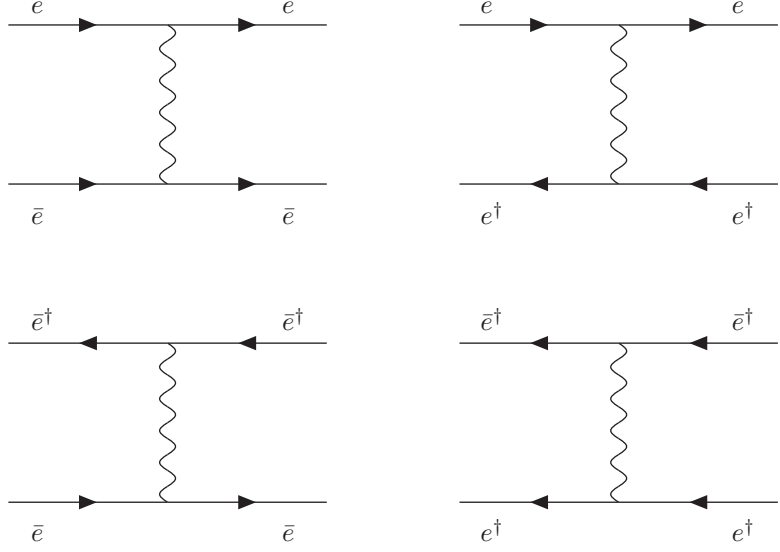

In our next example, we consider the computation of Bhabha scattering
in QED (that is, we consider photon exchange but neglect
$Z^0$-exchange). We denote the initial state electron and
positron momenta and helicities by 
($p_1,\lambda_1$) and ($p_2,\lambda
_2$) and the final state electron and positron momenta and helicities
by ($p_3,\lambda_3$) and ($p_4,\lambda_4$), respectively.
Neglecting the electron mass, we have in terms of the usual Mandelstam
variables $s,t,u$:
\beqa
&&p_1 \newcdot p_2 = p_3 \newcdot p_4 \equiv \BDpos\half s\, ,\\
&&p_1 \newcdot p_3 = p_2 \newcdot p_4 \equiv \BDneg \half t\, ,\\
&&p_1 \newcdot p_4 = p_2 \newcdot p_3 \equiv \BDneg \half u\,,
\eeqa
and $p_i^2=0$ for $i=1,\ldots, 4$.
There are eight distinct Feynman diagrams.
First, there are four $s$-channel diagrams, as shown in
\fig{fig:Bhabhalabels}
%%%%%%%%%%%%%%%%%%%%%%%%%%%
%%%%%%%%%%%%%%%%%%%%%%%%%%%%%%
with amplitudes that follow from the first and second Feynman rules of 
\fig{fig:SMintvertices}:
\beqa
i\mathcal{M}_s &=& \left ( \frac{ \BDneg i\metric^{\mu\nu}}{s} \right )
\Bigl [
(\BDneg ie\, x_1 \sigma_\mu y^\dagger_2)(\BDpos ie\, y_3 \sigma_\nu x^\dagger_4)
+
(\BDneg ie\, y^\dagger_1 \sigmabar_\mu x_2)(\BDpos ie\, y_3 \sigma_\nu  x^\dagger_4)
\nonumber \\ &&\!\!\!\!\!\!\!\!\!
\quad +
(\BDneg ie\, x_1 \sigma_\mu y^\dagger_2 )(\BDpos ie\, x^\dagger_3\sigmabar_\nu y_4)
+
(\BDneg ie\, y^\dagger_1 \sigmabar_\mu x_2)(\BDpos ie\, x^\dagger_3\sigmabar_\nu y_4)
\Bigr ],
\label{bhabhaschannel}
\eeqa
where $x_i\equiv x(\mathbold{\vec p_i},\lambda_i)$ and
$y_i\equiv y(\mathbold{\vec p_i},\lambda_i)$, for $i=1,4$.
The photon propagator in Feynman gauge is
$-i \metric^{\mu\nu}/(p_1+ p_2)^2 = \BDneg i \metric^{\mu\nu}/s$.
Here, we have chosen to write the external fermion spinors in the order
$1,2,3,4$. This
dictates in each term the use of either the $\sigmabar$ or
$\sigma$ forms of the Feynman rules of \fig{fig:SMintvertices}.
One can group the terms of \eq{bhabhaschannel} together more compactly:
\beq
i {\cal M}_s = e^2
\left ( \frac{\BDneg i \metric^{\mu\nu}}{s} \right )
\left (x_1 \sigma_\mu y^\dagger_2 + y^\dagger_1 \sigmabar_\mu x_2 \right )
\left (y_3 \sigma_\nu x^\dagger_4 + x^\dagger_3 \sigmabar_\nu y_4 \right ) .
\eeq
There are also four $t$-channel diagrams, as shown in
\fig{fig:Bhabhatchannel}.
The corresponding amplitudes for these four diagrams can be written:
\beqa
i \mathcal{M}_t &=& (-1) e^2
\left ( \frac{\BDneg i \metric^{\mu\nu}}{t} \right )
\left(
 x_1 \sigma_\mu x^\dagger_3   
+
  y^\dagger_1 \sigmabar_\mu y_3\right ) \left (
 x_2 \sigma_\nu  x^\dagger_4 +
  y^\dagger_2 \sigmabar_\nu y_4 \right )
 . \phantom{xxxxx}
\eeqa
Here, the overall factor of $(-1)$ comes from
Fermi-Dirac statistics, since the
external fermion wave functions are written in an odd permutation
$(1,3,2,4)$ of
the original
order $(1,2,3,4)$ established by the first term in
eq.~(\ref{bhabhaschannel}).

Fierzing each term using \eqst{twocompfierza}{twocompfierzc}, and
using \eqs{zonetwo}{barzonetwo}, the total amplitude can be written 
as:  
\beqa
\mathcal{M} &=& {\cal M}_s + {\cal M}_t =
         2 e^2 \biggl [
\frac{1}{s}
(x_1 y_3) (y^\dagger_2 x^\dagger_4)
+ \frac{1}{s} (y^\dagger_1 x^\dagger_3)(x_2 y_4)
\nonumber \\[4pt] &&
+\left (\frac{1}{s}+\frac{1}{t}\right )(y^\dagger_1 x^\dagger_4)(x_2 y_3)
+\left (\frac{1}{s}+\frac{1}{t}\right )(x_1 y_4)
(y^\dagger_2 x^\dagger_3)
\nonumber \\[4pt] &&
- \frac{1}{t} (x_1 x_2)(x^\dagger_3 x^\dagger_4)\,
- \frac{1}{t}(y^\dagger_1 y^\dagger_2)(y_3 y_4)
\biggr ]. \phantom{xxx}
\label{bhabhamat}
\eeqa
Squaring this amplitude and summing over spins, all of the cross terms
will vanish in the $m_e \ra 0 $ limit. This is because each
cross term will have an $x$ or an $ x^\dagger$ for some electron or
positron combined with a $y$ or a $y^\dagger$ for the same particle, and
the corresponding spin sum is proportional to $m_e$ [see
\eqs{yxsummed}{ydagxdagsummed}].  Hence, summing over final state
spins and averaging over initial state spins, the end result contains
only
the sum of the squares of the six terms in eq.~(\ref{bhabhamat}):
\beqa
&&
\frac{1}{4}
\sum_{\rm spins} |{\cal M}|^2 = 
e^4\sum_{ \lambda_1,
  \lambda_2, \lambda_3, \lambda_4} \biggl \lbrace 
\nonumber \\[4pt]
&&  
  \frac{1}{s^2} \left [  
  (x_1 y_3)(y^\dagger_3 x^\dagger_1) (y^\dagger_2x^\dagger_4)(x_4 y_2)
  +
  (y^\dagger_1x^\dagger_3)(x_3 y_1)(x_2 y_4)(y^\dagger_4 x^\dagger_2) \right ]
\nonumber \\[4pt]  
&&
+ \left (\frac{1}{s} + \frac{1}{t}
   \right )^2 \left[(y^\dagger_1 x^\dagger_4)(x_4 y_1) {(x_2 y_3 )(
  y^\dagger_3x^\dagger_2)}+ (x_1 y_4)(y^\dagger_4 x^\dagger_1)
  {(y^\dagger_2x^\dagger_3 )(
 x_3y_2)}
\right ] 
\nonumber \\[5pt]  
&&
+\frac{1}{t^2} \left[(x_1
  x_2)(x^\dagger_2 x^\dagger_1)(x^\dagger_3 x^\dagger_4)(x_4 x_3)+
  (y^\dagger_1
  y^\dagger_2)(y_2 y_1)(y_3 y_4)(y^\dagger_4 y^\dagger_3) \right ] \biggr \rbrace
\, . 
\eeqa
Here we have used eq.~(\ref{eq:conbil}) to get the complex square
of the fermion bilinears.
Performing these spin sums using
\eqs{xxdagsummed}{yydagsummed} and using the trace identities
\eq{trssbar}:
\beqa \frac{1}{4} \sum_{\rm spins} |{\cal
  M}|^2 &=& {8}e^4 \biggl [ \frac{p_2 \newcdot p_4\,  
 p_1 \newcdot  p_3}{s^2} +\frac{p_1 \newcdot p_2 \,
p_3 \newcdot p_4}{t^2}
+ \left (\frac{1}{s}+\frac{1}{t}\right )^2 p_1 \newcdot
p_4 \, p_2 \newcdot p_3 \biggr ]
\nonumber \\[3pt]
&=& 2e^4 \biggl [ \frac{t^2}{s^2}+\frac{s^2}{t^2} + \left (\frac{u}{s}
    +\frac{u}{t}\right )^2 \biggr ]\, .
\eeqa
Thus, the differential cross-section for Bhabha scattering is given
by:
\beq
\frac{d \sigma}{dt}
   = \frac{1}{16 \pi s^2} \Bigl(\frac{1}{4}\sum_{\rm spins} |{\cal
    M}|^2 \Bigr ) = \frac{2 \pi \alpha^2}{s^2} \biggl [ \frac{t^2}{s^2}
   +\frac{s^2}{t^2} + \left (\frac{u}{s}+\frac{u}{t}\right )^2
\biggr ].\phantom{xx}
\eeq

%%%%%%%%%%%%%%%%%%%%%%%%%%%%%%%%%%%%%%%%%%%%%%%%%%%%%%%%%%%%%%
\subsection[Neutral MSSM Higgs boson decays $\phi^0
\rightarrow f \fbar$, for $\phi^0 = h^0,H^0,A^0$]{Neutral
MSSM Higgs boson decays $\boldsymbol{\phi^0
\rightarrow f \fbar}$, for $\boldsymbol{\phi^0 = h^0,H^0,A^0}$} 
\setcounter{equation}{0}
\setcounter{figure}{0}
\setcounter{table}{0}

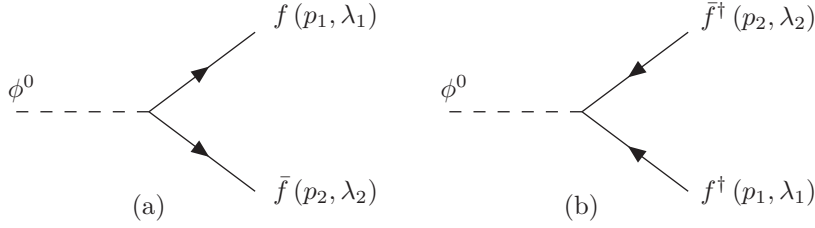
\begin{figure}[tb]
\begin{center}
\begin{picture}(150,80)(0,0)
\DashLine(60,40)(10,40)5
\ArrowLine(60,40)(100,70)
\ArrowLine(60,40)(100,10)
\Text(12,49)[]{$\phi^0$}
\Text(127,10)[]{$\bar f \,(p_2,\lam_2)$}
\Text(127,76)[]{$f \,(p_1,\lam_1)$}
\Text(60,4)[]{(a)}
\end{picture}
\hspace{0.25cm}
\begin{picture}(150,80)(0,0)
\DashLine(10,40)(60,40)5
\ArrowLine(100,70)(60,40)
\ArrowLine(100,10)(60,40)
\Text(12,49)[]{$\phi^0$}
\Text(127,10)[]{$f^\dagger \,(p_1,\lam_1)$}
\Text(127,76)[]{${\bar f}^\dagger \,(p_2,\lam_2)$}
\Text(60,4)[]{(b)}
\end{picture}
\end{center}
\caption{The Feynman diagrams for the decays $\phi^0 \ra f \fbar$,
where $\phi^0 = h^0, H^0, A^0$ are the neutral Higgs scalar bosons
of the MSSM, and $f$ is a Standard Model quark or lepton,
and $\fbar$ is the corresponding antiparticle.
The external fermions are labeled according to the 2-component
field names.}
\label{fig:hHffbardecay}
\end{figure}

In this subsection, we consider the decays of the neutral Higgs scalar
bosons $\phi^0 = h^0$, $H^0$, and $A^0$ of the MSSM
into Standard Model fermion-antifermion pairs. The relevant
tree-level Feynman diagrams are shown in \fig{fig:hHffbardecay}.
The final state fermion is assigned four-momentum $p_1$ and polarization
$\lambda_1$, and the antifermion is assigned four-momentum $p_2$
and polarization $\lambda_2$.
We will first work out the case that $f$ is a charge $-1/3$ quark or
a charged lepton, and later note the simple change needed for
charge $+2/3$ quarks. The second and fifth Feynman rules of 
\fig{nehiggsqq} tell us that
the amplitudes are:
\beqa
i \mathcal{M}_{a} &=&
-\frac{i}{\sqrt{2}}\, Y_f \, k_{d\phi^0}^* \, x^\dagger_1 x^\dagger_2
\, ,
\\
i \mathcal{M}_{b} &=&
-\frac{i}{\sqrt{2}}\, Y_f \, k_{d\phi^0} \, y_1 y_2
\, .
\eeqa
Here $Y_f$ is the Yukawa coupling of the fermion, $k_{d\phi^0}$ is
the Higgs mixing parameter from eq.~(\ref{eq:defkdphi0}), and
the external wave functions are denoted
$x_1 \equiv x(\boldsymbol{\vec p}_1,\lam_ 1)$,
$y_1 \equiv y(\boldsymbol{\vec p}_1,\lam_1)$ for the fermion and
$x_2 \equiv x(\boldsymbol{\vec p}_2,\lam_2)$,
$y_2 \equiv y(\boldsymbol{\vec p}_2,\lam_2)$ for the antifermion.
Squaring the total amplitude
$i {\cal M} = i {\cal M}_a + i {\cal M}_b$ using eq.~(\ref{eq:conbil})
results in:
\beqa
|{\cal M}|^2 &=&
\frac{1}{2} |Y_f|^2 \Bigl [
|k_{d\phi^0}|^2 ( y_1 y_2 \, y^\dagger_2 y^\dagger_1
+ x^\dagger_1 x^\dagger_2\, x_2 x_1)
\nonumber \\ &&
+ (k^*_{d\phi^0})^2 x^\dagger_1 x^\dagger_2\, y^\dagger_2 y^\dagger_1
+ (k_{d\phi^0})^2 y_1 y_2\, x_2 x_1
\Bigr ] .
\eeqa
Summing over the final state antifermion spin using
\eqst{xxdagsummed}{ydagxdagsummed} gives:
\beqa
\sum_{\lam_2} |{\cal M}|^2 &=&
\frac{1}{2} |Y_f|^2 \Bigl [
\BDpos |k_{d\phi^0}|^2 ( y_1 p_2 \newcdot \sigma y^\dagger_1
+ x^\dagger_1 p_2 \newcdot \sigmabar x_1)
\nonumber \\ &&
- (k^*_{d\phi^0})^2 m_f x^\dagger_1 y^\dagger_1
- (k_{d\phi^0})^2 m_f y_1  x_1
\Bigr ] .
\eeqa
Summing over the fermion spins in the same way yields:
\beqa
\sum_{\lam_1,\lam_2} |{\cal M}|^2 &=&
\frac{1}{2} |Y_f|^2 \Bigl \{
|k_{d\phi^0}|^2 ( 
{\rm Tr}[p_2 \newcdot \sigma p_1 \newcdot \sigmabar]
+ {\rm Tr}[p_2 \newcdot \sigmabar p_1 \newcdot \sigma])
\nonumber \\ &&\qquad 
- 2 (k^*_{d\phi^0})^2 m_f^2 - 2 (k_{d\phi^0})^2 m_f^2
\Bigr \} 
\\  
&=& |Y_f|^2 \left \{ \BDpos 2 |k_{d\phi^0}|^2 p_1 \newcdot p_2
- 2 {\rm Re}[(k_{d\phi^0})^2] m_f^2 \right \} 
\\  
&=& |Y_f|^2 \left \{ |k_{d\phi^0}|^2 (m^2_{\phi^0} - 2 m_f^2)
- 2 {\rm Re}[(k_{d\phi^0})^2] m_f^2 \right \} ,\phantom{xxx}
\eeqa
where we have used the trace identity eq.~(\ref{trssbar}) to obtain
the second equality. The corresponding expression for charge $+2/3$
quarks can be obtained by simply replacing $k_{d\phi^0}$ with
$k_{u\phi^0}$.  The total decay rates now follow from integration over
phase space 
\beq
\Gamma (\phi^0 \rightarrow f \fbar) =
\frac{N_c^f}{16 \pi m_{\phi^0}} \left (1 - 4 m_f^2/m_{\phi^0}^2
\right )^{1/2}
\sum_{\lam_1,\lam_2} |{\cal M}|^2 .
\eeq
The factor of $N_c^f = 3$ for quarks and 1 for leptons comes from the
sum over colors.

Results for special cases are obtained by putting in the relevant
values for the couplings and the mixing parameters from
eqs.~(\ref{eq:defkuphi0}) and (\ref{eq:defkdphi0}). In
particular,
for the CP-even Higgs bosons $h^0$ and $H^0$, $k_{d\phi^0}$ and
$k_{u\phi^0}$ are real, so
one obtains:
\beqa
\Gamma(h^0 \ra b \bbar)
&=& \frac{3}{16 \pi}\,Y_b^2 \, \sin^2\!\alpha \,
m_{h^0}
\left (1 - {4m_b^2/ m_{h^0}^2} \right )^{3/2},\phantom{xxxxxxx}
\label{hbbresult}
\\
\Gamma(h^0 \ra c \bar c)
&=& \frac{3}{16 \pi}\,Y_c^2 \, \cos^2\!\alpha \,
m_{h^0}
\left (1 - {4m_c^2/ m_{h^0}^2} \right )^{3/2},
\label{hccresult}
\\
\Gamma(h^0 \ra \tau^+ \tau^-)
&=& \frac{1}{16 \pi}\,Y_\tau^2 \, \sin^2\!\alpha \,
m_{h^0}
\left (1 - {4m_\tau^2/ m_{h^0}^2} \right )^{3/2},
\label{htautauresult}
\\
\Gamma(H^0 \ra t \tbar)   
&=& \frac{3}{16 \pi}\,Y_t^2 \, \sin^2\!\alpha \,
m_{H^0}
\left (1 - {4m_t^2/ m_{H^0}^2} \right )^{3/2},
\label{Httresult}
\\
\Gamma(H^0 \ra b \bbar)
&=& \frac{3}{16 \pi}\,Y_b^2 \, \cos^2\alpha \,
m_{H^0}
\left (1 - {4m_b^2/ m_{H^0}^2} \right )^{3/2},
\label{Hbbresult}
\eeqa  
etc., which check with the expressions in \app{C} of
ref.~[\refcite{HHG}].  For the CP-odd Higgs boson $A^0$, the mixing
parameters $k_{uA^0} = i \cos\!\beta_0$ and $k_{dA^0} = i \sin\!\beta_0$
are purely imaginary, so
\beqa \Gamma(A^0 \ra t \tbar) &=& \frac{3}{16\pi}
\,Y_t^2 \, \cos^2\!\beta_0 \, m_{A^0} \left (1 - {4m_t^2/
    m_{A^0}^2} \right )^{1/2},\phantom{xxxxxx}
\label{Attresult}
\\
\Gamma(A^0 \ra b \bbar)
&=& \frac{3}{16 \pi}\,Y_b^2 \, \sin^2\!\beta_0 \,
m_{A^0}
\left (1 - {4m_b^2/ m_{A^0}^2} \right )^{1/2},
\label{Abbresult}
\\
\Gamma(A^0 \ra \tau^+ \tau^-)
&=& \frac{1}{16 \pi}\,Y_\tau^2 \, \sin^2\!\beta_0 \, m_{A^0}
\left (1 - {4m_\tau^2/ m_{A^0}^2} \right )^{1/2} .
\label{Atautauresult}
\eeqa

The differing kinematic factors for the CP-odd Higgs decays
came about because of the different relative sign between the two Feynman
diagrams. For example, in the case of $h^0 \rightarrow b \bbar$,
the matrix element is
\beq
i {\cal M} = \frac{i}{\sqrt{2}} Y_b \sin\!\alpha \,(y_1 y_2
+ x^\dagger_1 x^\dagger_2),   
\eeq
while for $A^0 \rightarrow b \bbar$, it is
\beq
i {\cal M} =\frac{1}{\sqrt{2}} Y_b \sin\!\beta_0 \,(y_1 y_2 -
x^\dagger_1 x^\dagger_2).
\eeq
The differing relative sign between $y_1 y_2$ and $x^\dagger_1 x^\dagger_2$
follows from the imaginary pseudoscalar Lagrangian
coupling, which is complex conjugated in the second diagram.

%%%%%%%%%%%%%%%%%%%%%%%%%%%%%%%%%%%%%%%%%%%%%%%%%%%%%%%%%%%%%%%%%%%%%

\subsection[Neutralino decays
$\Ni \ra \phi^0 \Nj$, for $\phi^0 = h^0, H^0, A^0$]{Neutralino decays
$\boldsymbol{\Ni \ra \phi^0 \Nj}$, for $\boldsymbol{\phi^0 = h^0, H^0, A^0}$}
\setcounter{equation}{0}
\setcounter{figure}{0}
\setcounter{table}{0}

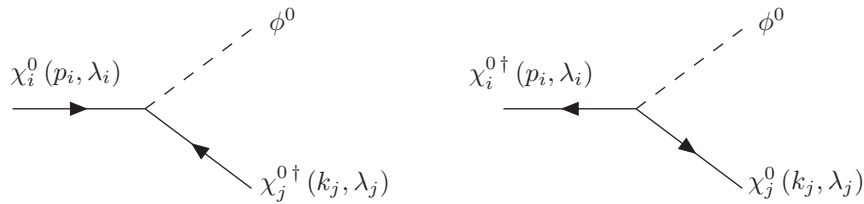
\begin{figure}[b!]   
\begin{center}
\begin{picture}(150,70)(5,12)
\ArrowLine(-80,40)(-30,40)
\DashLine(-30,40)(10,70)5
\ArrowLine(10,10)(-30,40)
\Text(-60,54)[]{$\chi^0_i\,(p_i,\lam_i)$}
\Text(38,12)[]{${\chi^{0\,\dagger}_j} \,(k_j,\lam_j)$}
\Text(22,73)[]{$\phi^0$}
\ArrowLine(155,40)(105,40)
\DashLine(155,40)(195,70)5
\ArrowLine(155,40)(195,10)
\Text(116,54)[]{${\chi^{0\,\dagger}_i}\, (p_i,\lam_i)$}
\Text(220,12)[]{$\chi^0_j \,(k_j,\lam_j)$}
\Text(207,73)[]{$\phi^0$}
\end{picture}
\end{center}
\caption{The Feynman diagrams for $\stilde N_i \ra \stilde N_j
 \phi^0 $ in the MSSM.}
\label{fig:neut1toneut2h}
\end{figure}

Next we consider the decay of a neutralino to a lighter neutralino and
neutral Higgs boson $\phi^0 = h^0$, $H^0$, or $A^0$.
The two tree-level Feynman
graphs are shown in \fig{fig:neut1toneut2h}, where we have also 
labeled the momenta and helicities.
%%%%%%%%%%%%%%%
%%%%%%%%%%%%%%%
We denote the masses for the
neutralinos and the Higgs boson as $m_{\Ni}$, $m_{\Nj},$ and   
$m_{\phi^0}$.
Using the first Feynman rule of \fig{Higgschichi}, the amplitudes
are respectively given by
\beqa
i{\cal M}_1 = -i Y \hspace{0.35pt} x_iy_j\,, \qquad\quad
i{\cal M}_2 = -i Y^* \hspace{0.35pt} y^\dagger_i x^\dagger_j\,,
\eeqa
where the coupling $Y \equiv Y^{\phi^0\chi^0_i\chi^0_j}$ is defined in
\eq{higgs-gauginos1},
and the external wave functions are
$x_i\equiv x(\boldsymbol {\vec p}_i,\lam_i)$,
$y^\dagger_i\equiv y^\dagger(\boldsymbol {\vec p}_i,\lam_i)$,
$y_j\equiv y(\boldsymbol{\vec k}_j,\lam_j)$, and
$x^\dagger_j\equiv x^\dagger(\boldsymbol{\vec k}_j,\lam_j)$.

Taking the square of the total matrix element using
eq.~(\ref{eq:conbil}) gives: 
\beqa |{\cal M}|^2 = |Y|^2 (x_i y_j
y^\dagger_j x^\dagger_i +
y^\dagger_i x^\dagger_j x_j y_i) + Y^2 x_i y_j x_j y_i +
Y^{*2} y^\dagger_i x^\dagger_j y^\dagger_j x^\dagger_i .  \phantom{xxx}
\eeqa
Summing over the final state neutralino spin using
\eqst{xxdagsummed}{ydagxdagsummed} yields
\beqa
\sum_{\lam_j} |{\cal M}|^2 &=&
\BDpos |Y|^2 (x_i k_j \newcdot \sigma x^\dagger_i +
 y^\dagger_i k_j \newcdot \sigmabar y_i) 
\nonumber \\[-8pt]  &&
 - Y^2 m_{\Nj} x_i y_i - Y^{*2}
m_{\Nj}  y^\dagger_i x^\dagger_i . \phantom{xx} 
\eeqa 
Averaging over the initial state
neutralino spins in the same way gives
\beqa
\frac{1}{2}
\sum_{\lam_i,\lam_j} |{\cal M}|^2 &=& \frac{1}{2} |Y|^2 ({\rm Tr}[ k_j
\newcdot \sigma p_i \newcdot \sigmabar] + {\rm Tr}[k_j \newcdot
\sigmabar p_i \newcdot \sigma]) + {\rm Re}[Y^2] m_{\Ni} m_{\Nj} {\rm
  Tr}[1]\nonumber   
\\[-5pt]
&=& \BDpos 2 |Y|^2 p_i\newcdot k_j +2 {\rm Re}[Y^2] m_{\Ni} m_{\Nj}
\nonumber \\
&=& |Y|^2 (m_{\Ni}^2 + m_{\Nj}^2 - m_{\phi^0}^2) + 2 {\rm Re}[Y^2]
m_{\Ni} m_{\Nj} , \eeqa where we have used eq.~(\ref{trssbar}) to
obtain the second equality.  The total decay rate is therefore
\beqa
\Gamma (\Ni \ra \phi^0\Nj) &=& \frac{1}{16 \pi m_{\Ni}^3 }
\lambda^{1/2} (m_{\Ni}^2, m\ls{\phi^0}^2 , m_{\Nj}^2 ) \biggl ( \frac{1}{2}
    \sum_{\lambda_i,\lambda_j} |{\cal M}|^2 \biggr ) \nonumber
\\
&=& 
\frac{ m_{\Ni}}{16\pi} \lambda^{1/2}(1,r_\phi,r_j)
\Bigl \{ |Y^{\phi^0\chi^0_i\chi^0_j}|^2 (1+r_j-r_\phi ) 
\nonumber \\ &&+2{\rm Re}
  \bigl[\bigl(Y^{\phi^0\chi^0_i\chi^0_j}\bigr)^2\bigr] \sqrt{r_j}
\Bigr \},\phantom{x} \label{N2phiN}
\eeqa
where the triangle function $\lambda^{1/2}$ is defined in
\eq{eq:deftrianglefunction},
$r_j\equiv {m_{\Nj}^2}/{m_{\Ni}^2}$ and $r_\phi \equiv
{m_{\phi^0}^2}/{m_{\Ni}^2}$.  The results for $\phi^0 = h^0,
H^0, A^0$ can now be obtained by using eqs.~(\ref{eq:defkuphi0}) and
(\ref{eq:defkdphi0}) in eq.~(\ref{higgs-gauginos1}).  In comparing 
\eq{N2phiN} with the original calculation in ref.~[\refcite{Gunion:1987yh}],
it is helpful to employ eqs.~(4.51) and (4.53) of ref.~[\refcite{gunhab}]. The
results agree.

%%%%%%%%%%%%%%%%%%%%%%%%%%%%%%%%%%%%%%%%%%%%%%%%%%%%%%%%%%%%%%

\subsection[$\Ni \ra Z^0 \Nj$]{$\boldsymbol{\Ni \ra Z^0 \Nj}$}
\setcounter{equation}{0}
\setcounter{figure}{0}
\setcounter{table}{0}

For this two-body decay there are two tree-level Feynman diagrams,
shown in \fig{fig:neut1toneut2Z} with the definitions of the
helicities and the momenta.
The two amplitudes are given by\footnote{When comparing with the
  4-component Feynman rule in ref.~[\refcite{HaberKane}] note that ${\cal
    O}_{ij}^{\prime\prime L}=-{\cal O}_{ij}^{\prime\prime R*}$
%    =-{\cal O}_{ji}^{\prime\prime}$
    [cf.~\eq{eq:defOLpp} above].}
\beqa
i{\cal M}_1&=& \BDneg i \frac{g}{c_W} {\cal O}_{ji}^{\prime\prime L}
              x_i \sigma^\mu x^\dagger_j \varepsilon^*_\mu\,,
\\
i{\cal M}_2&=& \BDpos i \frac{g}{c_W} {\cal O}_{ij}^{\prime\prime L}
               y^\dagger_i \sigmabar^\mu y_j \varepsilon^*_\mu\,,
\eeqa
where we have used the fifth Feynman rule of \fig{fig:CNvec} in its $-\sigma$ and $\sigmabar$
forms, and the external wave functions are
$x_i=x(\boldsymbol{\vec p}_i,\lam_i)$,
$ y^\dagger_i= y^\dagger(\boldsymbol{\vec p}_i,\lam_i)$,
$x^\dagger_j=x^\dagger(\boldsymbol{\vec k}_j,\lam_j)$,
$y_j=y(\boldsymbol{\vec k}_j,\lam_j)$, and
$\varepsilon^*_\mu = \varepsilon_\mu ({\boldsymbol{\vec k}}_Z,\lam_Z)^*$.
Noting that ${\cal O}_{ji}^{\prime\prime L} =
{\cal O}_{ij}^{\prime\prime L *}$ [see eq.~(\ref{eq:defOLpp})], and applying
eqs.~(\ref{eq:conbilsig}) and (\ref{eq:conbilsigbar}), we find that the
squared matrix element is:
\beqa
|{\cal M}|^2 &=& \frac{g^2}{c_W^2} \varepsilon^*_\mu \varepsilon_\nu
\biggl [
|{\cal O}_{ij}^{\prime\prime L}|^2
(x_i \sigma^\mu x^\dagger_j x_j \sigma^\nu x^\dagger_i +
 y^\dagger_i \sigmabar^\mu y_j  y^\dagger_j \sigmabar^\nu y_i)
\nonumber \\ &&
- \left ({\cal O}_{ij}^{\prime\prime L} \right )^2
 y^\dagger_i \sigmabar^\mu y_j x_j \sigma^\nu  x^\dagger_i
- \left ({\cal O}_{ij}^{\prime\prime L*} \right )^2
x_i \sigma^\mu  x^\dagger_j  y^\dagger_j \sigmabar^\nu y_i
\biggr ]\,.
\eeqa
Summing over the final state neutralino spin using
\eqst{xxdagsummed}{ydagxdagsummed} yields:
\beqa
\sum_{\lambda_j} |{\cal M}|^2
&=& \frac{g^2}{c_W^2} \varepsilon^*_\mu \varepsilon_\nu
\biggl [
\BDpos |{\cal O}_{ij}^{\prime\prime L}|^2
(x_i \sigma^\mu k_j \newcdot \sigmabar \sigma^\nu  x^\dagger_i +
 y^\dagger_i \sigmabar^\mu k_j \newcdot \sigma \sigmabar^\nu y_i)
\nonumber \\ &&
+ \left ({\cal O}_{ij}^{\prime\prime L} \right )^2
m_{\Nj}  y^\dagger_i \sigmabar^\mu \sigma^\nu  x^\dagger_i
+ \left ({\cal O}_{ij}^{\prime\prime L*} \right )^2
m_{\Nj} x_i \sigma^\mu \sigmabar^\nu y_i
\biggr ] .\phantom{xxx}
\eeqa
\begin{figure}[t]
\begin{center}
\begin{picture}(150,70)(0,5)
\ArrowLine(-80,40)(-30,40)
\Photon(-30,40)(10,70)54
\ArrowLine(-30,40)(10,10)
\Text(-60,54)[]{$\chi^0_i\,(p_i,\lam_i)$} 
\Text(38,12)[]{$\chi^0_j \,(k_j,\lam_j)$}
\Text(30,78)[]{$Z^0\,(k_Z,\lambda_Z)$}
\ArrowLine(170,40)(120,40)
\Photon(170,40)(210,70)54
\ArrowLine(210,10)(170,40)
\Text(123,54)[]{${\chi^{0\,\dagger}_i}\, (p_i,\lam_i)$}
\Text(210,3)[]{${\chi^{0\,\dagger}_j} \,(k_j,\lam_j)$}
\Text(211,76)[]{$Z^0\,(k_Z, \lambda_Z)$}
\end{picture}
\end{center}
\caption{The Feynman diagrams for $\stilde N_i \ra \stilde N_j
 Z^0 $ in the MSSM.}
\label{fig:neut1toneut2Z}
\end{figure}
Averaging over the initial state neutralino spin in the same way gives
\beqa
\frac{1}{2} \sum_{\lambda_i,\lambda_j} |{\cal M}|^2
&=& \frac{g^2}{2 c_W^2} \varepsilon^*_\mu \varepsilon_\nu
\biggl [
|{\cal O}_{ij}^{\prime\prime L}|^2
\Bigl ( {\rm Tr}[ \sigma^\mu k_j \newcdot \sigmabar \sigma^\nu p_i \newcdot
\sigmabar]
+
{\rm Tr}[\sigmabar^\mu k_j \newcdot \sigma \sigmabar^\nu p_i \newcdot
\sigma] \Bigr )
\nonumber \\ &&
- \left ({\cal O}_{ij}^{\prime\prime L} \right )^2
m_{\Ni} m_{\Nj} {\rm Tr} [\sigmabar^\mu \sigma^\nu ]
- \left ({\cal O}_{ij}^{\prime\prime L*} \right )^2
m_{\Ni} m_{\Nj} {\rm Tr} [\sigma^\mu \sigmabar^\nu ]
\biggr ] \nonumber
\\
&=&
\frac{2 g^2}{c_W^2} \varepsilon^*_\mu \varepsilon_\nu
\biggl \{|{\cal O}_{ij}^{\prime\prime L}|^2 \left (
k_j^\mu p_i^\nu
+ p_i^\mu k_j^\nu
- p_i \newcdot k_j \metric^{\mu\nu} \right )
\nonumber \\ &&
\BDminus
{\rm Re}\Bigl [\bigl ({\cal O}_{ij}^{\prime\prime L} \bigr )^2 \Bigr]
m_{\Ni} m_{\Nj} \metric^{\mu\nu}
\biggr \} , \phantom{xxx}
\eeqa
where in the last equality we have applied
\eqst{trssbar}{trsbarssbars}.
Using
\beq 
\sum_{\lambda_Z}\varepsilon^{\mu*}\varepsilon^\nu =
\BDneg g^{\mu\nu} + k_Z^\mu k_Z^\nu/{m_Z^2}\,,
\eeq
we obtain
\beqa
\frac{1}{2} \sum_{\lam_i,\lam_j,\lam_Z}|{\cal M}|^2
&=& \frac{2 g^2}{c_W^2}
\biggl \{ |{\cal O}_{ij}^{\prime\prime L}|^2 \left ( 
\BDpos p_i \newcdot k_j + 2 p_i \newcdot k_Z k_j \newcdot k_Z/m_Z^2 \right
)
\nonumber \\
&&
+ 3 m_{\Ni} m_{\Nj}
{\rm Re}\bigl [\bigl ({\cal O}_{ij}^{\prime\prime L} \bigr )^2 \bigr]
\biggr \}\,.
\phantom{xxx}
\eeqa
Using
$2k_j\newcdot k_Z = \BDpos m_{\Ni}^2 \BDminus m_{\Nj}^2 \BDminus m_Z^2$,
$2p_i\newcdot k_j = \BDpos m_{\Ni}^2 \BDplus m_{\Nj}^2 \BDminus m_Z^2$,
and
$2p_i\newcdot k_Z = \BDpos m_{\Ni}^2 \BDminus m_{\Nj}^2 \BDplus m_Z^2$,
we obtain the total decay width:
\beqa
&&\Gamma (\Ni \ra Z^0\Nj) = \frac{1}{16 \pi m_{\Ni}^3 }
\lambda^{1/2}\bigl(m_{\Ni}^2 ,m\ls{Z}^2, m_{\Nj}^2\bigr)
\biggl ( \frac{1}{2} \sum_{\lam_i,\lam_j,\lam_Z} |{\cal M}|^2 \biggr )
\phantom{xxxxx.}\\   
&&\quad
= \frac{g^2 m_{\Ni} }{16\pi\cw^2} \lambda^{1/2}(1,r_Z,r_j)
\biggl[|{\cal O}_{ij}^{\prime\prime L}|^2\left (1+r_j-2r_Z+(1-r_j)^2/r_Z
\right)
\nonumber \\ &&
\qquad
+6{\rm Re}\bigl[\bigl ({\cal O}_{ij}^{\prime\prime L} \bigr )^2\bigr]
\sqrt{r_j} \biggr],\phantom{xxxxx} \label{chi2chiZ}
\eeqa
where
\beq
r_j\equiv {m_{\Nj}^2}/{m_{\Ni}^2}\,,\qquad\quad
r_Z\equiv {m_{Z}^2}/{m_{\Ni}^2}\,,
\eeq
and the triangle function $\lambda^{1/2}$ is defined in
eq.~(\ref{eq:deftrianglefunction}). The result
obtained in \eq{chi2chiZ} agrees with
the original calculation in ref.~[\refcite{Gunion:1987yh}].

%%%%%%%%%%%%%%%%%%%%%%%%%%%%%%%%%%%%%%%%%%%%%%%%%%%%%%
\subsection[$e^-e^+ \ra \Ni\Nj$]{$\boldsymbol{e^-e^+ \ra \Ni\Nj}$}
\label{eechichi}
\setcounter{equation}{0}
\setcounter{figure}{0}
\setcounter{table}{0}

Next we consider the pair production of neutralinos via
$e^-e^+$ annihilation.
There are four Feynman graphs for
$s$-channel $Z^0$ exchange, shown in \fig{fig:ee2neutneut},
and four for $t$/$u$-channel selectron exchange, shown
in \fig{fig:ee2neutneut2}.
The momenta and polarizations are as labeled in the graphs.
We denote the neutralino masses as
$m_{{\widetilde N}_{i}},m_{{\widetilde N}_{j}}$ and the
selectron masses as $m_{\widetilde e_L}$ and $m_{\widetilde e_R}$. The
electron mass will again be neglected. The kinematic variables are
then given by
\beqa
s &=&
\BDpos 2p_1\newcdot p_2 = m_{\Ni}^2 + m_{\Nj}^2 \BDplus 2k_i\newcdot k_j ,
\\
t &=& m_{\Ni}^2 \BDminus  2p_1\newcdot k_i =
    m_{\Nj}^2 \BDminus  2p_2\newcdot k_j ,
\\
u &=& m_{\Ni}^2 \BDminus  2p_2\newcdot k_i =
      m_{\Nj}^2 \BDminus  2p_1\newcdot k_j .
\eeqa
\begin{figure}[!t]
\centerline{
\begin{picture}(300,210)(-170,30)
\ArrowLine(-180,225)(-152,197.5)
\ArrowLine(-152,197.5)(-180,170)
\Photon(-152,197.5)(-98,197.5)46  
\ArrowLine(-98,197.5)(-70,225)
\ArrowLine(-70,170)(-98,197.5)
\put(-180,230){$e\, (p_1,\lam_1)$}
\put(-180,161){${e}^\dagger\, (p_2,\lam_2)$}
\put(-87,230){$\chi^0_i \, (k_i,\lam_i)$}
\put(-85,157){${\chi^{0\,\dagger}_j} \,(k_j,\lam_j)$}
\put(-129,210){$Z^0$}
\ArrowLine(48,197.5)(20,225)
\ArrowLine(20,170)(48,197.5)
\Photon(48,197.5)(102,197.5)46
\ArrowLine(102,197.5)(130,225)
\ArrowLine(130,170)(102,197.5)
\put(0,230){${\bar e}^\dagger\,(p_1,\lam_1)$}
\put(0,159){$\bar e\,(p_2,\lam_2)$}
\put(90,230){$\chi^0_i\,(k_i,\lam_i)$}
\put(90,157){${\chi^{0\,\dagger}_j}\,(k_j,\lam_j)$}
\put(71,210){$Z^0$}
\ArrowLine(-180,115)(-152,87.5) 
\ArrowLine(-152,87.5)(-180,60)    
\Photon(-152,87.5)(-98,87.5)46
\ArrowLine(-70,115)(-98,87.5) 
\ArrowLine(-98,87.5)(-70,60)
\put(-180,120){$e\, (p_1,\lam_1)$}
\put(-180,51){${e}^\dagger\, (p_2,\lam_2)$}
\put(-87,120){${\chi^{0\,\dagger}_i}\, (k_i,\lam_i)$}
\put(-85,48){$\chi^0_j\, (k_j,\lam_j)$}
\put(-129,100){$Z^0$}
\ArrowLine(48,87.5)(20,115) 
\ArrowLine(20,60)(48,87.5)
\Photon(48,87.5)(102,87.5)46  
\ArrowLine(130,115)(102,87.5) 
\ArrowLine(102,87.5)(130,60)
\put(0,120){${\bar e}^\dagger\,(p_1,\lam_1)$}
\put(0,49){$\bar e\,(p_2,\lam_2)$}
\put(88,120){${\chi^{0\,\dagger}_i}\,(k_i,\lam_i)$}
\put(90,48){$\chi^0_j\,(k_j,\lam_j)$}
\put(71,100){$Z^0$}
\end{picture}
}
\caption{\label{fig:ee2neutneut} The four Feynman diagrams for $e^-e^+
\ra \Ni\Nj$ via $s$-channel $Z^0$ exchange.}
\end{figure}
%%%%%%%%%%%%%%%%%%%%%%%%%%%
\begin{figure}[tbp]
\centerline{
\begin{picture}(150,90)(0,0)
\ArrowLine(20,70)(55,70)
\ArrowLine(55,15)(20,15)
\DashArrowLine(55,70)(55,15)5
\ArrowLine(90,70)(55,70) 
\ArrowLine(55,15)(90,15) 
\put(2,78){$e\, (p_1,\lam_1)$}
\put(2,3){${e}^\dagger\, (p_2,\lam_2)$}  
\put(75,78){${\chi^{0\,\dagger}_i}\, (k_i,\lam_i)$}
\put(75,3){${\chi^0_j}\, (k_j,\lam_j)$}
\put(38,42.5){$\stilde e_L$}
\end{picture}
\hspace{0.4cm}
\begin{picture}(150,90)(0,0)
\ArrowLine(55,70)(20,70)
\ArrowLine(20,15)(55,15)
\DashArrowLine(55,15)(55,70)5
\ArrowLine(55,70)(90,70) 
\ArrowLine(90,15)(55,15) 
\put(2,78){${\bar e}^\dagger\,(p_1,\lam_1)$}
\put(2,3){$\bar e\,(p_2,\lam_2)$}  
\put(75,78){$\chi^0_i\,(k_i,\lam_i)$}
\put(75,3){${\chi^{0\,\dagger}_j}\,(k_j,\lam_j)$}
\put(38,42.5){$\stilde e_R^{\,*}$}
\end{picture}
}
\vspace{0.1cm}
\centerline{
\begin{picture}(150,100)(0,0)
\ArrowLine(20,70)(55,70)
\ArrowLine(55,15)(20,15)
\DashArrowLine(55,70)(55,15)5
\Line(55,70)(70.75,45.25) 
\ArrowLine(90,15)(74.25,39.75)  
\Line(72.5,42.5)(55,15)
\ArrowLine(72.5,42.5)(90,70)
\put(2,78){$e\,(p_1,\lam_1)$}
\put(2,3){${e}^\dagger\,(p_2,\lam_2)$}
\put(75,75){$\chi^0_i\,(k_i,\lam_i)$}
\put(75,1){${\chi^{0\,\dagger}_j}\,(k_j,\lam_j)$}
\put(38,42.5){$\stilde e_L$}
\end{picture}
\hspace{0.4cm}
\begin{picture}(150,100)(0,0)
\ArrowLine(55,70)(20,70)
\ArrowLine(20,15)(55,15)
\DashArrowLine(55,15)(55,70)5
\Line(70.75,45.25)(55,70) 
\ArrowLine(74.25,39.75)(90,15)  
\Line(55,15)(72.5,42.5)
\ArrowLine(90,70)(72.5,42.5)
\put(2,78){${\bar e}^\dagger\,(p_1,\lam_1)$}
\put(2,3){$\bar e\,(p_2,\lam_2)$}
\put(75,75){${\chi^{0\,\dagger}_i}\,(k_i,\lam_i)$}
\put(75,1){$\chi^0_j\,(k_j,\lam_j)$}
\put(38,42.5){$\stilde e_R^{\,*}$}
\end{picture}
}
\caption{\label{fig:ee2neutneut2} The four Feynman diagrams for $e^-e^+
\ra\Ni\Nj$ via $t$/$u$-channel selectron exchange.}
\end{figure}

By applying the third and fourth Feynman rules of Figure \ref{fig:SMintvertices}
and the fifth of Figure \ref{fig:CNvec},
we obtain for the sum of the $s$-channel diagrams
in \fig{fig:ee2neutneut},
\beqa
i{\cal M}_Z &=&
\frac{\BDneg i \metric^{\mu\nu}}{D_Z}
\biggl [\BDpos \frac{ig(s_W^2 -\half)}{c_W} x_1 \sigma_\mu  y^\dagger_2
\BDplus  \frac{igs_W^2}{c_W}  y^\dagger_1 \sigmabar_\mu x_2
\biggr ]
\nonumber \\ &&\qquad
\biggl [
\BDpos \frac{ig}{c_W} O_{ij}^{\prime\prime L} x^\dagger_i \sigmabar_\nu y_j
\BDminus  \frac{ig}{c_W} O_{ji}^{\prime\prime L} y_i \sigma_\nu  x^\dagger_j
\biggr ]\,,  
\phantom{xxxxx}
\label{eq:eeNNs}
\eeqa
where $O_{ij}^{\prime\prime}$ is given in eq.~(\ref{eq:defOLpp}),
and $D_Z\equiv s-m_Z^2+i\Gamma_Zm_Z$.  The
fermion spinors are denoted by $x_1 \equiv x(\boldsymbol{\vec
  p}_1,\lam_1)$, $ y^\dagger_2 \equiv  y^\dagger(\boldsymbol{\vec p}_2,\lam_2)$,
$ x^\dagger_i \equiv  x^\dagger(\boldsymbol{\vec k}_i,\lam_i)$, $y_j \equiv
y(\boldsymbol{\vec k}_j,\lam_j)$, etc.  The
matrix elements of the four diagrams have been combined by factorizing with respect to
the common boson propagator.  For the four $t$/$u$-channel
diagrams, we obtain, by applying the first two rules of \fig{fig:nqsq}:
\beqa
i {\cal M}^{(t)}_{\widetilde e_L} &=&
(-1)
\biggl [ \frac{i}{t - m_{\widetilde e_L}^2} \Bigr ]
\biggl [
\frac{ig}{\sqrt{2}} \Bigl (N_{i2}^* + \frac{s_W}{c_W} N_{i1}^* \Bigr )
\biggr ]     
\nonumber \\ &&
\biggl [
\frac{ig}{\sqrt{2}} \Bigl (N_{j2} + \frac{s_W}{c_W} N_{j1} \Bigr )
\biggr ]
x_1 y_i  y^\dagger_2  x^\dagger_j
,
\label{eq:eeNNtL}
\eeqa\beqa
i {\cal M}^{(u)}_{\widetilde e_L} &=&
%(+1)
\biggl [ \frac{i}{u - m_{\widetilde e_L}^2} \Bigr ]
\biggl [
\frac{ig}{\sqrt{2}} \Bigl (N_{j2}^* + \frac{s_W}{c_W} N_{j1}^* \Bigr )
\biggr ]
\nonumber \\ &&
\biggl [
\frac{ig}{\sqrt{2}} \Bigl (N_{i2} + \frac{s_W}{c_W} N_{i1} \Bigr )
\biggr ]
x_1 y_j  y^\dagger_2  x^\dagger_i
,
\label{eq:eeNNuL}
\eeqa\beqa
i {\cal M}^{(t)}_{\widetilde e_R} &=&
(-1)
%\biggl [ 
\frac{i}{t - m_{\widetilde e_R}^2} 
%\Bigr ]
\bigl ( 
-i\sqrt{2} g \frac{s_W}{c_W} N_{i1}
\bigr )
\bigl (
-i\sqrt{2} g \frac{s_W}{c_W} N_{j1}^*
\bigr )
%\nonumber \\ &&
 y^\dagger_1  x^\dagger_i x_2 y_j
,\phantom{xxxx}
\label{eq:eeNNtR}
\eeqa\beqa
i {\cal M}^{(u)}_{\widetilde e_R} &=&
%\biggl [ 
\frac{i}{u - m_{\widetilde e_R}^2} 
%\Bigr ]
\bigl (
-i\sqrt{2} g \frac{s_W}{c_W} N_{j1}
\bigr )
\bigl (
-i\sqrt{2} g \frac{s_W}{c_W} N_{i1}^*
\bigr )
 y^\dagger_1  x^\dagger_j x_2 y_i .\phantom{xxx}
\label{eq:eeNNuR}
\eeqa
The first factors of $(-1)$ in each of eqs.~(\ref{eq:eeNNtL})
and (\ref{eq:eeNNtR}) are present because the order of the
spinors in each case is an odd permutation of the ordering $(1,2,i,j)$
established by the $s$-channel contribution.
The other contributions have spinors in an even permutation of
that ordering.
 
The $s$-channel diagram contribution of eq.~(\ref{eq:eeNNs})
can be profitably rearranged
using the Fierz identities of
\eqs{twocompfierza}{twocompfierzb}.
Then, combining the result with the $t$/$u$-channel and $s$-channel
contributions, we have for the total:
\beq
{\cal M} =
c_1 x_1 y_j  y^\dagger_2  x^\dagger_i
+ c_2 x_1 y_i  y^\dagger_2  x^\dagger_j
+ c_3  y^\dagger_1  x^\dagger_i x_2 y_j
+ c_4  y^\dagger_1  x^\dagger_j x_2 y_i ,
\label{eq:eeNNniceform}
\eeq 
where
\beqa
c_1 &=& \frac{g^2}{c_W^2} \bigl [
(1 - 2 s_W^2) O_{ij}^{\prime\prime L}/D_Z 
\nonumber \\ &&  
- \half(c_W N_{i2} + s_W N_{i1}) (c_W N_{j2}^* + s_W N_{j1}^*)/
(u - m^2_{\widetilde e_L}) \bigr ],\phantom{xxxx}
\\
c_2 &=& \frac{g^2}{c_W^2} \bigl [
(2 s_W^2 - 1) O_{ji}^{\prime\prime L}/D_Z
\nonumber \\ &&  
+ \half(c_W N_{i2}^* + s_W N_{i1}^*) (c_W N_{j2} + s_W N_{j1})/
(t - m^2_{\widetilde e_L}) \bigr ],\phantom{xxxxx.}
\\
c_3 &=&
\frac{2 g^2 s_W^2}{c_W^2} \left [
-O_{ij}^{\prime\prime L}/D_Z
+  N_{i1} N_{j1}^*/
(t - m^2_{\widetilde e_R}) \right ]
,
\\
c_4 &=&
\frac{2 g^2 s_W^2}{c_W^2} \left [
O_{ji}^{\prime\prime L}/D_Z
-  N_{i1}^* N_{j1}/
(u - m^2_{\widetilde e_R}) \right ] .
\eeqa
Squaring the amplitude
and averaging over electron and positron spins, only terms involving
$x_1  x^\dagger_1$ or $y_1  y^\dagger_1$, and
$x_2  x^\dagger_2$ or $y_2  y^\dagger_2$ survive in
the massless electron limit.  Thus,
\beqa
\sum_{\lam_1,\lam_2} |{\cal M}|^2 &=&
\sum_{\lam_1,\lam_2} \biggl (
|c_1|^2  y^\dagger_j  x^\dagger_1 x_1 y_j x_i y_2  y^\dagger_2  x^\dagger_i
+ |c_2|^2  y^\dagger_i  x^\dagger_1 x_1 y_i x_j y_2  y^\dagger_2  x^\dagger_j
\nonumber \\ &&
\qquad
+ |c_3|^2 x_i y_1  y^\dagger_1  x^\dagger_i  y^\dagger_j  x^\dagger_2 x_2 y_j
+ |c_4|^2 x_j y_1  y^\dagger_1  x^\dagger_j  y^\dagger_i  x^\dagger_2 x_2 y_i
\nonumber \\ &&
\qquad
+ 2 {\rm Re}\bigl [c_1 c_2^*  y^\dagger_i  x^\dagger_1 x_1 y_j x_j y_2  
y^\dagger_2  x^\dagger_i
\bigr ]
\nonumber \\ &&
\qquad
+ 2 {\rm Re}\bigl [c_3 c_4^* x_j y_1  
y^\dagger_1  x^\dagger_i  y^\dagger_i  x^\dagger_2 x_2 y_j \bigr ]
\biggr )\phantom{xxxx} 
\eeqa\beqa
\phantom{xxx |{\cal M}|^2}
&=&
|c_1|^2  y^\dagger_j p_1\newcdot \sigmabar y_j\, x_i p_2\newcdot\sigma  x^\dagger_i
+ |c_2|^2  y^\dagger_i p_1\newcdot \sigmabar y_i\, x_j p_2\newcdot \sigma
 x^\dagger_j
\nonumber \\ &&
\quad
+ |c_3|^2 x_i p_1 \newcdot \sigma  x^\dagger_i\,
   y^\dagger_j p_2 \newcdot \sigmabar y_j
+ |c_4|^2 x_j p_1 \newcdot \sigma  x^\dagger_j\,
   y^\dagger_i p_2 \newcdot \sigmabar y_i
\nonumber \\ &&
\quad
+ 2 {\rm Re}\bigl [c_1 c_2^*  y^\dagger_i p_1 \newcdot \sigmabar y_j\,
                        x_j p_2 \newcdot \sigma  x^\dagger_i \bigr ]
\nonumber \\ &&
\quad
+ 2 {\rm Re}\bigl [c_3 c_4^* x_j p_1 \newcdot \sigma  x^\dagger_i \,
                         y^\dagger_i p_2 \newcdot \sigmabar y_j\bigr ]\,,
\eeqa
%where eqs.~(\ref{xxdagsummed}) and (\ref{yydagsummed}) have been
%used to do the spin sums to obtain the second equality.
after employing the results of \eqst{xxdagsummed}{ydagxdagsummed}.
  
We now perform the remaining spin sums using
\eqst{xxdagsummed}{ydagxdagsummed} again, obtaining:
\beqa
\sum_{\lam_1,\lam_2,\lam_i,\lam_j} |{\cal M}|^2 &=&
|c_1|^2 {\rm Tr}[p_1 \newcdot \sigmabar k_j \newcdot \sigma]
        {\rm Tr}[p_2 \newcdot \sigma k_i \newcdot \sigmabar]
\nonumber \\[-10pt] &&
+|c_2|^2 {\rm Tr}[p_1 \newcdot \sigmabar k_i \newcdot \sigma]
        {\rm Tr}[p_2 \newcdot \sigma k_j \newcdot \sigmabar]
\nonumber \\ &&
+|c_3|^2 {\rm Tr}[p_1 \newcdot \sigma k_i \newcdot \sigmabar]
        {\rm Tr}[p_2 \newcdot \sigmabar k_j \newcdot \sigma]
\nonumber \\ &&
+|c_4|^2 {\rm Tr}[p_1 \newcdot \sigma k_j \newcdot \sigmabar]
        {\rm Tr}[p_2 \newcdot \sigmabar k_i \newcdot \sigma]
\nonumber \\ &&
+ 2 {\rm Re}[c_1 c_2^*] m_{\Ni} m_{\Nj}
    {\rm Tr}[p_2 \newcdot \sigma p_1 \newcdot \sigmabar]
\nonumber \\ &&
+ 2 {\rm Re}[c_3 c_4^*] m_{\Ni} m_{\Nj}
    {\rm Tr}[p_1 \newcdot \sigma p_2 \newcdot \sigmabar] .
    \phantom{xxxx}
\eeqa
Applying the trace identity of eq.~(\ref{trssbar}) to this yields
\beqa
\sum_{\rm spins} |{\cal M}|^2 &=&
(|c_1|^2 + |c_4|^2) 4 p_1 \newcdot k_j\> p_2 \newcdot k_i
+ (|c_2|^2 + |c_3|^2) 4 p_1 \newcdot k_i\> p_2 \newcdot k_j
\phantom{xxx}
\nonumber \\[-6pt] &&
\quad
\BDplus 4 {\rm Re}[c_1 c_2^* + c_3 c_4^*] m_{\Ni} m_{\Nj}
p_1 \newcdot p_2
%\eeqa
\\
%\beqa
&=&
(|c_1|^2 + |c_4|^2) (u - m_{\Ni}^2)(u - m_{\Nj}^2)
\nonumber \\ &&
+ (|c_2|^2 + |c_3|^2) (t - m_{\Ni}^2)(t - m_{\Nj}^2)\phantom{xxxxxxxxxx}
\nonumber \\ &&
+ 2 {\rm Re}[c_1 c_2^* + c_3 c_4^*] m_{\Ni} m_{\Nj} s .
\label{eq:eeNNnicerform}
\eeqa
The differential cross-section then follows:
\beq
\frac{d\sigma}{dt} = \frac{1}{16\pi s^2}
\Bigl (\frac{1}{4} \sum_{\rm spins} |{\cal M}|^2 \Bigr ) .
\eeq 
This agrees with the first complete calculation presented in
ref.~[\refcite{eeNN}]. For the case of pure photino pair production, i.e.
$N_{i1}\ra c_W\delta_{i1}$ and $N_{i2} \ra s_W\delta_{i1}$ and for
degenerate selectron masses this also agrees with eq.~(E9) of the
erratum of [\refcite{HaberKane}]. 
   
Defining $\cos\theta={\boldsymbol{\hat p_1}}\newcdot{\boldsymbol{\hat
k_i}}$ (the cosine of the angle between the initial state electron and
one of the neutralinos in the center-of-momentum frame), the
Mandelstam variables $t,u$ are given by
\beqa
t &=& \frac{1}{2} \left
[m_{\Ni}^2 + m_{\Nj}^2 -s + \lam^{1/2}(s,m_{\Ni}^2,m_{\Nj}^2) \cos\theta
\right ], \label{eq:t}
\\
u &=& \frac{1}{2}
\left
[m_{\Ni}^2 + m_{\Nj}^2 -s - \lam^{1/2}(s,m_{\Ni}^2,m_{\Nj}^2) \cos\theta
\right ]\,,
\eeqa
where the triangle function $\lam^{1/2}$ is defined in
\eq{eq:deftrianglefunction}.
Taking into account the identical fermions in the final state when $i=j$,
the total cross-section is
\beq
\sigma = \frac{1}{1+\delta_{ij}} \int_{t_-}^{t_+} \frac{d\sigma}{dt} dt\,,
\eeq
where $t_-$ and $t_+$ are obtained by inserting $\cos\theta = \mp1$ in
\eq{eq:t}, respectively.

%%%%%%%%%%%%%%%%%%%%%%%%%%%%%%%%%%%%%%%%%%%%%%%%%%%%%%%%%%%%%%
\subsection[$e^-e^+ \ra \Ciminus \Cjplus$]
{$\boldsymbol{e^-e^+ \ra \Ciminus \Cjplus}$}
\setcounter{equation}{0}
\setcounter{figure}{0}
\setcounter{table}{0}

Next we consider the pair production of charginos in electron-positron
collisions. The $s$-channel Feynman diagrams are shown in
\fig{fig:ee2charchar}, where we have also introduced the notation
for the fermion momenta and polarizations.
The Mandelstam variables
are given by
\beqa
s &=&
\BDpos 2p_1\newcdot p_2 = m_{\Ci}^2 + m_{\Cj}^2 \BDplus 2k_i\newcdot k_j ,
\\
t &=& m_{\Ci}^2 \BDminus  2p_1\newcdot k_i =
    m_{\Cj}^2 \BDminus  2p_2\newcdot k_j , 
\\
u &=& m_{\Ci}^2 \BDminus  2p_2\newcdot k_i =
      m_{\Cj}^2 \BDminus  2p_1\newcdot k_j .
\eeqa
The negatively charged chargino
carries momentum and polarization $(k_i,\lam_i)$, while the
positively charged one carries $(k_j,\lam_j)$.
\begin{figure}[tb!]
\centerline{
\begin{picture}(300,230)(-155,30)
\ArrowLine(-160,225)(-132,197.5)
\ArrowLine(-132,197.5)(-160,170)
\Photon(-132,197.5)(-78,197.5)46
\ArrowLine(-78,197.5)(-50,225)
\ArrowLine(-50,170)(-78,197.5)
\put(-165,230){$e\, (p_1,\lam_1)$}
\put(-165,161){${e}^\dagger\, (p_2,\lam_2)$}
\put(-77,230){$\chi^-_i\, (k_i,\lam_i)$}
\put(-75,156){${\chi^{-\,\dagger}_j}\, (k_j,\lam_j)$}
\put(-116.5,210){$\gamma,Z^0$}
\ArrowLine(68,197.5)(40,225)
\ArrowLine(40,170)(68,197.5)
\Photon(68,197.5)(122,197.5)46
\ArrowLine(122,197.5)(150,225)
\ArrowLine(150,170)(122,197.5)
\put(25,230){${\bar e}^\dagger\, (p_1,\lam_1)$}
\put(25,159){$\bar e\, (p_2,\lam_2)$}
\put(113,230){$\chi^-_i\, (k_i,\lam_i)$}
\put(108,156){${\chi^{-\,\dagger}_j} \, (k_j,\lam_j)$}
\put(83.5,210){$\gamma,Z^0$}
\ArrowLine(-160,115)(-132,87.5) 
\ArrowLine(-132,87.5)(-160,60)  
\Photon(-132,87.5)(-78,87.5)46
\ArrowLine(-50,115)(-78,87.5) 
\ArrowLine(-78,87.5)(-50,60)
\put(-165,120){$e\, (p_1,\lam_1)$}
\put(-165,51){${e}^\dagger\, (p_2,\lam_2)$}
\put(-77,120){${\chi_i^{+\,\dagger}}\, (k_i,\lam_i)$}
\put(-75,48){$\chi^+_j\, (k_j,\lam_j)$}
\put(-116.5,100){$\gamma,Z^0$}
\ArrowLine(68,87.5)(40,115) 
\ArrowLine(40,60)(68,87.5)
\Photon(68,87.5)(122,87.5)46  
\ArrowLine(150,115)(122,87.5) 
\ArrowLine(122,87.5)(150,60)
\put(25,120){${\bar e}^\dagger\, (p_1,\lam_1)$}
\put(25,49){$\bar e\, (p_2,\lam_2)$}
\put(108,120){${\chi_i^{+\,\dagger}}\, (k_i,\lam_i)$} 
\put(113,48){$\chi_j^+\, (k_j,\lam_j)$}
\put(83.5,100){$\gamma,Z^0$}
\end{picture}
}
\caption{\label{fig:ee2charchar} Feynman diagrams for $e^-e^+
\ra \Ciminus\Cjplus$ via $s$-channel $\gamma$ and $Z^0$ exchange.}
\end{figure}

Using the first four Feynman rules of Figure \ref{fig:SMintvertices}
and the first four of Figure \ref{fig:CNvec},
the sum of the photon-exchange diagrams is given by:
\beqa
i{\cal M}_{\gamma} = \frac{\BDneg i\metric^{\mu\nu}}{s}
\bigl (
\BDneg ie \, x_1 \sigma_\mu  y^\dagger_2
\BDminus i e \,  y^\dagger_1 \sigmabar_\mu x_2
\bigr )
\bigl (
\BDpos ie\, \delta_{ij} y_i \sigma_\nu  x^\dagger_j
\BDplus ie\, \delta_{ij}  x^\dagger_i \sigmabar_\nu y_j
\bigr ),\phantom{xxxx}
\label{eq:eeCCg}
\eeqa
and the $Z$-exchange diagrams yield:
\beqa
i {\cal M}_Z=
\frac{\BDneg i \metric^{\mu\nu}}{D_Z}
\Bigl [
\BDpos \frac{ig}{c_W} (s_W^2 - \half) \, x_1 \sigma_\mu  y^\dagger_2
\BDplus \frac{igs_W^2}{c_W} \,  y^\dagger_1 \sigmabar_\mu x_2
\Bigr ]&&
\nonumber \\ 
\Bigl [
\BDneg \frac{ig}{c_W} O_{ji}^{\prime L} \, y_i \sigma_\nu  x^\dagger_j
\BDminus
\frac{ig}{c_W} O_{ji}^{\prime R} \,  x^\dagger_i \sigmabar_\nu y_j
\Bigr ],\qquad\>\>&&
\label{eq:eeCCZ}
\eeqa
where $D_Z\equiv s-m_Z^2+i\Gamma_Zm_Z$.
The $t$-channel Feynman diagram via
sneutrino exchange is shown in \fig{fig:ee2charchar2}. 
Applying the eighth rule of \fig{fig:cqsq} and its conjugate with arrows reversed, we find:
\beqa
i{\cal M}_{\widetilde \nu_e} =
(-1) \frac{i}{t-m^2_{\widetilde \nu_e}}
\bigl ( -i g V_{i1}^* x_1 y_i \bigr )
\bigl ( -i g V_{j1}  y^\dagger_2  x^\dagger_j \bigr ).
\eeqa
The Fermi-Dirac factor $(-1)$ in this equation arises because the spinors
appear in an order which is an odd permutation of the order used in all
of the $s$-channel diagram results.
\begin{figure}[t!]
\centerline{
\begin{picture}(400,70)(-300,29)
\ArrowLine(-200,105)(-125,105)
\DashArrowLine(-125,105)(-125,50)5
\ArrowLine(-125,50)(-200,50)
\ArrowLine(-50,105)(-125,105)
\ArrowLine(-125,50)(-50,50)
\put(-201,111){$e\,(p_1,\lam_1)$}  
\put(-201,56){${e}^\dagger\,(p_2,\lam_2)$}
\put(-58,111){${\chi^{+\,\dagger}_i}\,(k_i,\lam_i)$}
\put(-58,56){${\chi^{-\,\dagger}_j}\,(k_j,\lam_j)$}
\put(-144,74){$\stilde\nu_e$}
\end{picture}
}
\caption{\label{fig:ee2charchar2} The Feynman diagram for $e^-e^+
  \ra\Ciminus\Cjplus$ via the $t$-channel exchange of a
  sneutrino.}
\end{figure}

One can now apply the Fierz transformation identities
\eqst{twocompfierza}{twocompfierzc} to eqs.~(\ref{eq:eeCCg}) and
(\ref{eq:eeCCZ}) to remove the $\sigma$ and $\sigmabar$ matrices.
The result can be combined with the $t$-channel contribution to
obtain a total matrix element ${\cal M}$ with exactly the
same form as eq.~(\ref{eq:eeNNniceform}), but now with:
\beqa
c_1 &=& 2 \frac{e^2 \delta_{ij}}{s}
- \frac{g^2}{c_W^2 D_Z} (1 - 2 s_W^2) O^{\prime R}_{ji}
,
\\
c_2 &=& \frac{2 e^2 \delta_{ij}}{s}
- \frac{g^2}{c_W^2 D_Z} (1 - 2 s_W^2) O^{\prime L}_{ji}
+ \frac{g^2 V_{i1}^* V_{j1}}{t - m_{\widetilde \nu_e}^2}
,
\\
c_3 &=& \frac{2 e^2 \delta_{ij}}{s}
+ \frac{2g^2s_W^2}{c_W^2 D_Z} O^{\prime R}_{ji}\,,\\   
c_4 &=& \frac{2 e^2 \delta_{ij}}{s}
+ \frac{2g^2s_W^2}{c_W^2 D_Z} O^{\prime L}_{ji}
.
\eeqa
The rest of this calculation is identical in form to
\eqst{eq:eeNNniceform}{eq:eeNNnicerform}, so that
the result is:
\beqa
\sum_{\rm spins} |{\cal M}|^2 &=&
(|c_1|^2 + |c_4|^2) (u - m_{\Ci}^2)(u - m_{\Cj}^2)
\nonumber \\[-6pt] &&
+ (|c_2|^2 + |c_3|^2) (t - m_{\Ci}^2)(t - m_{\Cj}^2)\phantom{xxxx}
\nonumber \\ &&
\quad
+ 2 {\rm Re}[c_1 c_2^* + c_3 c_4^*] m_{\Ci} m_{\Cj} s \,.
\label{eq:eeCCnicerform}
\eeqa
The differential cross-section then follows:
\beq
\frac{d\sigma}{dt} = \frac{1}{16\pi s^2}
\biggl (\frac{1}{4} \sum_{\rm spins} |{\cal M}|^2 \biggr ) .
\eeq
As in the previous subsection, we define
$\cos\theta = {\boldsymbol{\hat p_1}}\newcdot {\boldsymbol{\hat k_i}}$
(where $\theta$ is the angle between the initial state
electron and $\widetilde C_i^-$ in the center-of-momentum frame).  The
Mandelstam variables $t,u$ are given by
\beqa
t &=& \frac{1}{2} \left
[m_{\Ci}^2 + m_{\Cj}^2 -s + \lam^{1/2}(s,m_{\Ci}^2,m_{\Cj}^2) \cos\theta
\right ], \label{eeCC-t}
\\
u &=& \frac{1}{2}
\left
[m_{\Ci}^2 + m_{\Cj}^2 -s - \lam^{1/2}(s,m_{\Ci}^2,m_{\Cj}^2) \cos\theta
\right ] .
\eeqa
The total cross-section can now be computed as
\beq
\sigma = \int_{t_-}^{t_+} \frac{d\sigma}{dt} dt\,,
\eeq
where $t_-$ and $t_+$ are obtained with $\cos\theta = -1$ and $+1$ in
eq.~(\ref{eeCC-t}), respectively. This agrees with the original
first complete calculation in ref.~[\refcite{Bartl:1985fk}]. An extended
calculation for the production of polarized charginos is given in
ref.~[\refcite{Choi:1998ut}].

%%%%%%%%%%%%%%%%%%%%%%%%%%%%%%%%%%%%%%%%%%%%%%%%%%%%%%%%%%%%%%

\subsection[$u\dbar \ra \Ciplus \Nj$]
{$\boldsymbol{u{\dbar} \ra \Ciplus \Nj}$}
\setcounter{equation}{0}
\setcounter{figure}{0}
\setcounter{table}{0}

Next we consider the associated production of a chargino and a neutralino
in quark, anti-quark collisions. The leading order Feynman diagrams are shown
in \fig{fig:ud2charneut}, where we have also
defined the momenta and the helicities. The corresponding
Mandelstam variables are
\begin{figure}[!t]
\centerline{
\begin{picture}(300,210)(-150,30)
\ArrowLine(-160,225)(-132,197.5)
\ArrowLine(-132,197.5)(-160,170)
\Photon(-132,197.5)(-78,197.5)46
\ArrowLine(-78,197.5)(-50,225)
\ArrowLine(-50,170)(-78,197.5)
\put(-160,230){$u \, (p_1,\lam_1)$}
\put(-160,158){${d}^\dagger\, (p_2,\lam_2)$}
\put(-77,230){${\chi_i^+}\, (k_i,\lam_i)$}
\put(-75,158){${\chi^{0\,\dagger}_j}\, (k_j,\lam_j)$}
\put(-109,210){$W^+$}
\ArrowLine(40,225)(68,197.5)
\ArrowLine(68,197.5)(40,170)
\Photon(68,197.5)(122,197.5)46
\ArrowLine(150,225)(122,197.5)
\ArrowLine(122,197.5)(150,170)
\put(35,230){$u\, (p_1,\lam_1)$}
\put(35,159){${d}^\dagger\, (p_2,\lam_2)$}
\put(110,233){${\chi_i^{-\,\dagger}}\,(k_i,\lam_i)$}
\put(110,155){${\chi^0_j}\,(k_j,\lam_j)$}
\put(91,210){$W^+$}
\ArrowLine(-160,115)(-105,115)  
\ArrowLine(-105,60)(-160,60)
\DashArrowLine(-105,115)(-105,60)5
\ArrowLine(-105,60)(-50,60)   
\ArrowLine(-50,115)(-105,115)
\put(-160,123){$u\, (p_1,\lam_1)$}
\put(-160,46){${d}^\dagger\, (p_2,\lam_2)$}
\put(-77,123){${\chi^{-\,\dagger}_i}\, (k_i,\lam_i)$}
\put(-75,45){${\chi^0_j}\, (k_j,\lam_j)$}
\put(-122,87.5){$\stilde d_L$}
\ArrowLine(40,115)(95,115) 
\ArrowLine(95,60)(40,60)
\DashArrowLine(95,115)(95,60)5
\ArrowLine(150,60)(125.25,84.75)
\Line(95,115)(119.75,90.25)
\ArrowLine(122.5,87.5)(150,115)
\Line(95,60)(122.5,87.5)
\put(35,120){$u\, (p_1,\lam_1)$}
\put(35,49){${d}^\dagger\, (p_2,\lam_2)$}
\put(110,120){${\chi^+_i}\, (k_i,\lam_i)$}
\put(110,46){${\chi^{0\,\dagger}_j}\, (k_j,\lam_j)$}
\put(80,87.5){$\stilde u_L$}
\end{picture}
}
\caption{\label{fig:ud2charneut} The four tree-level Feynman diagrams
for $u\dbar\ra\Ciplus\Nj$.}
\end{figure}
\beqa
s &=&
\BDpos 2p_1\newcdot p_2 = m_{\Ci}^2 + m_{\Nj}^2 \BDplus 2k_i\newcdot k_j ,
\\
t &=& m_{\Ci}^2 \BDminus  2p_1\newcdot k_i =
    m_{\Nj}^2 \BDminus  2p_2\newcdot k_j ,
\\
u &=& m_{\Ci}^2 \BDminus  2p_2\newcdot k_i =
      m_{\Nj}^2 \BDminus  2p_1\newcdot k_j .
\eeqa
The matrix elements for the $s$-channel diagrams are obtained by
applying the fifth Feynman rule of Figure \ref{fig:SMintvertices}
and the last two of Figure \ref{fig:CNvec}:
\beqa
i{\cal M}_s = \frac{\BDneg i \metric^{\mu\nu}}{s - m_W^2}
\Bigl (\BDpos \frac{ig}{\sqrt{2}} x_1 \sigma_\mu  y^\dagger_2 \Bigr )
\bigl (\BDpos ig O_{ji}^{L*}  x^\dagger_i \sigmabar_\nu y_j
\BDplus i g O^{R*}_{ji} y_i \sigma_\nu  x^\dagger_j \bigr ) .\phantom{xxxx}
\eeqa
The external spinors are denoted by
$x_1\equiv x(\boldsymbol{\vec p}_1,\lam_1)$,
$ y^\dagger_2\equiv  y^\dagger(\boldsymbol{\vec p}_2,\lam_2)$,
$ x^\dagger_i\equiv  x^\dagger(\boldsymbol{\vec k}_i,\lam_i)$,
$y_j\equiv y(\boldsymbol{\vec k}_j,\lam_j)$, etc.
The matrix elements for the $t$ and $u$ graphs follow from the
first rule of Figure \ref{fig:nqsq} and the first two of Figure \ref{fig:cqsq}:
\beqa
i{\cal M}_t &=& (-1)
\frac{i}{t - m_{\widetilde d_L}^2}
\bigl (-ig U_{i1}^* \bigr )
\Bigl (
\frac{ig}{\sqrt{2}}\bigl [N_{j2} - \frac{s_W}{3c_W} N_{j1} \bigr ]
\Bigr )
x_1 y_i  y^\dagger_2  x^\dagger_j\,,\phantom{xxx}
\label{eq:eeCNt}
\\
i{\cal M}_u &=&
\frac{i}{u - m_{\widetilde u_L}^2}
\bigl (-ig V_{i1} \bigr )
\Bigl (
\frac{ig}{\sqrt{2}}\bigl [-N_{j2}^* - \frac{s_W}{3c_W} N_{j1}^* \bigr ]
\Bigr )
x_1 y_j  y^\dagger_2  x^\dagger_i\,.
\eeqa
The first factor of $(-1)$ in eq.~(\ref{eq:eeCNt}) is required because
the order of the spinors $(1,i,2,j)$ is in an odd permutation of
the order $(1,2,i,j)$ used in the $s$-channel and
$u$-channel results.

Now we can use the Fierz relations eqs.~(\ref{twocompfierza}) and
(\ref{twocompfierzc}) to rewrite the $s$-channel amplitude in a form
without $\sigma$ or $\sigmabar$ matrices. Combining the result with
the $t$-channel and $u$-channel contributions yields a total ${\cal M}$
with exactly the same form as eq.~(\ref{eq:eeNNniceform}), but now with
\beqa
c_1 &=& -\sqrt{2} g^2 \left [\frac{O_{ji}^{L*}}{s-m_W^2} +
\left(\frac{1}{2} N_{j2}^* + \frac{s_W}{6c_W}
N_{j1}^*\right)\frac{V_{i1}}{u- m_{\widetilde
u_L}}
\right]
,
\\
c_2 &=& -\sqrt{2} g^2 \left [ \frac{O_{ji}^{R*}}{s-m_W^2} +
\left(\frac{1}{2} N_{j2}^* - \frac{s_W}{6c_W}
N_{j1}^*\right)\frac{U_{i1}^*}{t- m_{\widetilde
d_L}}
\right]
,
\\
c_3 &=& c_4 = 0.
\eeqa
The rest of this calculation is identical in form to
that of \eqst{eq:eeNNniceform}{eq:eeNNnicerform}, leading to:
\beqa
\sum_{\rm spins} |{\cal M}|^2 &=&
|c_1|^2 (u - m_{\Ci}^2)(u - m_{\Nj}^2)
+ |c_2|^2 (t - m_{\Ci}^2)(t - m_{\Nj}^2)
\nonumber \\ &&
+ 2 {\rm Re}[c_1 c_2^*] m_{\Ci} m_{\Nj} s .   
\eeqa   
{}From this, one obtains:
\beqa
\frac{d\sigma}{dt} = \frac{1}{16\pi s^2}
\biggl (\frac{1}{3\cdot4} \sum_{\rm spins} |{\cal M}|^2 \biggr),
\label{qqCN}
\eeqa
where we have included a factor of $1/3$ from the color average for
the incoming quarks.  As in the previous two subsections,
\eq{qqCN} can be expressed in terms of the
angle between the $u$ quark and the chargino in the center-of-momentum
frame, using
\beqa
t &=& \frac{1}{2} \left
[m_{\Ci}^2 + m_{\Nj}^2 -s + \lam^{1/2}(s,m_{\Ci}^2,m_{\Nj}^2) \cos\theta
\right ],
\\
u &=& \frac{1}{2}
\left
[m_{\Ci}^2 + m_{\Nj}^2 -s - \lam^{1/2}(s,m_{\Ci}^2,m_{\Nj}^2) \cos\theta
\right ] .
\eeqa  
This process occurs in proton-antiproton and proton-proton collisions,
where $\sqrt{s}$ is not fixed, and the angle $\theta$ is different
than the lab frame angle. The observable cross-section depends crucially
on experimental cuts. The result in \eq{qqCN} agrees with the  
computation in ref.~[\refcite{Beenakker:1999xh}].

%%%%%%%%%%%%%%%%%%%%%%%%%%%%%%%%%%%%%%%%%%%%%%%%%%%%%%%%%%%%%%
\subsection[Neutralino decay to photon and
Goldstino: $\stilde N_i \ra \gamma \widetilde G$]
{Neutralino decay to photon and
Goldstino: $\boldsymbol{\stilde N_i \ra \gamma \widetilde G}$}
\setcounter{equation}{0}
\setcounter{figure}{0}
\setcounter{table}{0}

\begin{figure}[tb!]
\begin{center}
\begin{picture}(300,75)(-85,5)
\ArrowLine(-90,40)(-40,40)
\Photon(-40,40)(0,70){5}{5}
\ArrowLine(-40,40)(0,10)
\Text(-74,54)[]{$\chi^0_i\,(p,\lam_{\widetilde N})$}
\Text(15,0)[]{$\stilde G \,(k_{\widetilde G},\lam_{\widetilde G})$}
\Text(15,82)[]{$\gamma\, (k_\gamma,\lambda_\gamma)$}
\ArrowLine(165,40)(115,40)
\Photon(165,40)(205,70)54
\ArrowLine(205,10)(165,40)
\Text(118,54)[]{${\chi^{0\,\dagger}_i}\, (p,\lam_{\widetilde N})$}
\Text(198,0)[]{${\stilde G}^\dagger \,(k_{\widetilde G},\lam_{\widetilde G})$}
\Text(198,82)[]{$\gamma\, (k_\gamma,\lambda_\gamma)$}
\end{picture}
\end{center}
\caption{The two Feynman diagrams for $\stilde N_i \ra \gamma\stilde G$
in supersymmetric models with a light Goldstino.}
\label{fig:neut1togammaG}
\end{figure}
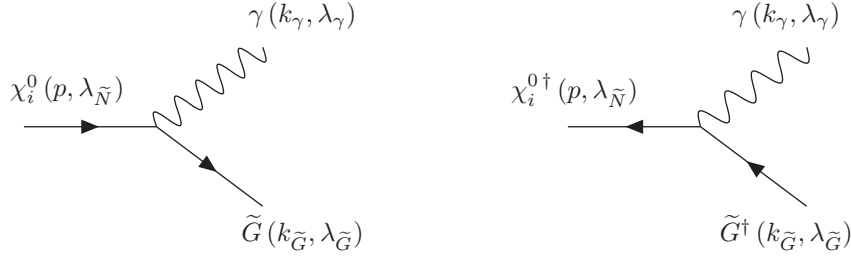

The Goldstino $\stilde G$ is a massless Weyl fermion that couples to the
neutralino and photon fields according to a
non-renormalizable Lagrangian term [\refcite{goldstinoint}]:
\beq
\mathscr{L}\, =\, \BDneg \frac{a_i}{2}
(
\chi_i^0
\sigma^\mu
\sigmabar^\rho
\sigma^\nu
\partial_\mu {\stilde G}^\dagger
) \,
( \partial_\nu A_\rho - \partial_\rho A_\nu ) + {\rm h.c.}
\eeq
Here $\chi_i^0$ is the left-handed 2-component 
fermion field that corresponds to the neutralino $\stilde N_i$ particle,
$\stilde G$ is the 2-component fermion field corresponding to the
(nearly) massless Goldstino,
and the effective coupling is
\beq
a_i \equiv \frac{1}{\sqrt{2} \langle F \rangle}
(N_{i1}^* \cos\theta_W + N_{i2}^* \sin\theta_W ) ,
\eeq
where $N_{ij}$ the mixing matrix for the neutralinos   
[see eq.~(\ref{eq:neutmix})],
and $\langle F \rangle$ is the $F$-term expectation value
associated with
supersymmetry breaking.
Therefore $\stilde N_i$ can decay to $\gamma$ plus $\stilde G$
through the diagrams shown in \fig{fig:neut1togammaG},
with amplitudes:
\beqa
i{\cal M}_{1} &=& \BDpos i \frac{a_i}{2} \>
x_{\widetilde N}
k_{\widetilde G} \newcdot \sigma
\left (
\varepsilon^* \newcdot \sigmabar \, k_\gamma \newcdot \sigma
-
k_\gamma \newcdot \sigmabar \,\varepsilon^* \newcdot \sigma
\right )  x^\dagger_{\widetilde G}
\, ,
\\
i{\cal M}_{2} &=& \BDneg i \frac{a_i^*}{2} \,
 y^\dagger_{\widetilde N}
k_{\widetilde G} \newcdot \sigmabar
\left (
\varepsilon^* \newcdot \sigma \,k_\gamma \newcdot \sigmabar
-
k_\gamma \newcdot \sigma \,\varepsilon^* \newcdot \sigmabar
\right )
y_{\widetilde G}
\, .
\eeqa
Here $x_{\widetilde N} \equiv x(\boldsymbol{\vec p},\lam_{\widetilde N})$,
$ y^\dagger_{\widetilde N} \equiv  y^\dagger (\boldsymbol{\vec p},\lam_{\widetilde N})$,
and
$ x^\dagger_{\widetilde G} \equiv
 x^\dagger (\boldsymbol{\vec k}_{\widetilde G},\lam_{\widetilde G})$,
$y_{\widetilde G} \equiv
y (\boldsymbol{\vec k}_{\widetilde G},\lam_{\widetilde G})$,
and $\varepsilon^* =
\varepsilon^* (\boldsymbol{\vec k}_{\gamma},\lam_{\gamma})$
are the external wave function factors for the neutralino, Goldstino,
and photon, respectively.

Using the on-shell condition $k_\gamma \newcdot \varepsilon^* = 0$,
we have
$
k_\gamma \newcdot \sigma \varepsilon^* \newcdot\sigmabar   
= -\varepsilon^*\newcdot\sigma k_\gamma\newcdot\sigmabar
$ and
$
k_\gamma \newcdot \sigmabar \varepsilon^* \newcdot\sigma
 =  
-\varepsilon^* \newcdot\sigmabar k_\gamma\newcdot\sigma
$ from
eqs.~(\ref{eq:ssbarsym})
and (\ref{eq:sbarssym}).
So we can rewrite the total amplitude as
\beq
{\cal M} = {\cal M}_1 + {\cal M}_2 =
x_{\widetilde N} A  x^\dagger_{\widetilde G} +  y^\dagger_{\widetilde N} B y_{\widetilde G}\,,
\eeq
where
\beqa
A &=&
\BDpos a_i \,k_{\widetilde G}\newcdot\sigma\,  \varepsilon^*\newcdot\sigmabar\,
    k_\gamma \newcdot \sigma,
\\
B &=&
\BDneg
a_i^* \, k_{\widetilde G}\newcdot\sigmabar\,  \varepsilon^*\newcdot\sigma\,
    k_\gamma \newcdot \sigmabar .
\eeqa
The complex square of the matrix element is therefore
\beqa
|{\cal M}|^2 &=&
x_{\widetilde N} A  x^\dagger_{\widetilde G} x_{\widetilde G} \hat{A}  x^\dagger_{\widetilde N}
+  y^\dagger_{\widetilde N} B y_{\widetilde G}  y^\dagger_{\widetilde G} \hat{B} y_{\widetilde N}
\nonumber \\ &&
+ x_{\widetilde N} A  x^\dagger_{\widetilde G}  y^\dagger_{\widetilde G} \hat{B} y_{\widetilde N}
+  y^\dagger_{\widetilde N} B y_{\widetilde G} x_{\widetilde G} \hat{A}  x^\dagger_{\widetilde N}
,
\phantom{xxx}
\eeqa
where $\hat{A}$ and $\hat{B}$ are obtained from $A$ and $B$
by reversing the order of the $\sigma$ and $\sigmabar$ matrices
and taking the complex conjugates of $a_i$ and $\varepsilon$
[cf.~\eq{ccspinorbilinears} and the associated text].

Summing over the Goldstino spins using \eqst{xxdagsummed}{ydagxdagsummed}
now yields:
\beqa
\sum_{\lambda_{\widetilde G}}
|{\cal M}|^2 &=&
\BDpos
x_{\widetilde N} A k_{\widetilde G} \newcdot \sigmabar \hat{A}  x^\dagger_{\widetilde N}
\BDplus
 y^\dagger_{\widetilde N} B  k_{\widetilde G} \newcdot \sigma \hat{B} y_{\widetilde N}.
\eeqa
(The $A,\hat B$ and $\hat A,B$ cross terms vanish because
of $m_{\widetilde G} = 0$.)
Averaging over the neutralino spins using
eqs.~(\ref{xxdagsummed}) and (\ref{yydagsummed}), we find
\beqa
\frac{1}{2} \sum_{\lambda_{\widetilde N}, \lambda_{\widetilde G}}
|{\cal M}|^2 &=&
\frac{1}{2}
{\rm Tr}[A k_{\widetilde G} \newcdot \sigmabar \hat{A} p \newcdot \sigmabar ]
+ \frac{1}{2}
{\rm Tr}[B  k_{\widetilde G} \newcdot \sigma \hat{B} p \newcdot \sigma]
\nonumber \\
&=&
\frac{1}{2} |a_i|^2 {\rm Tr} [
\varepsilon^* \newcdot \sigmabar\,
k_\gamma \newcdot \sigma\,
k_{\widetilde G} \newcdot \sigmabar\,
k_\gamma \newcdot \sigma\,
\varepsilon \newcdot \sigmabar\,
k_{\widetilde G} \newcdot \sigma\,
p \newcdot \sigmabar \,
k_{\widetilde G} \newcdot \sigma]
\nonumber \\ 
&&
+ (\sigma \leftrightarrow \sigmabar) .\phantom{xxx}
\eeqa
We now use
\beqa
k_\gamma \newcdot \sigma\,
k_{\widetilde G}  \newcdot \sigmabar\,
k_\gamma \newcdot \sigma\,
&=&
\BDpos 2 k_{\widetilde G} \newcdot k_\gamma\,
k_\gamma \newcdot \sigma,
\\
k_{\widetilde G} \newcdot \sigma\,
p \newcdot \sigmabar\,
k_{\widetilde G} \newcdot \sigma\,
&= &
\BDpos 2 k_{\widetilde G} \newcdot p\,
k_{\widetilde G} \newcdot \sigma ,
\eeqa
which follow from eq.~(\ref{eq:simplifyssbars}), and the corresponding
identities with $\sigma \leftrightarrow \sigmabar$,
to obtain:
\beqa
\frac{1}{2} \sum_{\lambda_{\widetilde N}, \lambda_{\widetilde G}}
|{\cal M}|^2 &=&
2 |a_i|^2 (k_{\widetilde G} \newcdot k_\gamma)
(k_{\widetilde G} \newcdot p) 
{\rm Tr} [
\varepsilon^* \newcdot \sigmabar\,
k_\gamma \newcdot \sigma\,
\varepsilon \newcdot \sigmabar\,
k_{\widetilde G} \newcdot \sigma]
\nonumber \\ &&
+ (\sigma \leftrightarrow \sigmabar) .
\eeqa
Applying the photon spin-sum identity
\beq \label{photonspinsum}
\sum_{\lambda_\gamma} \varepsilon^\mu \varepsilon^{\nu *} =
\BDneg \metric^{\mu\nu}\,,
\eeq
and the trace identities eq.~(\ref{trssbarssbar}) and (\ref{trsbarssbars}),
we get
\beqa
\frac{1}{2} \sum_{\lambda_\gamma, \lambda_{\widetilde N}, \lambda_{\widetilde G}}
|{\cal M}|^2 &=&
\BDpos 16 |a_i|^2 (k_{\widetilde G} \newcdot k_\gamma)^2
(k_{\widetilde G} \newcdot p)
= 2 |a_i|^2 m_{\widetilde N_i}^6\,.\phantom{xxxx}
\eeqa
So, the decay rate is [\refcite{NGgamma,Dimopoulos:1996vz}]:
\beqa
\Gamma(\stilde N_i \ra \gamma \GG ) &=&
\frac{1}{16 \pi m_{\stilde N_i}} \left ( \frac{1}{2}
\sum_{\lambda_\gamma, \lambda_{\widetilde N}, \lambda_{\widetilde G}}
|{\cal M}|^2\right )
\nonumber \\ 
&=& |N_{i1} \cos\theta_W + N_{i2}\sin\theta_W|^2
\frac{m_{\stilde N_i}^5}{16 \pi |\langle F \rangle|^2}
.
\eeqa

%%%%%%%%%%%%%%%%%%%%%%%%%%%%%%%%%%%%%%%%%%%%%%%%%%%%%%%%%%%%%%
\subsection[Gluino pair production from gluon fusion:  
$gg \ra \stilde g \stilde g$]{Gluino pair production from gluon fusion:
$\boldsymbol{gg \ra \stilde g \stilde g}$}
\setcounter{equation}{0}
\setcounter{figure}{0}
\setcounter{table}{0}

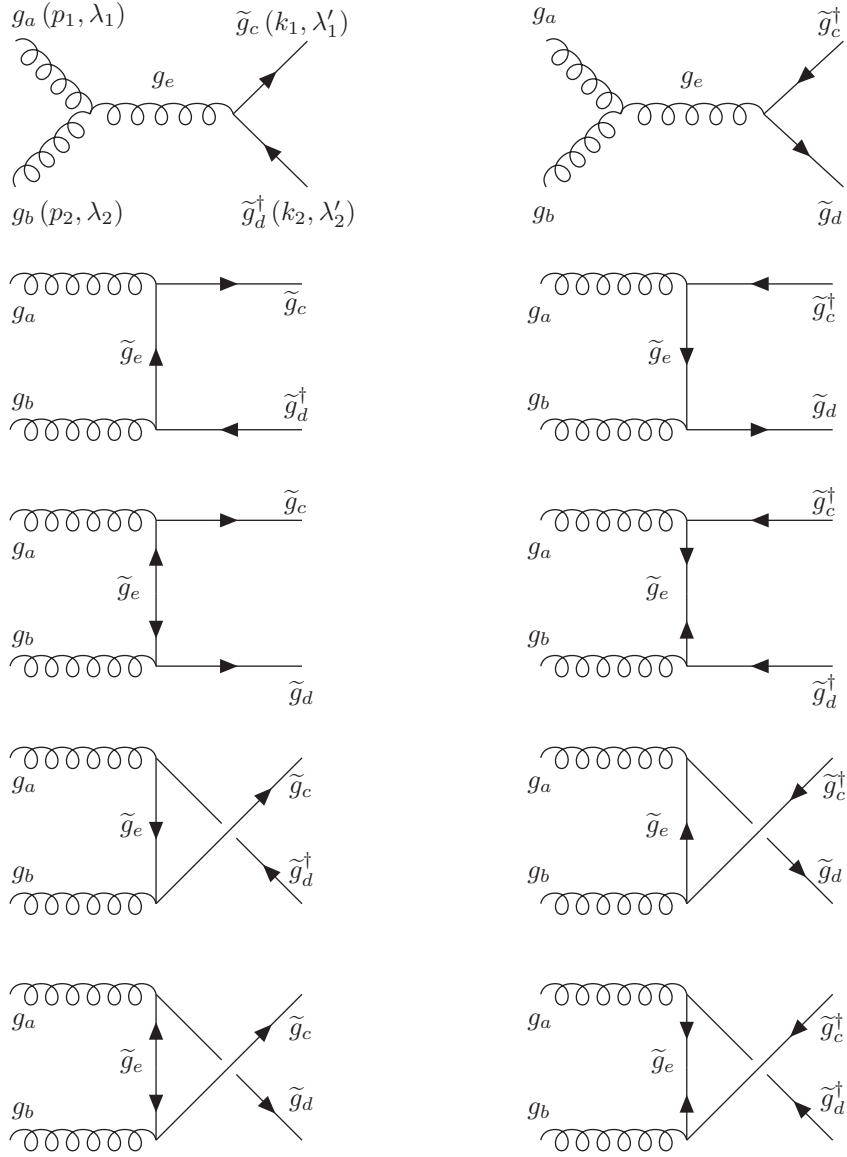
\begin{figure}[tp]
\begin{center}
\begin{picture}(300,88)(-152,157)  
\Gluon(-160,225)(-132,197.5){4}{4}
\Gluon(-160,170)(-132,197.5){4}{4}
\Gluon(-132,197.5)(-78,197.5){4}{5}
\ArrowLine(-78,197.5)(-50,225)
\ArrowLine(-50,170)(-78,197.5)
\put(-162,233){$g_a \,(p_1,\lam_{1})$}
\put(-162,158){$g_b \,(p_2,\lam_{2})$}
\put(-77,230){$\stilde g_c \,(k_1,\lam_1')$}
\put(-75,158){$\stilde g_d^\dagger \,(k_2,\lam_2')$}
\put(-109,210){$g_e$}
\Gluon(40,225)(68,197.5){4}{4}
\Gluon(40,170)(68,197.5){4}{4}
\Gluon(68,197.5)(122,197.5){4}{5}
\ArrowLine(152,225)(122,197.5)
\ArrowLine(122,197.5)(152,170)
\put(35,234){$g_a$}
\put(35,159){$g_b$}
\put(143,230){$\stilde g_c^\dagger$}
\put(143,158){$\stilde g_d$}
\put(91,210){$g_e$}
\end{picture}
\begin{picture}(300,83)(-150,165)
\Gluon(-160,225)(-105,225){4}{6}
\Gluon(-160,170)(-105,170){4}{6}
\ArrowLine(-105,170)(-105,225)
\ArrowLine(-105,225)(-50,225) 
\ArrowLine(-50,170)(-105,170) 
\put(-160,213){$g_a$}
\put(-160,180){$g_b$}
\put(-57,216){$\stilde g_c$}  
\put(-57,177){$\stilde g_d^\dagger$}
\put(-119,197.5){$\stilde g_e$}
\Gluon(40,225)(95,225){4}{6}
\Gluon(40,170)(95,170){4}{6}
\ArrowLine(150,225)(95,225)
\ArrowLine(95,225)(95,170)
\ArrowLine(95,170)(150,170)
\put(35,213){$g_a$}
\put(35,180){$g_b$}
\put(143,214){$\stilde g_c^\dagger$}
\put(143,177){$\stilde g_d$}
\put(81,197.5){$\stilde g_e$}
\end{picture}
\end{center}
\begin{center}
\begin{picture}(300,82)(-150,165)
\Gluon(-160,225)(-105,225){4}{6}
\Gluon(-160,170)(-105,170){4}{6}
\ArrowLine(-105,197.5)(-105,225)
\ArrowLine(-105,197.5)(-105,170)
\ArrowLine(-105,225)(-50,225)
\ArrowLine(-105,170)(-50,170)
\put(-160,213){$g_a$}
\put(-160,180){$g_b$}
\put(-57,230){$\stilde g_c$}
\put(-55,158){$\stilde g_d$}
\put(-120,197.5){$\stilde g_e$}
\Gluon(40,225)(95,225){4}{6}
\Gluon(40,170)(95,170){4}{6} 
\ArrowLine(95,170)(95,197.5)
\ArrowLine(95,225)(95,197.5)
\ArrowLine(150,225)(95,225)
\ArrowLine(150,170)(95,170)
\put(35,213){$g_a$}
\put(35,180){$g_b$}
\put(143,230){$\stilde g_c^\dagger$}
\put(143,158){${\stilde g_d^\dagger}$}
\put(80,197.5){$\stilde g_e$}  
\end{picture}
\end{center}
\begin{center}
\begin{picture}(300,82)(-150,55)
\Gluon(-160,115)(-105,115){4}{6}
\Gluon(-160,60)(-105,60){4}{6} 
\ArrowLine(-105,115)(-105,60)
\Line(-105,115)(-80.25,90.25)
\ArrowLine(-50,60)(-74.75,84.75)
\Line(-77.5,87.5)(-105,60)
\ArrowLine(-77.5,87.5)(-50,115)
\put(-160,104){$g_a$}
\put(-160,70){$g_b$}
\put(-55,102){${\stilde g_c}$}
\put(-55,70){${\stilde g^\dagger_d}$}
\put(-119,87.5){$\stilde g_e$}
\Gluon(40,115)(95,115){4}{6}   
\Gluon(40,60)(95,60){4}{6}
\ArrowLine(95,60)(95,115)
\ArrowLine(125.25,84.75)(150,60)
\Line(95,115)(119.75,90.25)
\ArrowLine(150,115)(122.5,87.5)
\Line(95,60)(122.5,87.5)
\put(35,104){$g_a$}
\put(35,70){$g_b$}
\put(147,102){${\stilde g^\dagger_c}$}
\put(145,70){${\stilde g_d}$}
\put(80,87.5){$\stilde g_e$}
\end{picture}
\end{center}
\begin{center}
\begin{picture}(300,82)(-150,55)
\Gluon(-160,115)(-105,115){4}{6}
\Gluon(-160,60)(-105,60){4}{6}
\ArrowLine(-105,87.5)(-105,60)  
\ArrowLine(-105,87.5)(-105,115)
\Line(-105,115)(-80.25,90.25)
\ArrowLine(-74.75,84.75)(-50,60)
\Line(-77.5,87.5)(-105,60)  
\ArrowLine(-77.5,87.5)(-50,115)
\put(-160,104){$g_a$}
\put(-160,70){$g_b$}
\put(-55,100){${\stilde g_c}$}
\put(-55,73){${\stilde g_d}$}
\put(-119,87.5){$\stilde g_e$}
\Gluon(40,115)(95,115){4}{6}
\Gluon(40,60)(95,60){4}{6}
\ArrowLine(95,60)(95,87.5)
\ArrowLine(95,115)(95,87.5)
\ArrowLine(150,60)(125.25,84.75)
\Line(95,115)(119.75,90.25)
\ArrowLine(150,115)(122.5,87.5)  
\Line(95,60)(122.5,87.5)
\put(35,104){$g_a$}
\put(35,70){$g_b$}
\put(146,100){${\stilde g^\dagger_c}$}
\put(146,73){${\stilde g^\dagger_d}$}
\put(82,87.5){$\stilde g_e$}   
\end{picture}
\end{center}
\caption{\label{fig:gg2ginogino} The ten Feynman diagrams for $gg
\ra\stilde g\stilde g$. The momentum and spin polarization assignments
are indicated on the first diagram.}
\end{figure}

In this subsection we will compute the cross-section for the process
$gg \rightarrow \stilde g \stilde g$. The relevant Feynman diagrams
are shown in \fig{fig:gg2ginogino}.
The initial state gluons have $SU(3)_c$ adjoint representation
indices $a$ and $b$, with momenta $p_1$ and $p_2$ and
polarization vectors
$\varepsilon_1^\mu = \varepsilon^\mu(\boldsymbol{\vec p}_1,\lam_1)$ and
$\varepsilon_2^\mu =
\varepsilon^\mu(\boldsymbol{\vec p}_2,\lam_2)$, respectively.
The final state gluinos carry adjoint representation
indices $c$ and $d$,
with
momenta $k_1$ and $k_2$ and wave function spinors
$ x^\dagger_1 =  x^\dagger (\boldsymbol{\vec k}_1,\lam_1')$ or
$y_1 = y (\boldsymbol{\vec k}_1,\lam_1')$ 
and
$ x^\dagger_2 =  x^\dagger (\boldsymbol{\vec k}_2,\lam_2')$ or
$y_2 = y (\boldsymbol{\vec k}_2,\lam_2')$, respectively.

The Feynman rule for the gluon coupling to gluinos
in the supersymmetric
extension of QCD was given in Figure \ref{fig:ggluinogluino}.
For the two $s$-channel amplitudes, we obtain:
\beqa
i{\cal M}_{s} &=&
\left (
-g_3 f^{abe} [ \metric_{\mu\nu} (p_1 - p_2)_\rho
+ \metric_{\nu\rho} (p_1 + 2 p_2)_\mu
- \metric_{\mu\rho} (2p_1 + p_2)_\nu ]
\right )
\nonumber \\ &&
%\times\,
\left ( \frac{\BDneg i \metric^{\rho\kappa}}{s} \right )
\varepsilon_1^\mu \varepsilon_2^\nu
\left [
(\BDneg g_3 f^{cde})\,
 x^\dagger_1 \sigmabar_\kappa y_2
\,+\, (\BDpos g_3 f^{dce})\,
y_1 \sigma_\kappa  x^\dagger_2
\right ]
.
\eeqa
The first factor is the Feynman rule for the three-gluon  
interaction of standard QCD, and the second factor is the gluon
propagator. The next four ($t$-channel) diagrams have a total
amplitude:
\beqa  
i {\cal M}_t &=&
\bigl (\BDneg g_3 f^{cea} \varepsilon_1^\mu \bigr )
\bigl (\BDneg g_3 f^{edb}\varepsilon_2^\nu \bigr )
\,
 x^\dagger_1 \sigmabar_\mu \biggl [
\frac{i (k_1 - p_1) \newcdot \sigma}{(k_1 - p_1)^2 \BDminus
m_{\widetilde g}^2}
\biggr ] \sigmabar_\nu y_2
\nonumber \\ &&
+
\bigl (\BDpos g_3 f^{eca} \varepsilon_1^\mu \bigr )
\bigl (\BDpos g_3 f^{deb}\varepsilon_2^\nu \bigr )
\,
y_1 \sigma_\mu \biggl [
\frac{i (k_1 - p_1) \newcdot \sigmabar}{(k_1 - p_1)^2 \BDminus
m_{\widetilde g}^2}
\biggr ] \sigma_\nu  x^\dagger_2
\nonumber \\ &&
+
\bigl (\BDneg g_3 f^{cea} \varepsilon_1^\mu \bigr )
\bigl (\BDpos g_3 f^{deb}\varepsilon_2^\nu \bigr )
\,
 x^\dagger_1 \sigmabar_\mu \biggl [
\frac{\BDpos i m_{\widetilde g}}{(k_1 - p_1)^2 \BDminus
m_{\widetilde g}^2}
\biggr ] \sigma_\nu  x^\dagger_2  
\nonumber \\ &&
+
\bigl (\BDpos g_3 f^{eca} \varepsilon_1^\mu \bigr )
\bigl (\BDneg g_3 f^{edb}\varepsilon_2^\nu \bigr )
\,
y_1 \sigma_\mu \biggl [
\frac{\BDpos i m_{\widetilde g}}{(k_1 - p_1)^2 \BDminus
m_{\widetilde g}^2}
\biggr ] \sigmabar_\nu y_2 .\phantom{xxxxx}
\label{eq:ggggt}
\eeqa
Finally, the $u$-channel Feynman diagrams result in:
\beqa
i {\cal M}_u &=&
\bigl (\BDneg g_3 f^{eda} \varepsilon_1^\mu \bigr )
\bigl (\BDneg g_3 f^{ceb} \varepsilon_2^\nu \bigr )
\,
 x^\dagger_1 \sigmabar_\nu \biggl [
\frac{i (k_1 - p_2) \newcdot \sigma}{(k_1 - p_2)^2 \BDminus
m_{\widetilde g}^2}
\biggr ] \sigmabar_\mu y_2
\nonumber \\ &&
+
\bigl (\BDpos g_3 f^{dea}\varepsilon_1^\mu \bigr )
\bigl (\BDpos g_3 f^{ecb} \varepsilon_2^\nu \bigr )
\,
y_1 \sigma_\nu \biggl [
\frac{i (k_1 - p_2) \newcdot \sigmabar}{(k_1 - p_2)^2 \BDminus
m_{\widetilde g}^2}
\biggr ] \sigma_\mu  x^\dagger_2
\nonumber \\ && 
+
\bigl (\BDpos g_3 f^{dea}\varepsilon_1^\mu \bigr )  
\bigl (\BDneg g_3 f^{ceb} \varepsilon_2^\nu \bigr )
\,
 x^\dagger_1 \sigmabar_\nu \biggl [
\frac{\BDpos i m_{\widetilde g}}{(k_1 - p_2)^2 \BDminus
m_{\widetilde g}^2}
\biggr ] \sigma_\mu  x^\dagger_2  
\nonumber \\ &&
+
\bigl (\BDneg g_3 f^{eda}\varepsilon_1^\mu \bigr )
\bigl (\BDpos g_3 f^{ecb} \varepsilon_2^\nu \bigr )
\,
y_1 \sigma_\nu \biggl [
\frac{\BDpos i m_{\widetilde g}}{(k_1 - p_2)^2 \BDminus
m_{\widetilde g}^2}
\biggr ] \sigmabar_\mu y_2 .\phantom{xxxxx}
\label{eq:ggggu}
\eeqa
 
We choose to work with {\em real}\, transverse polarization vectors
$\varepsilon_1$, $\varepsilon_2$.
These vectors must both be orthogonal to the
initial state collision axis in the center-of-momentum frame.
Hence,
\beqa
\varepsilon_1 \newcdot \varepsilon_1 &=&
\varepsilon_2 \newcdot \varepsilon_2 = \BDneg 1\,,
\\[-3pt]
\varepsilon_1 \newcdot p_1 &=&
\varepsilon_2 \newcdot p_1 =
\varepsilon_1 \newcdot p_2 =
\varepsilon_2 \newcdot p_2 = 0,
\\[-3pt]
\varepsilon_1 \newcdot k_2 &=& -\varepsilon_1 \newcdot k_1,
\\[-3pt]
%\qquad\qquad\quad
\varepsilon_2 \newcdot k_2 &=& -\varepsilon_2 \newcdot k_1 ,
\eeqa
for each choice of $\lambda_1, \lambda_2$.
The sums over gluon polarizations are performed using:
\beqa
\sum_{\lambda_1} \varepsilon_1^\mu \varepsilon_1^\nu
=
\sum_{\lambda_2} \varepsilon_2^\mu \varepsilon_2^\nu
=
\BDneg \metric^{\mu\nu} + {2\left (p_1^\mu p_2^\nu
+ p_2^\mu p_1^\nu \right )}/{s} .
\label{eq:ggggsumpol}
\eeqa
In QCD processes with two or more external gluons, the term
$2\left (p_1^\mu p_2^\nu+ p_2^\mu p_1^\nu \right )/s$
in \eq{eq:ggggsumpol} cannot in general be dropped [\refcite{cutler}].
This is to be contrasted to the photon polarization sum
[cf.~\eq{photonspinsum}], where this latter term
can always be neglected (due to a Ward identity of
quantum electrodynamics).

Before taking the complex square of the amplitude,  it is convenient to
rewrite the last two terms in each of
eqs.~(\ref{eq:ggggt}) and (\ref{eq:ggggu}) by using the identities
[see eq.~(\ref{onshellfour})]:
\beqa
m_{\widetilde g}  x^\dagger_1  = \BDpos y_1 (k_1\newcdot \sigma)\,,
\qquad\qquad
m_{\widetilde g} y_1  = \BDpos  x^\dagger_1 (k_1  \newcdot\sigmabar)\,.
\eeqa
Using eqs.~(\ref{eq:simplifyssbars}) and (\ref{eq:simplifysbarssbar}),
the resulting total matrix element is then
reduced to
a sum of terms that each contain exactly one $\sigma$ or $\sigmabar$
matrix.
We define convenient factors:
\beqa
G_s &\equiv& g_3^2 f^{abe} f^{cde}/s,
\\
G_t &\equiv& g_3^2 f^{ace} f^{bde}/(t - m_{\widetilde g}^2),
\\
G_u &\equiv& g_3^2 f^{ade} f^{bce}/(u - m_{\widetilde g}^2) .
\eeqa
where the usual Mandelstam variables are:
\beqa
s &=& \BDpos (p_1 + p_2)^2 = \BDpos (k_1 + k_2)^2
,
\\
t &=& \BDpos (k_1 - p_1)^2 = \BDpos (k_2 - p_2)^2
,
\\   
u &=& \BDpos (k_1 - p_2)^2 = \BDpos (k_2 - p_1)^2.
\eeqa
Then the total amplitude is  (noting that the gluon polarizations
$\varepsilon_1, \varepsilon_2$ were
chosen real):
\beqa
{\cal M} = {\cal M}_s + {\cal M}_t + {\cal M}_u
=  x^\dagger_1 a \newcdot \sigmabar y_2 +
            y_1 a^* \newcdot \sigma  x^\dagger_2,
\eeqa
where
\beqa
a^\mu &\equiv&
- (G_t + G_s) \varepsilon_1 \newcdot \varepsilon_2 \,p_1^\mu
- (G_u - G_s) \varepsilon_1 \newcdot \varepsilon_2 \,p_2^\mu
- 2 G_t k_1\newcdot \varepsilon_1 \,\varepsilon_2^\mu
\nonumber \\ &&
- 2 G_u k_1\newcdot \varepsilon_2 \,\varepsilon_1^\mu
- i \epsilon^{\mu\nu\rho\kappa}
\varepsilon_{1\nu} \varepsilon_{2\rho} (G_t p_1 - G_u p_2)_\kappa .
\label{eq:ggggdefa}
\eeqa

Squaring the amplitude using eqs.~(\ref{eq:conbilsig})
and (\ref{eq:conbilsigbar}), we get:
\beqa
|{\cal M}|^2 &=&
 x^\dagger_1 a \newcdot \sigmabar y_2
 y^\dagger_2 a^* \newcdot \sigmabar x_1  
+
y_1 a^* \newcdot \sigma  x^\dagger_2
x_2 a \newcdot \sigma  y^\dagger_1
\nonumber \\ &&
+
 x^\dagger_1 a \newcdot \sigmabar y_2
x_2 a \newcdot \sigma  y^\dagger_1
+
y_1 a^* \newcdot \sigma  x^\dagger_2
 y^\dagger_2 a^* \newcdot \sigmabar x_1 .
\phantom{xxxx} 
\eeqa
Summing over the gluino spins using
\eqst{xxdagsummed}{ydagxdagsummed}, we find:
\beqa
\sum_{\lambda'_1,\lambda'_2} |{\cal M}|^2 &=&
{\rm Tr}[
a \newcdot \sigmabar k_2 \newcdot \sigma
a^* \newcdot \sigmabar k_1 \newcdot \sigma]
+
{\rm Tr}[
a^* \newcdot \sigma k_2 \newcdot \sigmabar
a \newcdot \sigma k_1 \newcdot \sigmabar]
\nonumber \\[-9pt] &&
- m^2_{\widetilde g} {\rm Tr}[a \newcdot \sigmabar a \newcdot \sigma]
- m^2_{\widetilde g} {\rm Tr}[a^* \newcdot \sigma a^* \newcdot \sigmabar ].
\eeqa
Performing the traces with \eqst{trssbar}{trsbarssbars} then yields:
\beqa
\sum_{\lambda'_1,\lambda'_2} |{\cal M}|^2
&=&
8\, {\rm Re}[a \newcdot k_1 a^* \newcdot k_2]
- 4 a \newcdot a^* \, k_1 \newcdot k_2
\nonumber \\[-6pt] &&
- 4 i \epsilon^{\mu\nu\rho\kappa} k_{1\mu} k_{2\nu} a_\rho a_\kappa^*
\BDminus 4 m_{\widetilde g}^2 {\rm Re}[a^2]
.
\eeqa
Inserting the explicit form for $a^\mu$ [\eq{eq:ggggdefa}]
into the above result:
\beqa
\sum_{\lambda'_1,\lambda'_2} |{\cal M}|^2 &=&
2 (t - m_{\widetilde g}^2)(u - m_{\widetilde g}^2)
[(G_t + G_u)^2 
\nonumber \\[-9pt] &&
+ 4 (G_s + G_t) (G_s - G_u)
(\varepsilon_1 \newcdot \varepsilon_2)^2]
\BDplus
16 (G_t + G_u) [G_s (t-u) \phantom{xxx}
\nonumber \\ &&
+ G_t (t-m_{\widetilde g}^2)
+ G_u (u-m_{\widetilde g}^2)]
(\varepsilon_1 \newcdot \varepsilon_2)
(k_1 \newcdot \varepsilon_1)
(k_1 \newcdot \varepsilon_2)
\phantom{xx}
\nonumber \\ &&
-32 (G_t + G_u)^2
(k_1 \newcdot \varepsilon_1)^2
(k_1 \newcdot \varepsilon_2)^2 .
\eeqa
The sums over gluon polarizations can be done using
eq.~(\ref{eq:ggggsumpol}), which implies:
\beqa
&&   
\sum_{\lambda_1,\lambda_2} 1 = 4,
%\\ &&
\qquad\qquad\quad
\sum_{\lambda_1,\lambda_2} (\varepsilon_1 \newcdot \varepsilon_2)^2 = 2,
\\
&&\sum_{\lambda_1,\lambda_2}
(\varepsilon_1 \newcdot \varepsilon_2)
(k_1 \newcdot \varepsilon_1)
(k_1 \newcdot \varepsilon_2) =
\BDpos m_{\widetilde g}^2 \BDminus (t - m_{\widetilde g}^2)(u - m_{\widetilde g}^2)/s
,\phantom{xxxx}
\\
&&\sum_{\lambda_1,\lambda_2}
(k_1 \newcdot \varepsilon_1)^2
(k_1 \newcdot \varepsilon_2)^2 =
\left (
m_{\widetilde g}^2 - (t - m_{\widetilde g}^2)(u - m_{\widetilde g}^2)/s
\right )^2 .
\eeqa
Summing over colors using $f^{abe} f^{cde} f^{abe'} f^{cde'} =
2 f^{abe} f^{cde} f^{ace'} f^{bde'} = N_c^2 (N_c^2 -1) = 72$,
\beqa
&&
\sum_{\rm colors} G_s^2 = \frac{72 g_3^4}{s^2} ,
\qquad\quad\quad\!\qquad\,\,\,
\sum_{\rm colors} G_t^2 = \frac{72 g_3^4}{(t - m^2_{\widetilde g})^2} ,
\\
&&
\sum_{\rm colors} G_u^2 = \frac{72 g_3^4}{(u - m^2_{\widetilde g})^2} ,
\qquad\quad\quad\!
\sum_{\rm colors} G_s G_t = \frac{36 g_3^4}{s(t - m^2_{\widetilde g})} ,
\\
&&
\sum_{\rm colors} G_s G_u = -\frac{36 g_3^4}{s(u - m^2_{\widetilde g})} ,
\quad\quad\!
\sum_{\rm colors} G_t G_u =
\frac{36 g_3^4}{(t - m^2_{\widetilde g})(u - m^2_{\widetilde g})} .
\phantom{xxxxxxx}
\eeqa
Putting all the factors together, and averaging over the initial
state colors and spins, we have:
\beqa
\frac{d\sigma}{dt} &=&
\frac{1}{16\pi s^2}
\Biggl (
\frac{1}{64}\sum_{\rm colors} 
\frac{1}{4}\sum_{\rm spins} |{\cal M}|^2
\Biggr ) \nonumber
\\[6pt]
&=& \frac{9 \pi \alpha_s^2}{4 s^4}
\Biggl [
2 (t - m_{\widetilde g}^2)(u - m_{\widetilde g}^2)
-3 s^2 - 4 m_{\widetilde g}^2 s
\nonumber \\ &&
+ \frac{s^2(s+2 m_{\widetilde g}^2)^2}{
(t - m_{\widetilde g}^2)(u - m_{\widetilde g}^2)}
- \frac{4 m_{\widetilde g}^4 s^4}{(t -
 m_{\widetilde g}^2)^2(u - m_{\widetilde g}^2)^2}
\Biggr ] ,
\phantom{xxxx}
\eeqa
which agrees with the result of refs.~[\refcite{ggggpapers,Dawson:1983fw}] (after some
rearrangement).
In the center-of-momentum frame,
the Mandelstam variable $t$ is
related to the scattering
angle $\theta$ between an initial state gluon and a final state gluino by:
\beqa
t = m_{\widetilde g}^2 + \frac{s}{2}\Bigl (
\cos\theta \, \sqrt{1- 4 m_{\widetilde g}^2/s} - 1
\Bigr )\,.
\label{t-gggg}
\eeqa
Since the final state has identical particles, the total
cross-section can now be obtained by:
\beqa
\sigma = \frac{1}{2} \int_{t_-}^{t_+} \frac{d\sigma}{dt} dt\,,
\eeqa
where $t_\pm$ are obtained by inserting $\cos\theta = \pm 1$ into
\eq{t-gggg}.

\section{Conclusion}
\setcounter{equation}{0}
\setcounter{figure}{0}
\setcounter{table}{0}
\renewcommand{\theequation}{\arabic{section}.\arabic{equation}}
\renewcommand{\thefigure}{\arabic{section}.\arabic{figure}}
\renewcommand{\thetable}{\arabic{section}.\arabic{table}}

The preceding notes have some important omissions; there is nothing in the way of history or proper attribution, 
no derivations or proofs,
and no discussion of anomaly cancellation or other loop diagrams involving fermions. For
these and many more details see ref.~[\refcite{DHM}], on which these notes are based.

Even more glaring, of course, is the lack of mention of the current status of the
search for supersymmetry, the ostensible subject. 
There are a couple of reasons for this.
First, the experimental progress is so rapid that anything I could write would be obsolete on a time scale of weeks. Second, at this writing (April 2012),
there is nothing at all to discuss as far as hints of possible superpartner 
discovery signals. I think most 
theorists who have worked on supersymmetry extensively are surprised by this; either we have been wrong 
all along about supersymmetry at the weak scale, or else we were just too optimistic about its early 
discovery. 
If indeed there is no supersymmetry to be had within the reach of the LHC, then perhaps there will be 
some other surprises, hopefully of the kind that nobody has dreamed of yet. 
That would be a great outcome;
much-needed humility lessons for many of us older folks, and
new hard puzzles to be worked out by the young!

Personally, I remain guardedly optimistic about supersymmetry, however. 
The reason is that the hierarchy problem associated with the smallness of 
the electroweak scale is still there. The squared mass parameter of the 
Higgs field is quadratically sensitive, through radiative corrections, to 
every other larger mass scale to which it couples, directly or 
indirectly. Note that there are good hints for several such mass scales, 
besides the Planck scale. The quantization of weak hypercharge, the way 
that the fermion representations of the 
Standard Model fit into $SU(5)$ and $SO(10)$ multiplets, and the 
renormalization group running convergence of gauge couplings all hint at 
some sort of full or partial unification of forces, the scale of which 
(if it exists) must be very 
high to evade proton decay and other bounds. Of course, this might be 
just a 
coincidence, and the hierarchy problem definitely should not be viewed as 
hinging on the existence of unification. Other affirmative, and perhaps 
stronger, hints of the existence of mass scales far above the electroweak 
scale include: the presence of neutrino masses, which are most naturally 
explained with the seesaw mechanism; the puzzle of the origin of 
baryogenesis, which cannot be explained in the Standard Model alone 
because of the lack of sufficient CP violation; the solution of the 
strong CP problem, which can be explained by axions, but only if the 
Peccei-Quinn breaking scale is very high; and the existence of dark 
matter.

There is a hint of a $\sim$125 GeV Higgs boson, which is compatible with 
LHC-scale supersymmetry (even in its minimal form, if the top squarks are 
rather heavy or highly mixed). Moreover, a large range of heavier Higgs 
bosons not compatible with supersymmetry have now been ruled out by the 
LHC, if they are at all Standard-Model like. The lack of LHC hints for 
exotic physics at this writing suggests that none of the {\em other} 
theories that have been proposed to address the hierarchy problem seem to 
be in any better shape than supersymmetry is. So, my best guess is that 
the superpartners are still out there and will eventually be found at the 
LHC. Fortunately, we shall see.

\section*{Acknowledgments} I am grateful to Herbi Dreiner and Howie 
Haber, my collaborators in ref.~[\refcite{DHM}], on which these notes are 
entirely based. I thank the program co-directors Konstantin Matchev and 
Tim Tait and the local organizers Tom DeGrand and K.T.~Mahanthappa, for 
the opportunity to lecture at the beautifully organized TASI 2011, as 
well as the students for their attention and insightful questions. This 
work was supported in part by the National Science Foundation grant 
number PHY-1068369.

%%%%%%%%%%%%%%%%%%%%%%%%%%%%%%%%%%%%%%%%%%%%%%%%%%%%%%%%%%%%%%

\end{document}